\documentclass[11pt,a4paper]{article}
\usepackage{jheppub,multirow}
\usepackage{float}
\usepackage{graphicx}
\usepackage{url}
\usepackage{cancel}
\usepackage{slashed}
\usepackage{placeins}
\usepackage{feynman}
\usepackage{soul} % For strikeout, using  \st{Hellow world}
\usepackage{bm}
\usepackage{amsmath}
\usepackage{amssymb}
\usepackage{listings}
\usepackage[toc, page]{appendix}
\usepackage{color}
\usepackage{booktabs} % for tables
\usepackage{pgf}
\usepackage{empheq}
\usepackage{tikz-feynman}
\usepackage{tikz}

\usepackage{lipsum}
\usepackage{tabularx}
 \usetikzlibrary{decorations.pathmorphing}
  \usetikzlibrary{decorations.markings}
\tikzset{snake it/.style={decorate, decoration=snake}}
\usetikzlibrary{shapes.geometric, arrows}
\tikzstyle{process} = [rectangle, minimum width=3cm, minimum height=1cm, text centered, draw=black, fill=orange!30]
\tikzstyle{arrow} = [thick,->,>=stealth]
\usetikzlibrary{arrows.meta,bending}
\usepackage{subfigure}
\usepackage{pgf}
\usetikzlibrary{arrows}
\usetikzlibrary{patterns}
\renewcommand{\vec}[1]{\textbf{#1}}
 \usepackage{braket}
 \usepackage[inline]{enumitem} 
\usepackage{pifont}
\usepackage{array}
\newcolumntype{P}[1]{>{\centering\arraybackslash}p{#1}}
\usepackage{amsthm}
\usepackage{hyperref}
 \hypersetup{
    colorlinks=true,
    linkcolor=blue,
    filecolor=black,      
    urlcolor=blue,
}
\usepackage{tensor}
\usepackage{slashed} %Slash notation
\def\G1{{\bf \gamma^{(1)}_N}}

\DeclareMathOperator*{\argmin}{arg\,min}

\definecolor{violet}{cmyk}{0,1,0,0.2}
\definecolor{nnorange}{RGB}{245,199,179}
\definecolor{nngreen}{RGB}{124, 202, 154}
\definecolor{palegreen}{RGB}{187,223,209}
\definecolor{nnblue}{RGB}{77, 76, 246}

\usepackage[font=small,labelfont=bf,labelsep=period]{caption}
\usepackage{tabularx}
\newcolumntype{C}[1]{>{\centering\arraybackslash}p{#1}}

\newcommand{\be}{\begin{equation}}
\newcommand{\ee}{\end{equation}}
\newcommand{\bea}{\begin{eqnarray}}
\newcommand{\eea}{\end{eqnarray}}
\newcommand{\bi}{\begin{itemize}}
\newcommand{\ei}{\end{itemize}}
\newcommand{\ben}{\begin{enumerate}}
\newcommand{\een}{\end{enumerate}}

\newcommand{\smefit}{{\sc SMEFiT}}
\newcommand{\nnpdf}{{\sc NNPDF4.0}}

\usepackage{framed}

\numberwithin{equation}{section}
\numberwithin{figure}{section}
\numberwithin{table}{section}

\title{A critical study of the Monte Carlo replica method}
\author[a]{Mark N. Costantini,}
\author[b]{Maeve Madigan,}
\author[a]{Luca Mantani,}
\author[a]{James M. Moore}
\affiliation[a]{DAMTP, University of Cambridge, Wilberforce Road,
  Cambridge, CB3 0WA, United Kingdom}
\affiliation[b]{Institut f\"ur Theoretische Physik, Universit\"at Heidelberg, Philosophenweg 16, D-69120, Heidelberg, Germany}
\emailAdd{mnc33@cam.ac.uk}
\emailAdd{madigan@thphys.uni-heidelberg.de}
\emailAdd{luca.mantani@maths.cam.ac.uk}
\emailAdd{jmm232@cam.ac.uk}

\abstract{We present a detailed mathematical study of the Monte Carlo replica method as applied in the global fitting literature
from the high-energy physics theory community. For the first time, we provide a rigorous derivation of the parameter distributions
implied by the method, and show that, whilst they agree with Bayesian posteriors for linear models, they disagree otherwise. We proceed to numerically quantify the disagreement between the Monte Carlo replica method and the Bayesian method in the 
context of two phenomenologically relevant scenarios: fits of the SMEFT Wilson coefficients, and fits of PDFs (albeit in a toy scenario).
In both scenarios, we find that uncertainty estimates of the quantities of interest are discrepant between the two approaches when non-linearity is relevant.
Our findings motivate future investigation of Bayesian methodologies for global PDF fits, especially in the context of simultaneous determination of PDFs and SMEFT Wilson coefficients.}

\keywords{Bayesian Statistics, Global Fits, Parton Distributions, SMEFT}
\subheader{\small ...}

\begin{document}

\maketitle
\flushbottom

\section{Introduction}
Obtaining reliable interval estimates for model parameters is one of the most fundamental problems in statistics. The usual starting point is the observation of data and its modelling with a parametric theory. The objective is then to estimate a region in parameter space that has a statistical meaning. In textbook discussions, this is usually effected by obtaining \textit{confidence regions} (in frequentist statistics) or \textit{credible regions} (in Bayesian statistics).

This paper studies an inference methodology which is often compared to the Bayesian paradigm, namely the \textit{Monte Carlo (MC) replica method}. This method has been deployed in various fits of the Wilson coefficients in
the Standard Model Effective Field Theory (SMEFT)~\cite{Giani:2023gfq,Ethier:2021bye,Ethier:2021ydt,Hartland:2019bjb,Biekoetter:2018ypq}, and to fits of the parton distribution
functions (PDFs) which parametrise hadron structure~\cite{Ball:2008by, Ball:2010de, Ball:2011uy, Ball:2012cx, NNPDF:2014otw, NNPDF:2017mvq, NNPDF:2021njg, Cocuzza:2022hse, Hunt-Smith:2023sdz, Hunt-Smith:2024khs, Ball:2022uon} (and, indeed, in simultaneous extractions of PDFs and SMEFT Wilson coefficients~\cite{Carrazza:2019sec, Greljo:2021kvv, Iranipour:2022iak, Kassabov:2023hbm, Hammou:2023heg, Costantini:2024xae}).

The Monte Carlo replica method has a very intuitive basis, described in detail in Sect.~\ref{sec:multi}. Given an observation drawn from some known distribution (almost exclusively a multivariate Gaussian in the high-energy physics literature), we can use it to simulate further data samples. Each of these samples is subsequently used to estimate the parameters of a theoretical model employed for describing the data. Through the collection of estimates, one intends to propagate uncertainties from the data to the parameter space.

Despite this intuitive basis, it is important to note that - to the best of the authors' knowledge - little has been said in the literature about the mathematical foundations of the method, apart from the case of a linear model (see Ref.~\cite{DelDebbio:2021whr}). There are similarities between the Monte Carlo replica method and the \textit{parametric bootstrap} described in the mathematics literature; however, Monte Carlo seems to be a singular case of the method corresponding to an instance where a single initial observation is used to simulate the underlying distribution.\footnote{We are grateful for a discussion with the statistician Alastair Young, an expert in bootstrap methods, for clarifying this point.} The literature typically deals only with the limiting case where many initial observations of the 
same quantity are used to simulate the underlying distribution; further detail can be found in Ref.~\cite{Hall1992TheBA,Shao1996TheJA}.

On the other hand, there has been significant effort in the physics literature to numerically benchmark the Monte Carlo replica method
against other frameworks.  For example, in the context of PDF fitting, Ref.~\cite{Hunt-Smith:2022ugn}
compares the use of the Monte Carlo replica method with other methods of PDF error determination in the context of a
toy fit. Only two PDF flavours are included, and they are fitted to a single deep inelastic scattering (DIS) dataset, 
where the corresponding theory predictions are linear in the PDFs. The authors 
find perfect agreement between the Monte Carlo replica method and the other standard methods of PDF error determination
considered; however, as we shall explain in Section~\ref{sec:multi} and Section~\ref{sec:applications}, this is to be 
expected when using only DIS data, and does not necessarily generalise to the inclusion of data from proton-proton
collisions. Additionally, the benchmarking exercise between the Monte Carlo replica method and the \textit{Hessian method} for estimating
PDF uncertainties carried out in Ref.~\cite{Ball:2009zz} suggests that there is good agreement between these two approaches.
% however, it does not comment on whether either approach agrees with a fully Bayesian methodology.
 Further benchmarking will be possible given the advent of new Bayesian PDF fits, which have started appearing in the literature in recent years; see for example Ref.~\cite{Capel:2024qkm}, in which the authors present a novel parton density determination code to analyse collider data within a Bayesian framework, and Ref.~\cite{Candido:2024hjt} in which Gaussian processes for the solution of the PDF fitting inverse problem are discussed.

In contrast, in the case of benchmarking Monte Carlo versus Bayesian SMEFT fits, significant tension was observed in Ref.~\cite{Kassabov:2023hbm}. Indeed, a careful discussion in App. E of this work highlighted that the issue was methodological 
by providing a complete calculation of the Monte Carlo `posterior distribution' in a very simple toy case. To the authors' knowledge, this is the first time in the literature that disagreement was explicitly shown and explained between the two approaches.

This work aims to clarify the Monte Carlo replica method in much more generality, on two fronts. Firstly, we provide the first rigorous mathematical
derivation of the form of the `Monte Carlo posterior distribution', and we show that it is typically inequivalent to the posterior
distribution that would have been inferred from a Bayesian method. Secondly, we apply our results to further numerical
benchmarks in the high-energy physics literature, namely global fits of the SMEFT Wilson coefficients
and, separately, of the PDFs of the proton. In general, we find that interval estimates obtained from the Monte Carlo replica method do not
agree with those produced by Bayesian methods.\footnote{This has one notable exception, as we shall show in Sect.~\ref{sec:multi}, which is the case of linear models; it can be shown that Monte Carlo and Bayesian fits of a linear model coincide, provided that a sufficiently wide uniform prior is used for the Bayesian fit.}

The structure of the paper is as follows. In Sect.~\ref{sec:multi}, we introduce the Monte Carlo replica method and derive the 
`Monte Carlo posterior' distribution for model parameters, as compared to the parameter distributions that would be implied by a Bayesian method. We proceed to give some toy examples of the comparison in analytically tractable cases, including the important example of a linear model, where the Monte Carlo replica method and Bayesian method coincide in a simple way. Subsequently, in Sect.~\ref{sec:applications}, we compare the Monte Carlo replica method with a Bayesian method in the context of two phenomenologically relevant scenarios. We begin by discussing a top-sector fit of the SMEFT Wilson coefficients, displaying significant disagreement between the Wilson coefficient distributions implied by the Monte Carlo replica method and a Bayesian method. We continue by discussing fits of the PDFs of the proton using a simplified toy model for the PDFs based on linear interpolation. We begin with a fit only to deep inelastic scattering data, where the two inference methods coincide perfectly. However, we follow up with a discussion of a fit to hadronic-only (proton-proton) data, in which the PDFs enter theory predictions quadratically; here, we find significant underestimation of PDF uncertainties (at worst by around $90$\%) in the low-$x$ region when the Monte Carlo replica method is used as compared to a Bayesian method. Finally, we perform a fit using the complete DIS plus hadronic dataset, still observing an underestimation of uncertainties (at worst by around $60$\%). Instead, good agreement is found in the mid to high-$x$ region in all of the toy fits.
We summarise the results of this work in Sect.~\ref{sec:conclusions}.

\section{The mathematics of the Monte Carlo replica method}
\label{sec:multi}
In this section, the main mathematical results of our work are presented, accompanied by toy examples. The take-home message is that, in general, the Monte Carlo replica method does not reproduce Bayesian credible regions. In particular, it is quite difficult to assess a priori whether the outcome of the fit is affected by pathological behaviours or whether it approximates the Bayesian posterior distribution. The notable exception to this is the case of linear theories, in which the posterior distribution is Gaussian and well reproduced by the Monte Carlo methodology. Given this consideration, should the reader choose, they may proceed to Section~\ref{sec:applications}, where the application of these results to significant examples in high-energy physics is presented, bypassing the technical details outlined here.

We begin in Sect.~\ref{subsec:bayes_inf} by reminding the reader of a Bayesian framework for statistical inference, the paradigm with which the Monte Carlo replica method is most often compared, before introducing the Monte Carlo replica method itself. We explain how both methods are applied to obtain uncertainty estimates on theory parameters in the case that data are distributed according to a multivariate normal distribution; this is the case to which the Monte Carlo replica method is almost universally applied within the high-energy physics literature. In Sect.~\ref{subsec:mc_posterior}, we proceed to give a detailed mathematical calculation of the parameter distributions implied by the Monte Carlo replica method, contrasting them with distributions obtained from the Bayesian method. In Sect.~\ref{subsec:toy_examples}, we give a selection of toy examples, both analytic and numerical, displaying the behaviour in a range of settings.

\subsection{Bayesian vs Monte Carlo inference}
\label{subsec:bayes_inf}
In traditional statistical inference, there are two standard paradigms for constructing uncertainty estimates on parameters, the \textit{Bayesian} and \textit{frequentist} frameworks. The Bayesian approach assumes that the parameters $\vec{c}$ of interest are themselves random variables; a Bayesian \textit{$100\alpha$\% credible region} is then a region of parameter space within which the parameter has a probability $\alpha$ of lying, given some observed data. On the other hand, the frequentist approach assumes that the parameters $\vec{c}$ are fixed; a frequentist \textit{$100\alpha$\% confidence region} is a region constructed from the observed data in such a manner to ensure that if the data and confidence region were regenerated $100$ times, then $100\alpha$ of the confidence intervals would cover the true parameter value. These methods sound extremely similar at face value, but they are not equivalent in general; agreement is only guaranteed in the large sample limit by the Bernstein-von Mises theorem~\cite{Vaart_1998}.

The Bayesian approach is often used in global fits in the high-energy physics theory literature since it allows us to efficiently constrain a high-dimensional parameter space, while making use of the pre-processed information provided by experiments.  This typically consists of a single ‘measurement’ of the data, usually a central value of a measured cross-section and some uncertainty estimate (or covariance matrix in the multi-dimensional case), see for example Refs.~\cite{Giani:2023gfq,Ethier:2021bye,Ethier:2021ydt,Ellis:2020unq,Castro:2016jjv,DeBlas:2019ehy}.  
%In the frequentist framework, obtaining reliable confidence intervals typically relies on working in the large sample limit in which the profile likelihood may be used as a test statistic~\cite{Cowan:2010js}, see for example Refs.~\cite{ATL-PHYS-PUB-2022-037,CMS:2023xyc,Elmer:2023wtr}.
Further, the Monte Carlo replica method itself is most frequently compared with the Bayesian paradigm; see for instance Refs.~\cite{Giani:2023gfq,Hartland:2019bjb,DelDebbio:2021whr, Ball:2022uon}, and particularly Ref.~\cite{Hunt-Smith:2022ugn}, where the Monte Carlo replica method is described as a \textit{`frequentist approach to obtaining a Bayesian posterior'}. Therefore, in this work we shall choose to benchmark the Monte Carlo replica method exclusively against the Bayesian practice.\\

\paragraph{Bayesian method.} In order to fix notation for the discussion of the Monte Carlo replica method in the sequel, let us describe the Bayesian approach in more detail for a specific setup. Let us suppose that experimental data comprising $N_{\text{dat}}$ datapoints is distributed according to a multivariate normal distribution:\footnote{In principle, other distributions could be considered, however, in the high-energy physics literature the multivariate normal is often used when applying the Monte Carlo replica method. Further, the Monte Carlo replica method requires that we can efficiently sample from the distribution from which the data is drawn, so a normal is a natural candidate with which to begin working. Although, it is worth noting that in NNPDF and SIMUnet a \textit{truncated} multivariate normal is technically used, due to rejection of non-positive pseudodata~\cite{NNPDF:2021uiq,Costantini:2024xae}.}
\begin{equation}
\vec{d} \sim \mathcal{N}(\vec{t}(\vec{c}), \Sigma),
\label{eq:data_distribution}
\end{equation}
where $\Sigma$ is an $N_{\text{dat}} \times N_{\text{dat}}$ experimental covariance matrix (of which we assume perfect knowledge), and $\vec{t} : \mathbb{R}^{N_{\text{param}}} \rightarrow \mathbb{R}^{N_{\text{dat}}}$ is a smooth theory prediction function, taking as argument a vector of $N_{\text{param}}$ unknown theory parameters $\vec{c} \in \mathbb{R}^{N_{\text{param}}}$. Given an observation $\vec{d}_0$ of the experimental data, our aim is to recover a reliable estimate of the region in which $\vec{c}$ lies.

Since in Bayesian statistics, $\vec{c}$ itself is assumed to be a random variable, it has some associated \textit{prior probability density} $p(\vec{c})$ reflecting our knowledge of $\vec{c}$ prior to the experimental observation. Bayes' theorem then tells us that after the observation, the probability density of $\vec{c}$ given $\vec{d}_0$ is:
\begin{equation}
p(\vec{c} | \vec{d}_0) \propto p(\vec{c}) \cdot p(\vec{d}_0 | \vec{c}) = p(\vec{c}) \exp\left( - \frac{1}{2} (\vec{d}_0 - \vec{t}(\vec{c}))^T \Sigma^{-1} (\vec{d}_0 - \vec{t}(\vec{c})) \right),
\label{eq:bayesian_definition}
\end{equation}
where we have inserted the probability density for a multivariate Gaussian $p(\vec{d}_0|\vec{c})$ according to Eq.~\eqref{eq:data_distribution}. For convenience, the argument of the exponential in the posterior is usually written in terms of the \textit{$\chi^2$-statistic} evaluated on the data,
\begin{equation}
\chi^2_{\vec{d}_0}(\vec{c}) := (\vec{d}_0 - \vec{t}(\vec{c}))^T \Sigma^{-1} (\vec{d}_0 - \vec{t}(\vec{c})).
\label{eq:chi2_statistic}
\end{equation}
The Bayesian posterior distribution is used to construct uncertainty estimates for the parameter $\vec{c}$; in particular, a \textit{$100\alpha$\% credible region} is defined to be a region $R$ such that:
\begin{equation}
\label{eq:multicredible}
N\int\limits_{R} p(\vec{c}) \exp\left( -\frac{1}{2} \chi^2_{\vec{d}_0}(\vec{c}) \right) = \alpha,
\end{equation}
where $N$ is the appropriate normalisation constant in the proportionality relation Eq.~\eqref{eq:bayesian_definition}. Such regions can be constructed efficiently using numerical algorithms, including Markov Chain Monte Carlo (MCMC) methods such as Nested Sampling (NS)~\cite{Neal2011ProbabilisticIU,216920b6-7e32-3767-8d02-cac746ef1295,Ashton:2022grj,10.1063/1.1835238,10.1214/06-BA127,Feroz:2008xx}.

\paragraph{The Monte Carlo replica method.} In the Monte Carlo replica method, we begin by introducing the \textit{pseudodata distribution} $\vec{d}_p \sim \mathcal{N}(\vec{d}_0, \Sigma)$, which is intended to approximate the actual distribution from which the measurement $\vec{d}_0$ was drawn. We then define the corresponding `\textit{best-fit parameter}' values to be those which minimise the $\chi^2$-statistic evaluated on the pseudodata:\footnote{This equation does not necessarily hold exactly in practice; often, numerical implementations of the Monte Carlo replica method also use a random training-validation split when finding the minimum, together with a cross-validation stopping. We further discuss this in App.~\ref{app:training_validation_splits}.}

\begin{equation}
\vec{c}_p(\vec{d}_p) := \argmin_{\vec{c}} \chi^2_{\vec{d}_p}(\vec{c}) = \argmin_{\vec{c}}\ (\vec{d}_p - \vec{t}(\vec{c}))^T \Sigma^{-1} (\vec{d}_p - \vec{t}(\vec{c})),
\label{eq:monte_carlo_definition}
\end{equation}
essentially defining $\vec{c}_p(\vec{d}_p)$ as a function of a random variable $\vec{d}_p$. The distribution of the \textit{Monte Carlo replicas} $\vec{c}_p(\vec{d}_p)$ is now usually interpreted in the same way as the Bayesian posterior~\cite{Hunt-Smith:2022ugn,Giani:2023gfq,NNPDF:2021njg}; we shall call this distribution the \textit{Monte Carlo posterior} to make the comparison clear.

It is worth noting that, in practice, the determination of the minimiser is performed numerically; in particular, this means that the values obtained for $\vec{c}_p(\vec{d}_p)$ may depend on the numerical setup (e.g. choice of optimiser, initialisation, learning rate). This is also true for a Bayesian analysis when numerical methods are used. Ideally, we should strive to reduce this `methodological uncertainty' as much as possible, so that the uncertainty on the parameters is dominated by the information coming from the data; justification that we have indeed achieved this in our study is presented in App.~\ref{sec:numerical_mc}.

In the following section, we shall give a detailed calculation of the Monte Carlo posterior, and compare it to the Bayesian posterior.

\subsection{Calculation of the Monte Carlo posterior}
\label{subsec:mc_posterior}
Naturally, it is important to ask whether the distribution of the Monte Carlo replicas $\vec{c}_p(\vec{d}_p)$ presented in Eq.~\eqref{eq:monte_carlo_definition} is comparable to the distribution obtained via the Bayesian approach in Eq.~\eqref{eq:bayesian_definition}, since these methods are so often treated equivalently. In this section, we present a detailed calculation of the distribution of $\vec{c}_p(\vec{d}_p)$, using basic methods from probability theory, in order to facilitate this comparison.

We begin by noting that $\vec{c}_p(\vec{d}_p)$ is a function of the random variable $\vec{d}_p$, which ensures that its distribution is given by the general formula:
\begin{equation}
p(\vec{c}) \propto \int d^d\vec{d}_p\ \delta( \vec{c} - \vec{c}_p(\vec{d}_p)) \exp\left( - \frac{1}{2}(\vec{d}_p - \vec{d}_0)^T \Sigma^{-1} (\vec{d}_p - \vec{d}_0) \right),
\label{eq:initial_mc}
\end{equation}
where an integral without limits denotes an integral over the entire space (in this case, $\mathbb{R}^{N_{\text{dat}}}$). Importantly, this formula assumes that $\vec{c}_p(\vec{d}_p)$ is a single-valued function of the pseudodata $\vec{d}_p$; that is, there are not multiple equivalent minima of the $\chi^2$-statistic given in Eq.~\eqref{eq:monte_carlo_definition}. This is not a deficiency of the mathematical approach; in the case that $\vec{c}_p(\vec{d}_p)$ is only discretely multi-valued, the posterior is simply upgraded to:
\begin{equation}
p(\vec{c}) \propto  \int d^d\vec{d}_p\ \sum_{i=1}^{N_{\text{multi}}(\vec{d}_p)}\delta\left( \vec{c} - \vec{c}_p^{(i)}(\vec{d}_p)\right) \exp\left( - \frac{1}{2}(\vec{d}_p - \vec{d}_0)^T \Sigma^{-1} (\vec{d}_p - \vec{d}_0) \right),
\label{eq:multivalued_mc}
\end{equation}
where $\vec{c}_p^{(i)}(\vec{d}_p)$ is the $i$th branch of $\vec{c}_p(\vec{d}_p)$ arising from the pseudodata $\vec{d}_p$. In particular, we will need this case in one of the toy examples in the sequel. In the more serious case that $\vec{c}_p(\vec{d}_p)$ has some \textit{continuum} of multiple values for a given piece of pseudodata, we say that the problem contains a \textit{flat direction}. We discuss this in Appendix~\ref{sec:numerical_mc}. Therefore, for simplicity, we shall now proceed to assume that $N_{\text{branch}}(\vec{d}_p) \equiv 1$, safe in the knowledge that generalisation is possible. We now manipulate the integral Eq.~\eqref{eq:initial_mc} to present it in a form where it can be compared to the Bayesian posterior.

\paragraph{Coordinates on pseudodata space.} We begin by introducing a new system of coordinates on the space of pseudodata, allowing us to reduce the integral to a convenient form. Currently, the pseudodata space is parametrised directly in terms of $\vec{d}_p$, but a natural alternative parametrisation arises for a pseudodata point $\vec{d}_p$ in terms of its corresponding best-fit parameter values $\vec{c}_p$, and a vector $\pmb{\lambda}$ describing the displacement of $\vec{d}_p$ from the theory prediction of the best-fit parameter values, $\vec{t}(\vec{c}_p)$. These coordinates $(\vec{c}_p, \pmb{\lambda})$ will allow us (in the generic case) to `absorb' the delta functions in Eq.~\eqref{eq:initial_mc}.

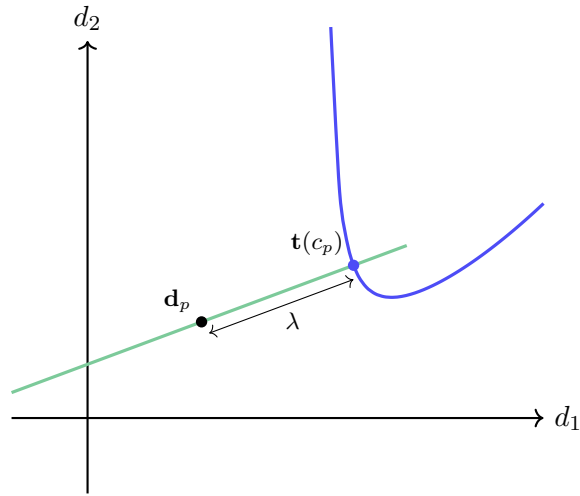
\begin{figure}
\centering
\begin{tikzpicture}
\draw[thick,->] (-1,1) -- (6,1) node[right] {$d_1$};
\draw[thick,->] (0,0) -- (0,6) node[above] {$d_2$};
\draw[very thick,color=nnblue,scale=1,domain=3.2:6,smooth,variable=\x] plot ({\x},{1 + 0.1*(\x-2)^4/((\x - 3)^2) });
\draw[very thick,color=nngreen,scale=1,domain=-1:4.2,smooth,variable=\x] plot ({\x},{0.375*(\x - 1) + 2.0875});
\draw[scale=1,domain=1.6:3.5,smooth,variable=\x,<->] plot ({\x},{0.375*(\x - 1) + 1.9});
\node[below] at (2.7,2.5375) {\footnotesize $\lambda$};
\node[circle,fill,inner sep=1.5pt] at (1.5,2.275) {};
\node[above left] at (1.5,2.275) {\footnotesize $\vec{d}_p$};
\node[color=nnblue,circle,fill,inner sep=1.5pt] at (3.5,3.025) {};
\node[above left] at (3.5,3.025) {\footnotesize $\vec{t}(c_p)$};
\end{tikzpicture}
	\caption{A visualisation of the construction of `natural' coordinates on pseudodata space, in the case where $N_{\text{dat}} = 2$ and $N_{\text{param}} = 1$, and where $\Sigma = \sigma^2 I$ is an uncorrelated covariance matrix (so that `distance' on pseudodata space coincides with Euclidean distance). In this case, the theory $\vec{t} : \mathbb{R}^2 \rightarrow \mathbb{R}$ may be viewed as constructing a parametric curve $\vec{t}(c)$ in the plane, shown in blue. The `natural' coordinates on the point $\vec{d}_p$ in pseudodata space are constructed in terms of the corresponding best-fit parameter values $c_p$, and the displacement $\vec{w} = M(c_p)\lambda$ from $\vec{t}(c_p)$ to $\vec{d}_p$, where in this case $\lambda \in \Lambda(c_p) \subseteq \mathbb{R}$; see the main text for details. The set $\Lambda(c_p)$ is indicated in the figure by the green solid line. It consists of all points whose closest point on the theory curve $\vec{t} : \mathbb{R} \rightarrow \mathbb{R}^2$ is $\vec{t}(c_p)$ (since $\Sigma = \sigma^2 I$ is Euclidean in this figure; in the general case, we mean `closest' with respect to the distance measure induced from the inner product induced by $\Sigma$). \label{fig:natural_coords}}
\end{figure}

To construct such coordinates, we begin by noting that the function $\vec{c}_p(\vec{d}_p)$ must satisfy the system of equations:
\begin{equation}
\vec{0} = \left( \frac{\partial \vec{t}}{\partial \vec{c}} \right)^T (\vec{c}_p(\vec{d}_p)) \Sigma^{-1} (\vec{d}_p - \vec{t}(\vec{c}_p(\vec{d}_p))),
\end{equation}
since $\vec{c}_p(\vec{d}_p)$ is a minimiser of the $\chi^2$-statistic evaluated on pseudodata, and hence is a stationary point of the $\chi^2$-statistic evaluated on pseudodata. In particular, this implies the relation:
\begin{equation}
\vec{d}_p = \vec{t}(\vec{c}_p(\vec{d}_p)) + \Sigma \vec{w}, \qquad \text{where} \qquad \vec{w} \in \textrm{Ker}\left( \left( \frac{\partial \vec{t}}{\partial \vec{c}}\right)^T(\vec{c}_p(\vec{d}_p)) \right).
\label{eq:pseudodata_form}
\end{equation}
That is, a \textit{necessary} condition on pseudodata to lead to the best-fit parameter values $\vec{c}_p$ is that it takes the form $\vec{d}_p = \vec{t}(\vec{c}_p) + \Sigma \vec{w}$, for some $\vec{w}$ in the above kernel. Importantly though, it is \textit{not} a \textit{sufficient} condition; we have only shown that pseudodata $\vec{d}_p = \vec{t}(\vec{c}_p) + \Sigma \vec{w}$ for a given $\vec{w}$ leads to $\vec{c}_p$ being a \textit{stationary} point of the $\chi^2$-statistic evaluated on $\vec{d}_p$ - it could be a maximum or a saddle. Thus, in order to guarantee that pseudodata of the form Eq.~\eqref{eq:pseudodata_form} leads to $\vec{c}_p(\vec{d}_p)$ as a minimum, we must further restrict to an appropriate range of $\vec{w}$.

Now, rank-nullity implies that the dimension of the kernel is at least $N_{\text{dat}} - N_{\text{param}}$,
\begin{equation}
\dim\left( \textrm{Ker}\left( \left(\frac{\partial \vec{t}}{\partial \vec{c}}\right)^T(\vec{c}_p)\right) \right) \geq N_{\text{dat}} - N_{\text{param}}.
\end{equation}
Let us write the dimension of the kernel as $N_{\perp}(\vec{c}_p)$ for short, denoting directions `orthogonal' to the theory surface. Introducing a basis $\vec{w}_1, ..., \vec{w}_{N_{\perp}(\vec{c}_p)}$ for the kernel, we can package the basis as a single matrix:
\begin{equation}
M(\vec{c}_p) := \begin{pmatrix} \vec{w}_1 & \vec{w}_2 & \cdots & \vec{w}_{N_\perp(\vec{c}_p)} \end{pmatrix},
\end{equation}
which allows us to express $\vec{d}_p$ in the form:
\begin{equation}
\vec{d}_p = \vec{t}(\vec{c}_p(\vec{d}_p)) + \Sigma M(\vec{c}_p(\vec{d}_p)) \pmb{\lambda},
\end{equation}
where $\pmb{\lambda} \in \mathbb{R}^{N_{\perp}(\vec{c}_p(\vec{d}_p))}$ is a set of coordinates for the `orthogonal' directions to the theory surface near $\vec{c}_p(\vec{d}_p)$. Recall, $\pmb{\lambda}$ must be additionally restricted to an allowed range which results in this pseudodata giving $\vec{c}_p(\vec{d}_p)$ as a \textit{minimiser} of the $\chi^2$-statistic; let us write this range as $\Lambda(\vec{c}_p(\vec{d}_p))$. 

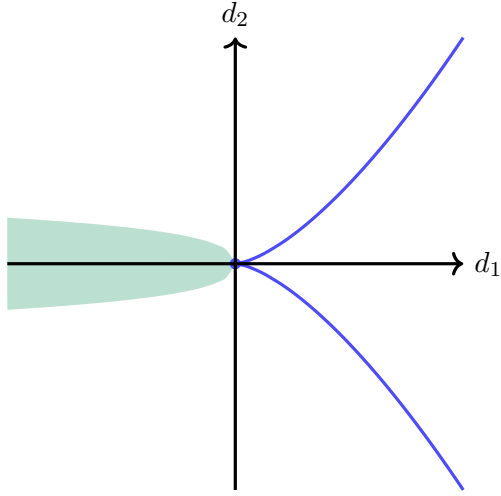
\begin{figure}
\centering
\begin{tikzpicture}
\draw[very thick,color=nnblue,domain=0:1,scale=3,smooth,variable=\x] plot ({\x},{\x^1.5) });
\draw[very thick,color=nnblue,domain=0:1,smooth,scale=3,variable=\x] plot ({\x},{-\x^1.5) });
\fill[color=palegreen,scale=1,domain=-1:-0.0001,smooth,scale=3,variable=\x] plot ({\x},{-2\x/((1-18*\x)^0.5 - 1)^0.5 + (((1-18*\x)^0.5 - 1)/9)^(3/2) + 2((1-18*\x)^0.5 - 1)^0.5 / 9}) -- (-1,0);
\fill[color=palegreen,scale=-3,domain=0.0001:1,smooth,variable=\x] plot ({\x},{-(2\x/((1+18*\x)^0.5 - 1)^0.5 - (((1+18*\x)^0.5 - 1)/9)^(3/2)  - 2((1+18*\x)^0.5 - 1)^0.5 / 9)}) -- (1,0);
\node[color=nnblue,circle,fill,inner sep=1.5pt] at (0,0) {};
\draw[very thick,->] (-3,0) -- (3,0) node[right] {$d_1$};
\draw[very thick,->] (0,-3) -- (0,3) node[above] {$d_2$};
\end{tikzpicture}
\caption{An example of a point on a theory surface $\vec{t}(c) = (c^2, c^3)^T$ (with Euclidean covariance $\Sigma = \sigma^2 I$) whose derivative does not have full rank at $c=0$. In this case, the `natural' coordinates $(c, \lambda)$ constructed on the remainder of this space do not apply at this point; instead, the required coordinates are $(\lambda_1, \lambda_2)$ with $c$ neglected (the `local parametrisation' function in this case is simply a constant, $f : \{\emptyset\} \rightarrow \mathbb{R}$ given by $f(\emptyset) = 0$). The space $\Lambda(0)$ is a two-dimensional subset of the left half-plane, indicated in green in the figure. At $c=0$, there will be a delta function singularity in the Monte Carlo posterior, which is not present in the Bayesian posterior.}
\label{fig:delta_function_cusp}
\end{figure}

We would like to state that the `natural' coordinates on pseudodata space are now $(\vec{c}_p, \pmb{\lambda})$ indicating where on the theory surface we are (the $\vec{c}_p$ coordinate), and then how far away we are in an orthogonal direction (the $\pmb{\lambda}$ coordinate). 
An example of the construction of these coordinates is shown in Fig.~\ref{fig:natural_coords}.
However, since $(\partial \vec{t}/\partial \vec{c})^T$ can have less than full rank, the number of these coordinates can exceed $N_{\text{dat}}$; near points where this Jacobian matrix has less than full rank, some of the $\vec{c}_p$ coordinates are redundant (see Fig.~\ref{fig:delta_function_cusp} for an example). Therefore, near any given $\vec{c} \in \mathbb{R}^{N_{\text{param}}}$ where the Jacobian matrix $(\partial \vec{t} / \partial \vec{c})^T(\vec{c})$ has rank $N_{\parallel}(\vec{c})$, we shall assume that we can introduce some smooth parametrisation:\footnote{Whilst this assumption may seem reasonable, there exist examples of exceptionally pathological multivariable functions which fail to have locally constant rank except at isolated points. The authors consider these unlikely to occur in a physical problem.}
\begin{equation}
\vec{f} : \mathbb{R}^{N_{\parallel}(\vec{c})} \rightarrow \mathbb{R}^{N_{\text{param}}}
\end{equation} 
which removes this redundance, allowing us to define a bijective correspondence between pseudodata $\vec{d}_p$ and coordinates $(\vec{u}, \pmb{\lambda})$ (at least locally in $\vec{u}$):
\begin{equation}
\vec{d}_p = \vec{t}(\vec{f}(\vec{u})) + \Sigma M(\vec{f}(\vec{u})) \pmb{\lambda}.
\end{equation}
In the case that $N_{\parallel}(\vec{c}) = N_{\text{param}}$, we are in the happy case where the Jacobian matrix has full rank, and we can simply take $\vec{f}(\vec{u}) = \vec{u}$, essentially using the coordinates $(\vec{c}_p, \pmb{\lambda})$ as initially desired. Otherwise, this more complicated construction is needed.

\paragraph{Change of coordinates in Eq.~\eqref{eq:initial_mc}.} Let us now consider changing coordinates in Eq.~\eqref{eq:initial_mc} near $\vec{c}$ to the `natural' system $(\vec{u}, \pmb{\lambda})$ that we defined above. The Jacobian of the transformation $\vec{d}_p \mapsto (\vec{u}, \pmb{\lambda})$ is given by:
\begin{equation}
\left| \det\left( \frac{\partial \vec{t}}{\partial \vec{c}}(\vec{f}(\vec{u})) \frac{\partial \vec{f}}{\partial \vec{u}} + \frac{\partial \left( \Sigma M \pmb{\lambda} \right)}{\partial \vec{c}} (\vec{f}(\vec{u})) \frac{\partial \vec{f}}{\partial \vec{u}} \bigg| \Sigma M(\vec{f}(\vec{u})) \right) \right|,
\end{equation}
and hence the integral Eq.~\eqref{eq:initial_mc} reduces to:
\begin{align}
&\int d^{N_\parallel(\vec{c})}\vec{u}\ \delta\left( \vec{c} - \vec{f}(\vec{u})\right) \int\limits_{\Lambda(\vec{c})} d^{N_{\perp}(\vec{c})}\pmb{\lambda} \left| \det\left( \frac{\partial \vec{t}}{\partial \vec{c}}(\vec{f}(\vec{u})) \frac{\partial \vec{f}}{\partial \vec{u}} + \frac{\partial \left( \Sigma M \pmb{\lambda} \right)}{\partial \vec{c}} (\vec{f}(\vec{u})) \frac{\partial \vec{f}}{\partial \vec{u}} \bigg| \Sigma M(\vec{f}(\vec{u})) \right) \right| \notag \\[1.5ex]
&\quad\qquad \cdot   \exp\left( - \frac{1}{2}(\vec{t}(\vec{c}) + \Sigma M(\vec{c}) \pmb{\lambda} - \vec{d}_0)^T \Sigma^{-1} (\vec{t}(\vec{c}) + \Sigma M(\vec{c}) \pmb{\lambda} - \vec{d}_0) \right).
\end{align}
Expanding the exponent, this can be manipulated into the form:
\begin{align}
&\exp\left( -\frac{1}{2} \chi^2_{\vec{d}_0}(\vec{c})\right)\notag \\[1.5ex]
&\quad\qquad\cdot \int d^{N_\parallel(\vec{c})}\vec{u}\ \delta\left( \vec{c} - \vec{f}(\vec{u})\right) \int\limits_{\Lambda(\vec{c})} d^{N_{\perp}(\vec{c})}\pmb{\lambda} \left| \det\left( \frac{\partial \vec{t}}{\partial \vec{c}}(\vec{f}(\vec{u})) \frac{\partial \vec{f}}{\partial \vec{u}} + \frac{\partial \left( \Sigma M \pmb{\lambda} \right)}{\partial \vec{c}} (\vec{f}(\vec{u})) \frac{\partial \vec{f}}{\partial \vec{u}} \bigg| \Sigma M(\vec{f}(\vec{u})) \right) \right| \notag \\[1.5ex]
&\quad\qquad \cdot   \exp\left( - \frac{1}{2} \pmb{\lambda}^T M(\vec{c})^T \Sigma M(\vec{c}) \pmb{\lambda} + \pmb{\lambda}^T M(\vec{c})^T (\vec{d}_0 - \vec{t}(\vec{c})) \right),
\label{eq:final_mc_posterior}
\end{align}
which allows comparison with the Bayesian posterior in Eq.~\eqref{eq:bayesian_definition}. We see that the approaches are in general very different: the Bayesian posterior of Eq.~\eqref{eq:bayesian_definition} and the Monte Carlo posterior above are equivalent \textit{if and only if} the prior in the Bayesian approach is chosen according to the complicated $\vec{c}$-dependent multiplicative factor featuring in the above equation.

In the case that $(\partial \vec{t} / \partial \vec{c})^T$ has full rank at $\vec{c}$, the general formula given in Eq.~\eqref{eq:final_mc_posterior} significantly simplifies. We can take $\vec{f}(\vec{u}) = \vec{u}$, and $N_{\parallel}(\vec{c}) = N_{\text{param}}$, $N_{\perp}(\vec{c}) = N_{\text{dat}} - N_{\text{param}}$, which allows us to absorb all of the delta functions in Eq.~\eqref{eq:final_mc_posterior}, reducing it to:
\begin{align}
&\exp\left( -\frac{1}{2} \chi^2_{\vec{d}_0}(\vec{c})\right) \int\limits_{\Lambda(\vec{c})} d^{N_{\text{dat}} - N_{\text{param}}}\pmb{\lambda} \left| \det\left( \frac{\partial \vec{t}}{\partial \vec{c}}  + \frac{\partial \left( \Sigma M \pmb{\lambda} \right)}{\partial \vec{c}}  \bigg| \Sigma M(\vec{c}) \right) \right| \notag \\[1.5ex]
&\quad\qquad \cdot   \exp\left( - \frac{1}{2} \pmb{\lambda}^T M(\vec{c})^T \Sigma M(\vec{c}) \pmb{\lambda} + \pmb{\lambda}^T M(\vec{c})^T (\vec{d}_0 - \vec{t}(\vec{c})) \right).
\label{eq:final_simplified_mc_posterior}
\end{align}

\paragraph{Summary.} We have shown that, in the general multivariable case for a general theory prediction, the Bayesian posterior and the Monte Carlo posterior coincide only for a very judicious choice of Bayesian prior (which seems unlikely to be motivated). Choosing such a prior, the resulting coincident posteriors have the following features:
\begin{itemize}
	\item At points $\vec{c}$ where the Jacobian matrix $(\partial \vec{t}/\partial \vec{c})^T$ has full-rank, the posterior is a `scaled' version of the Bayesian posterior obtained under the assumption of a sufficiently wide uniform prior.  Note that the overall scale factor is $\vec{c}$-dependent and may lead to shape differences.  The integral factor that multiplies the Bayesian posterior depends on the size of the region $\Lambda(\vec{c})$ of the pseudodata which gives rise to the best-fit value $\vec{c}$ in the Monte Carlo approach.

\item At points $\vec{c}$ where the Jacobian matrix $(\partial \vec{t}/\partial \vec{c})^T$ does not have full-rank, the posterior has delta function singularities. The strength of these singularities again corresponds to the size of the region $\Lambda(\vec{c})$ of the pseudodata which gives rise to the best-fit value $\vec{c}$ in the Monte Carlo approach.
\end{itemize}

\noindent In the subsequent section, we shall exhibit toy examples showcasing these phenomena, before extending to examples in the high-energy physics literature in Sect.~\ref{sec:applications}.

%\newpage
%\subsection{Special cases of the Monte Carlo posterior}
%Consider a region where the theory can be well-approximated by the linear Taylor formula:
%\begin{equation}
%\vec{t}(\vec{c}) \approx \vec{t}(\vec{c}_0) + \left( \frac{\partial \vec{t}}{\partial \vec{c}} \right)^T(\vec{c}_0) (\vec{c} - \vec{c}_0).
%\end{equation}
%Here, we have approximately constant Jacobian matrix:
%\begin{equation}
%\left( \frac{\partial \vec{t}}{\partial \vec{c}} \right)^T(\vec{c}) \approx \left( \frac{\partial \vec{t}}{\partial \vec{c}} \right)^T(\vec{c}_0),
%\end{equation}
%and hence we may take $M(\vec{c})$ independent of $\vec{c}$ in this region. 

\subsection{Toy examples: linear, quadratic and circular theories}
\label{subsec:toy_examples}

In this section, we present a selection of toy examples, demonstrating the ideas of the previous section more concretely. We give three examples: (i) a linear theory, where the Bayesian and Monte Carlo posteriors coincide in a simple way (they are coincident provided the Bayesian prior is a sufficiently wide uniform distribution); (ii) a quadratic theory of one datapoint, which exhibits the delta function behaviour discussed above; (iii) a circular theory, which does not exhibit the delta function behaviour but nonetheless demonstrates an unusual scaling behaviour because of the integral factors in Eq.~\eqref{eq:final_simplified_mc_posterior}. 

A further analytic example is given in App.~\ref{sec:extra_analytic}, namely a calculation of the Monte Carlo posterior in the case of a purely quadratic theory of multiple datapoints.

\paragraph{Example 1 - Linear theory.} The most basic example is the case of a linear theory, $\vec{t}(\vec{c}) = \vec{t}_0 + \vec{t}_{\text{lin}} \vec{c}$, where $\vec{t}_0 \in \mathbb{R}^{N_{\text{dat}}}$ and $\vec{t}_{\text{lin}}$ is an $N_{\text{dat}} \times N_{\text{param}}$ matrix. In this case, the relevant Jacobian matrix is:
\begin{equation}
\left( \frac{\partial \vec{t}}{\partial \vec{c}} \right)^T = \vec{t}_{\text{lin}}^T,
\end{equation}
which is independent of $\vec{c}$. In particular, the matrix $M(\vec{c})$ may be constructed independently of $\vec{c}$, obeying the property $M(\vec{c})^T \vec{t}_{\text{lin}} = 0$ by definition. Assuming that $\vec{t}_{\text{lin}}^T$ is of full-rank (else we in fact get a flat direction), we may use Eq.~\eqref{eq:final_simplified_mc_posterior} to calculate the Monte Carlo posterior:
\begin{align}
&\exp\left( -\frac{1}{2} \chi^2_{\vec{d}_0}(\vec{c})\right) \int\limits_{\Lambda(\vec{c})} d^{N_{\text{dat}} - N_{\text{param}}}\pmb{\lambda} \left| \det\left( \vec{t}_{\text{lin}}  \bigg| \Sigma M \right) \right|   \exp\left( - \frac{1}{2} \pmb{\lambda}^T M^T \Sigma M \pmb{\lambda} + \pmb{\lambda}^T M^T (\vec{d}_0 - \vec{t}_0) \right).
\end{align}
Observe that the $\vec{c}$-dependence has entirely dropped out from the Monte Carlo `scale factor', except for in the integration range $\Lambda(\vec{c})$. To compute this range, we must determine for which values $\pmb{\lambda}$ the pseudodata $\vec{t}(\vec{c}) + \Sigma M \pmb{\lambda}$ leads to $\vec{c}$ minimising the $\chi^2$-statistic evaluated on this pseudodata. The $\chi^2$-statistic evaluated on this pseudodata is given by:
\begin{align}
\chi^2(\vec{c}') &= \left( \vec{t}(\vec{c}') - \vec{t}(\vec{c}) - \Sigma M \pmb{\lambda}\right)^T \Sigma^{-1} \left( \vec{t}(\vec{c}') - \vec{t}(\vec{c}) - \Sigma M \pmb{\lambda} \right) \nonumber\\[1.5ex]
&= \left( \vec{t}_{\text{lin}} (\vec{c}' - \vec{c}) - \Sigma M \pmb{\lambda}\right)^T \Sigma^{-1} \left( \vec{t}_{\text{lin}} (\vec{c}' - \vec{c}) - \Sigma M \pmb{\lambda} \right)\\[1.5ex]
&= (\vec{c}' - \vec{c})^T \vec{t}_{\text{lin}}^T \Sigma^{-1} \vec{t}_{\text{lin}} (\vec{c}' - \vec{c}) + \pmb{\lambda}^T M^T \Sigma M \pmb{\lambda}. \nonumber
\end{align}
This expression is minimised by $\vec{c}'$ such that $\vec{t}_{\text{lin}}(\vec{c}' - \vec{c}) = \vec{0}$, since $\Sigma^{-1}$ is positive definite. But assuming that $\vec{t}_{\text{lin}}$ is full-rank, this has a unique solution $\vec{c}' = \vec{c}$. Thus $\Lambda(\vec{c}) = \mathbb{R}^{N_{\text{dat}} - N_{\text{param}}}$ for all values of $\vec{c}$, indicating that the Monte Carlo `scale factor' is $\vec{c}$-independent.

It follows that the Monte Carlo posterior for this linear theory is proportional to:
\begin{equation}
\exp\left( - \frac{1}{2} \chi^2_{\vec{d}_0}(\vec{c}) \right),
\end{equation}
which is simply the Bayesian posterior obtained from a sufficiently wide uniform prior. Thus we have shown: \textit{for a linear theory, the Bayesian and Monte Carlo approaches coincide} (provided that the Bayesian approach uses a sufficiently wide uniform prior). This result will be very important in the phenomenological study of PDFs in Sect.~\ref{subsec:pdf_fits}.

\paragraph{Example 2 - Quadratic theory, one datapoint.} Consider $t(c) = t_0 + t_{\text{lin}} c + t_{\text{quad}}c^2$, a quadratic theory in one parameter, with $t_{\text{quad}} > 0$. In this case, the Jacobian matrix is:
\begin{equation}
\frac{\partial t}{\partial c} = t_{\text{lin}} + 2c t_{\text{quad}},
\end{equation}
which has full rank unless $c = -t_{\text{lin}}/2t_{\text{quad}}$. Thus for all $c \neq -t_{\text{lin}}/2t_{\text{quad}}$, we can choose $M(c)$ empty, which seems to give the Monte Carlo posterior for $c \neq -t_{\text{lin}}/2t_{\text{quad}}$ as:
\begin{equation}
\left| \det\left( \frac{\partial t}{\partial c} \right) \right| \exp\left( - \frac{1}{2} \chi^2_{d_0}(c) \right)  = |2 c t_{\text{quad}} + t_{\text{lin}}| \exp\left( - \frac{1}{2} \chi^2_{d_0}(c) \right) .
\end{equation}
However, it is important to observe that in this case, the function $c_p(d_p)$ is in fact multi-valued; we can see this by writing the theory prediction as:
\begin{equation}
t(c) = t_{\text{quad}} \left( c + \frac{t_{\text{lin}}}{2 t_{\text{quad}}} \right)^2 + t_0 - \frac{t_{\text{lin}}^2}{4t_{\text{quad}}}.
\label{eq:one_datapoint_complete_square}
\end{equation}
This shows that there is a symmetry in the parameter $c$ about $c = - t_{\text{lin}}/2t_{\text{quad}}$; two equivalent values of $c$ give the same prediction provided that $c \neq -t_{\text{lin}}/2t_{\text{quad}}$. Thus we must additionally include a factor of $2$ in the Monte Carlo posterior accounting for the multi-valuedness; this follows from the more general form of the posterior given in Eq.~\eqref{eq:multivalued_mc}.

In the remaining case, $c = -t_{\text{lin}}/2t_{\text{quad}}$, we take $M(-t_{\text{lin}}/2t_{\text{quad}}) = 1$, and parametrise the lower rank surface with $f : \{\emptyset\} \rightarrow \mathbb{R}$ by $f(\emptyset) = -t_{\text{lin}}/2t_{\text{quad}}$. Then, the Monte Carlo posterior about $c = -t_{\text{lin}}/2t_{\text{quad}}$ takes the form:
\begin{equation}
\exp\left( -\frac{1}{2}\chi^2_{d_0}(c)\right) \delta\left( c + \frac{t_{\text{lin}}}{2t_{\text{quad}}}\right) \int\limits_{\Lambda(-t_{\text{lin}}/2t_{\text{quad}})} d\lambda \ \sigma^2 \exp\left( -\frac{1}{2} \sigma^2 \lambda^2 + \lambda (d_0 - t(c)) \right),
\end{equation}
where $\sigma^2$ is the one-by-one covariance matrix. The set $\Lambda(-t_{\text{lin}}/2t_{\text{quad}})$ consists of all $\lambda$ such that the pseudodata $t(-t_{\text{lin}}/2t_{\text{quad}}) + \sigma^2\lambda$ leads to $-t_{\text{lin}}/2t_{\text{quad}}$ as a minimiser of the $\chi^2$-statistic on this pseudodata. We note that the $\chi^2$-statistic on this pseudodata is given by:
\begin{align}
\chi^2(c') &= \frac{1}{\sigma^2}\left( t(c') - t\left( -\frac{t_{\text{lin}}}{t_{\text{quad}}} \right) - \sigma^2 \lambda \right)^2\\[1.5ex]
&= \frac{1}{\sigma^2} \left( t_{\text{quad}} \left( c' + \frac{t_{\text{lin}}}{2t_{\text{quad}}}\right)^2 - \sigma^2 \lambda \right)^2,
\end{align}
using the form of the theory prediction given in Eq.~\eqref{eq:one_datapoint_complete_square}. If $\lambda > 0$, then we can solve the quadratic in the bracket to obtain a value of $c'$ different from $-t_{\text{lin}}/2t_{\text{quad}}$ which gives a $\chi^2$-statistic of zero to the pseudodata. Otherwise, if $\lambda \leq 0$, the minimiser is given by $c' = -t_{\text{lin}}/2t_{\text{quad}}$. Hence we see that $\Lambda(-t_{\text{lin}}/2t_{\text{quad}}) = (-\infty, 0]$. 

Putting everything together, it follows that the complete Monte Carlo posterior is given by:
\begin{align}
\exp\left( -\frac{1}{2}\chi^2_{d_0}(c) \right) \left[ \delta\left( c + \frac{t_{\text{lin}}}{2t_{\text{quad}}}\right) \int\limits_{-\infty}^{0} d\lambda \ \sigma^2 \exp\left( -\frac{1}{2} \sigma^2 \lambda^2 + \lambda (d_0 - t(c)) \right) + 2|2 c t_{\text{quad}} + t_{\text{lin}}|\right],
\end{align}
which (through an appropriate substitution in the integral) matches the formula given in Appendix E of Ref.~\cite{Kassabov:2023hbm}, where a very brief discussion of the statistical validity of the Monte Carlo method was first presented.\footnote{It should be noted that Eq. (E.11) in the reference Ref.~\cite{Kassabov:2023hbm} actually contains an \textit{erratum}, namely the division by $|2ct_{\text{quad}} + t_{\text{lin}}|$ in the second term of the equation should actually be a multiplication, as presented here.}

Importantly, this demonstrates a singular behaviour of the Monte Carlo posterior at the point $c = -t_{\text{lin}}/2t_{\text{quad}}$ as compared to the Bayesian posterior. Indeed, this can result in a significant bias in the central value and a significant underestimation of the uncertainties for the parameter $c$, since the posterior is highly concentrated around $c = -t_{\text{lin}}/2t_{\text{quad}}$. This phenomenon is showcased in the phenomenologically relevant cases of the SMEFT in Sect.~\ref{subsec:smeft_fits} and of PDFs in Sect.~\ref{subsec:pdf_fits}.

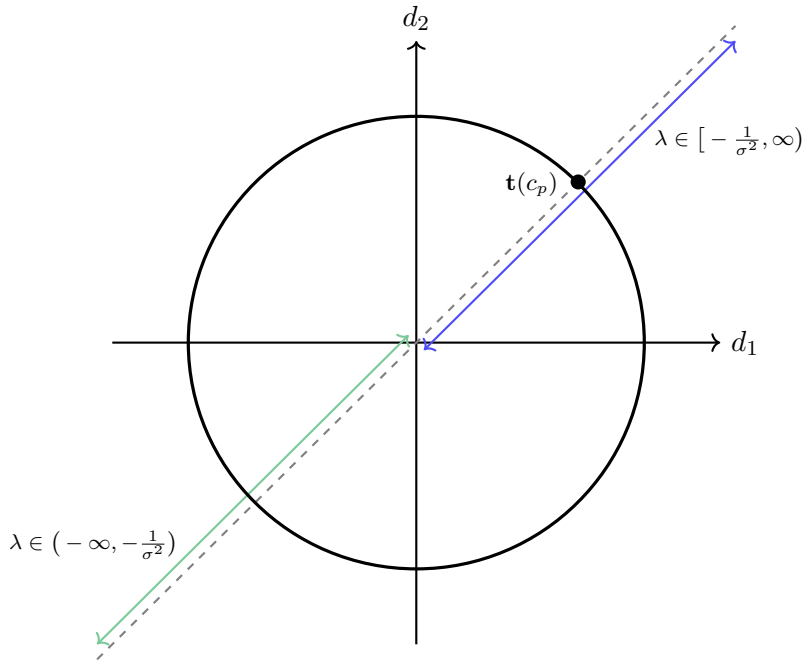
\begin{figure}
\centering
\begin{tikzpicture}
\draw[thick,->] (-4,0) -- (4,0) node[right] {$d_1$};
\draw[thick,->] (0,-4) -- (0,4) node[above] {$d_2$};
%lines and lambdas
\draw[thick, color=nnblue, scale=1,domain=0.1:4.2,smooth,variable=\x,<->] plot ({\x},{\x-0.2});
        \node[below right] at (3,3) {\scriptsize $\lambda \in \big[-\frac{1}{\sigma^{2}}, \infty\big)$};
\draw[thick, color=nngreen, scale=1,domain=-4.2:-0.1,smooth,variable=\x,<->] plot ({\x},{\x+0.2});
        \node[above left] at (-3,-3) {\scriptsize $\lambda \in \big(-\infty, -\frac{1}{\sigma^{2}}\big)$};
\draw[thick, color=gray, scale=1,domain=-4.2:4.2,dashed,variable=\x] plot ({\x},{\x});
%Big circle
\draw[very thick, fill=none](0,0) circle (3.0) node  {};
%t(cp)
\node[color=black,circle,fill,inner sep=2pt] at (2.13,2.13) {};
\node[left] at (2,2.1) {\footnotesize $\vec{t}(c_p)$};
\end{tikzpicture}
\caption{A depiction of the circular theory described in Example 3.  The dashed grey line indicates the possible values of $\vec{d}_p$ described by the range of $\lambda$ shown. Pseudodata of the form $\vec{d}_p = \vec{t}(c_p) + \lambda \sigma^2 \vec{t}(c_p) = (1 + \lambda \sigma^2)\vec{t}(c_p)$ has best-fit parameter value $c_p$ if and only if $\lambda \in [-1/\sigma^2, \infty)$, the green portion of the dashed line indicated in the figure. On the other hand, pseudodata of this form has best-fit parameter value $c_p + \pi$ otherwise.}
\label{fig:circular_theory}
\end{figure}

\paragraph{Example 3 - Circular theory.} Consider a theory $\vec{t} : \mathbb{R} \rightarrow \mathbb{R}^2$ given by:
\begin{equation}
\vec{t}(c) = t_0\begin{pmatrix} \cos(c) \\ \sin(c) \end{pmatrix},
\end{equation}
with Euclidean covariance matrix $\Sigma = \sigma^2 I$. In this case, the relevant Jacobian matrix is:
\begin{equation}
\frac{\partial \vec{t}}{\partial c} = t_0\begin{pmatrix} -\sin(c) \\ \cos(c) \end{pmatrix},
\end{equation}
which is always of full-rank. We see that we may take $M(c) = \vec{t}(c)$, and hence the Monte Carlo posterior is given for all $c$ by:
\begin{align}
\exp\left( -\frac{1}{2} \chi^2_{\vec{d}_0}(c) \right) \int\limits_{\Lambda(c)} d\lambda\ t_0^2 \sigma^2 (1 + \lambda \sigma^2) \exp\left( -\frac{1}{2} t_0^2 \sigma^2 \lambda^2 +  \lambda \left( \vec{t}(c) \cdot \vec{d}_0 - t_0^2 \right) \right).
\end{align}
To determine the integration range $\Lambda(c)$, we can use some geometry. Since we are working with a Euclidean covariance matrix, $\Sigma = \sigma^2 I$, the Monte Carlo best-fit parameter according to a particular pseudodata value $\vec{d}_p$ is simply the closest point on the theory surface to the pseudodata value $\vec{d}_p$. Now, the closest point on the theory surface to the pseudodata $\vec{t}(c) + \lambda \sigma^2 \vec{t}(c)$ is $c$ (modulo $2\pi$) for $\lambda \in [-1/\sigma^2, \infty)$, and $c + \pi$ (modulo $2\pi$) otherwise; see Fig.~\ref{fig:circular_theory} for a more detailed description. In particular, this implies that $\Lambda(c) = [-1/\sigma^2, \infty)$. This allows us to directly calculate the Monte Carlo posterior, via a change of variables:
\begin{equation}
u = t_0 \sigma \lambda - \frac{(\vec{t}(c) \cdot \vec{d}_0 - t_0^2)}{t_0 \sigma}.
\end{equation}
This yields the final Monte Carlo posterior as:
\begin{align}
&\exp\left( -\frac{1}{2} \chi^2_{\vec{d}_0}(c) \right) \exp\left( \frac{(\vec{t}(c) \cdot \vec{d}_0 - t_0^2)^2}{2t_0^2 \sigma^2}\right) \int\limits_{-\frac{\vec{t}(c) \cdot \vec{d}_0}{t_0\sigma}}^{\infty} du\ \left( \sigma^2 u + \frac{\sigma \vec{t}(c) \cdot \vec{d}_0}{t_0}\right) \exp\left( -\frac{1}{2} u^2 \right)\notag\\[1.5ex]
&\quad\propto\exp\left( -\frac{1}{2} \chi^2_{\vec{d}_0}(c) \right) \exp\left( \frac{(\vec{t}(c) \cdot \vec{d}_0 - t_0^2)^2}{2t_0^2 \sigma^2}\right) \left( \exp\left( -\frac{(\vec{t}(c) \cdot \vec{d}_0)^2}{2t_0^2 \sigma^2} \right) + \frac{\vec{t}(c) \cdot \vec{d}_0}{\sqrt{2} t_0 \sigma} \sqrt{\pi} \textrm{erfc}\left( -\frac{\vec{t}(c) \cdot \vec{d}_0}{\sqrt{2} t_0 \sigma} \right) \right),
\end{align}
where $\textrm{erfc}$ is the complementary error function, defined by:
\begin{equation}
\textrm{erfc}(z) = \frac{2}{\sqrt{\pi}} \int\limits_{z}^{\infty} dt\ e^{-t^2} \, .
\end{equation}
This formula can be simplified by absorbing the `Bayesian' part of the posterior, yielding a Monte Carlo posterior which is proportional to the compact form:
\begin{equation}
1 + \frac{\vec{t}(c) \cdot \vec{d}_0}{\sqrt{2} t_0 \sigma} \sqrt{\pi} \exp\left( \frac{(\vec{t}(c) \cdot \vec{d}_0)^2}{2t_0^2 \sigma^2}\right) \textrm{erfc}\left( -\frac{\vec{t}(c) \cdot \vec{d}_0}{\sqrt{2} t_0 \sigma} \right). 
\end{equation}

\begin{figure}
\centering
    \begin{subfigure}
        \centering
        \includegraphics[scale=0.45]{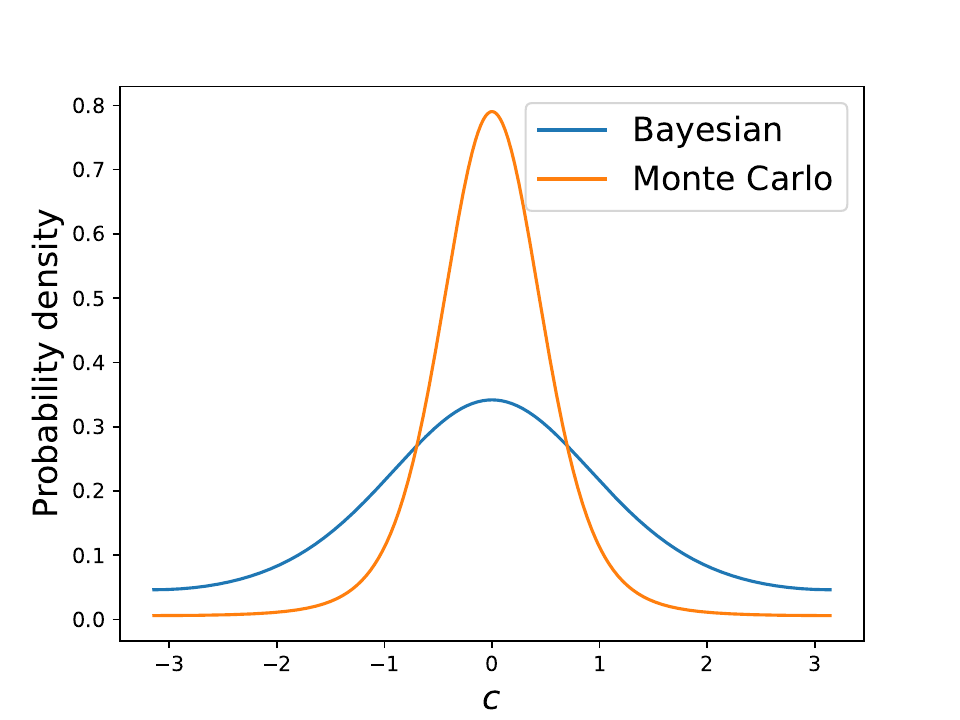} 
%        \caption{Generic} \label{fig:sigma_nnpdf40}
    \end{subfigure}
    \hfill
    \begin{subfigure}
        \centering
        \includegraphics[scale=0.45]{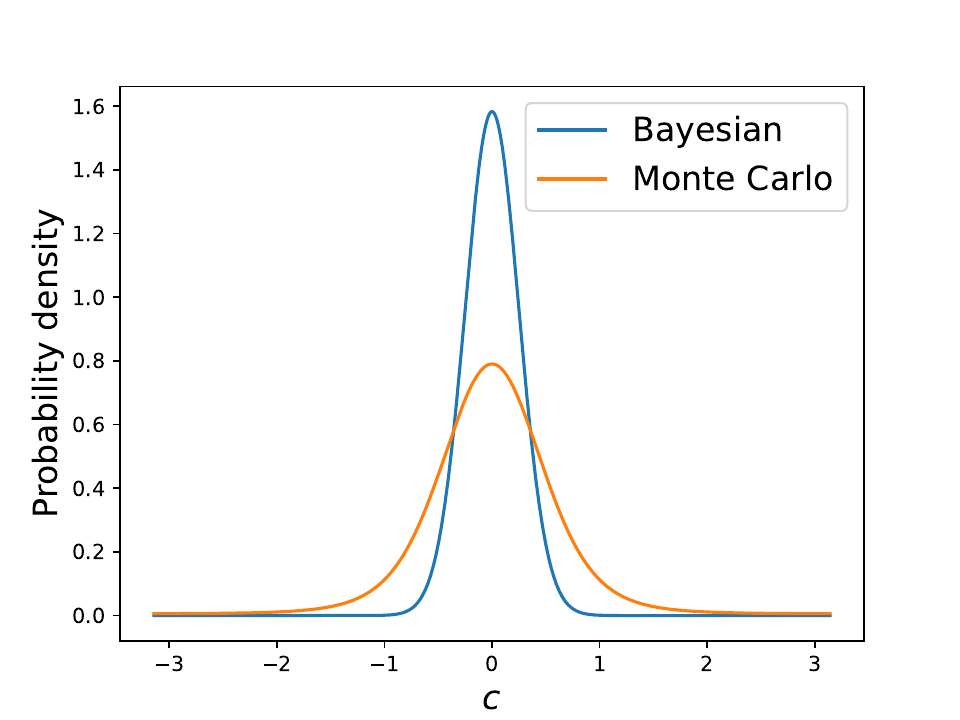} 
%        \caption{Competitors} \label{fig:timing2}
    \end{subfigure}
\caption{A comparison of the Monte Carlo and Bayesian posterior distributions for the circular theory described in Example 3. In the first panel, we show the distributions when $|\vec{d}_0| \gg t_0$ (in particular, with $\vec{d}_0 = (2,0)$, $t_0 = 0.5$ and $\sigma=1$); in this case, the Monte Carlo posterior is much more peaked compared to the Bayesian posterior. On the other hand, in the second panel we show the distributions when $|\vec{d}_0| \ll t_0$ (in particular, with $\vec{d}_0 = (2,0)$, $t_0 = 8$ and $\sigma=1$); in this case, the opposite behaviour is observed.}
\label{fig:circular_theory_distributions}
\end{figure}

Note that this formula for the Monte Carlo posterior does not contain any singular behaviour; however, it still only agrees with the Bayesian approach if an extremely specific, unmotivated prior is chosen. If instead we choose a sufficiently wide uniform prior for the Bayesian method, we can readily compare the two posteriors through some plots to exhibit the behaviour. In Fig.~\ref{fig:circular_theory_distributions} we show a comparison of the two posteriors in two cases: (i) when $|\vec{d}_0| \gg t_0$; (ii) when $|\vec{d}_0| \ll t_0$. In particular, we see that the Monte Carlo approach can both underestimate the parameter uncertainty that the Bayesian approach would imply, and overestimate this uncertainty. This is important, because it is not immediately obvious from the formula Eq.~\eqref{eq:final_simplified_mc_posterior} that both possibilities can occur. 

Further, this motivates the phenomenological study in the sequel; we emphasise the general point that \textit{Eq.~\eqref{eq:final_simplified_mc_posterior} is sufficiently complicated that the authors do not have a full understanding of its analytic behaviour}. Therefore, in realistic scenarios, our strategy in benchmarking the agreement of the Bayesian and Monte Carlo posteriors is to compute both and check.

\section{Applications in high energy physics}
\label{sec:applications}
In this section, we apply the mathematical discussion from Section~\ref{sec:multi} to some phenomenological examples in high-energy physics. In particular, we contrast the Bayesian and Monte Carlo posteriors in the case of fits of SMEFT Wilson coefficients in Sect.~\ref{subsec:smeft_fits}, before performing the same exercise in the context of uncertainty estimation for the PDFs of the proton in Sect.~\ref{subsec:pdf_fits}.

In summary, in Sect.~\ref{subsec:smeft_fits} we show that current global SMEFT fits cannot reliably make use of the Monte Carlo replica method. We find that uncertainties on SMEFT Wilson coefficients produced by the Monte Carlo replica method are discrepant from those estimated by a Bayesian method. Similarly, in Sect.~\ref{subsec:pdf_fits} we show that, in a heavily simplified model parametrisation, global PDF fits also suffer from the same issue. 
We demonstrate that when data from proton-proton collisions are included, the Monte Carlo replica method applied to our toy model leads to significantly smaller uncertainty estimates for the low-$x$ PDFs, as compared to a Bayesian approach.

\subsection{SMEFT fits}
\label{subsec:smeft_fits}

The \textit{Standard Model Effective Field Theory} (SMEFT) treats the SM as a low-energy effective limit of an ultraviolet theory. As such, it extends the SM Lagrangian by a series of non-renormalisable operators built from the SM fields and respecting the SM symmetries:
\begin{equation}
\mathcal{L}_{\text{SMEFT}} = \mathcal{L}_{\text{SM}} + \sum_{d=5}^{\infty} \sum_{i=1}^{N_d} \frac{c_d^{(i)} \mathcal{O}_d^{(i)}}{\Lambda^{d - 4}},
\end{equation}
where $\Lambda$ is some characteristic energy scale of New Physics (i.e. the scale at which we expect the SM to break down, becoming unreliable as an effective theory), and the sum over $d$ is a sum over the dimension of the operators. At each dimension $d$, there are $N_d$ independent operators, $\mathcal{O}_d^{(i)}$, indexed by $i=1,...,N_d$, with corresponding couplings $c_d^{(i)}$. The couplings $c_d^{(i)}$ are called \textit{Wilson coefficients}, and parametrise deviations from the SM.  See Ref.~\cite{Brivio:2017vri} for a review.
At $d=5$, neutrino measurements place strong constraints on $c_5^{(i)}$; however, global fits of the dimension six operators have become a topic of interest in both theoretical and experimental literature in recent years, see for example Refs.~\cite{Giani:2023gfq,Ethier:2021bye,Ethier:2021ydt,Hartland:2019bjb,Biekoetter:2018ypq,Han:2004az,daSilvaAlmeida:2018iqo,Ellis:2014jta,Almeida:2021asy,Biekotter:2018ohn,Kraml:2019sis,Ellis:2018gqa,Corbett:2012ja,Buckley:2015lku,Brivio:2019ius,Bissmann:2019gfc,Kassabov:2023hbm,Elmer:2023wtr,Allwicher:2022gkm,Boughezal:2022nof,Ellis:2020unq,Bartocci:2023nvp,Brivio:2022hrb}.

%In particular, the \smefit{} collaboration recently performed a global fit of the dimension-6 SMEFT to the electroweak, top and Higgs sectors~\cite{Ethier:2021bye}. In that work, and in their recent code paper~\cite{Giani:2023gfq}, they advertise the fact that their framework allows for two different fit options: a \textit{Bayesian fit}, based on the Nested Sampling algorithm, and a \textit{Monte Carlo fit}, which makes use of the Monte Carlo replica method. The \smefit{} collaboration is aware of the limitations of the Monte Carlo replica method and has now deprecated the feature, stressing that it can be reliably used only for linear fits.

Cross-section predictions from the dimension-six SMEFT generically take the form:
\begin{equation}
\vec{t}(\vec{c}) = \vec{t}_{\text{SM}} + \vec{t}_{\text{lin}} \vec{c} + \vec{t}_{\text{quad}} (\vec{c} \otimes \vec{c}),
\end{equation} 
where $\vec{t}_{\text{SM}}$ is a $N_{\text{dat}} \times 1$ vector of SM predictions, $\vec{t}_{\text{lin}}$ is an $N_{\text{dat}} \times N_{\text{op}}$ matrix of linear SMEFT predictions (with $N_{\text{op}}$ the number of SMEFT operators in the fit), and $\vec{t}_{\text{quad}}$ is an $N_{\text{dat}} \times N_{\text{op}}^2$ matrix of quadratic SMEFT predictions. The use of the \textit{Kronecker product} $\otimes$ of two vectors allows us to express the quadratic predictions in a fully vectorial form.

Importantly, this is a non-linear theory, and as we saw in Example 2 of Sect.~\ref{subsec:toy_examples} this can lead to discrepancies between the Monte Carlo posterior and the Bayesian posterior (unless a specific highly non-trivial prior is chosen). The quadratic behaviour of the SMEFT theory predictions additionally implies that there may be delta function singularities as we saw in Example 2 of Sect.~\ref{subsec:toy_examples}.  We show that this is likely to be the case in the global fit carried out below.

In the remainder of this section, we will compare the Bayesian and Monte Carlo posterior distributions of the SMEFT Wilson coefficients constrained specifically by top sector observables.
A comparison between the Bayesian and Monte Carlo posteriors was previously discussed in Appendix E of Ref.~\cite{Kassabov:2023hbm}, for the simple case of one Wilson coefficient constrained by a single differential measurement.  
There it was found that the choice of methodology led to a marked difference in the posterior obtained: the Bayesian posterior was significantly wider than the Monte Carlo posterior, which exhibited the `spiked' behaviour induced by the presence of a delta function as discussed in Sec.~\ref{subsec:mc_posterior}.
In this section, we assess whether this discrepancy remains at the level of a global fit.

To do so, we perform an analysis of the 175 measurements used in the simultaneous PDF and SMEFT fit of Ref.~\cite{Kassabov:2023hbm}, encompassing $t \bar{t}$, $t \bar{t} + X$, single top, single top $+X$, $t \bar{t} t \bar{t}$, $t \bar{t} b \bar{b}$ and top decay observables.  These observables will be used to determine the posterior distributions of 25 Wilson coefficients of the dimension-6 SMEFT, the notation for which can be found in Table B.1 of~\cite{Kassabov:2023hbm}.  Datapoints and theory predictions are taken directly from Ref.~\cite{Kassabov:2023hbm}, and we include the quadratic effect of the dimension-6 coefficients on our SMEFT theory predictions.  The \smefit{} code~\cite{Giani:2023gfq} is used to compare the Nested Sampling and Monte Carlo replica methodologies due to the availability of both methodologies in the public code. Note that the \smefit{} collaboration is aware of the limitations of the Monte Carlo replica method and has now deprecated the feature, stressing that it can be reliably used only for linear fits.

\begin{figure}[htb!]
\centering
\includegraphics[scale=0.2]{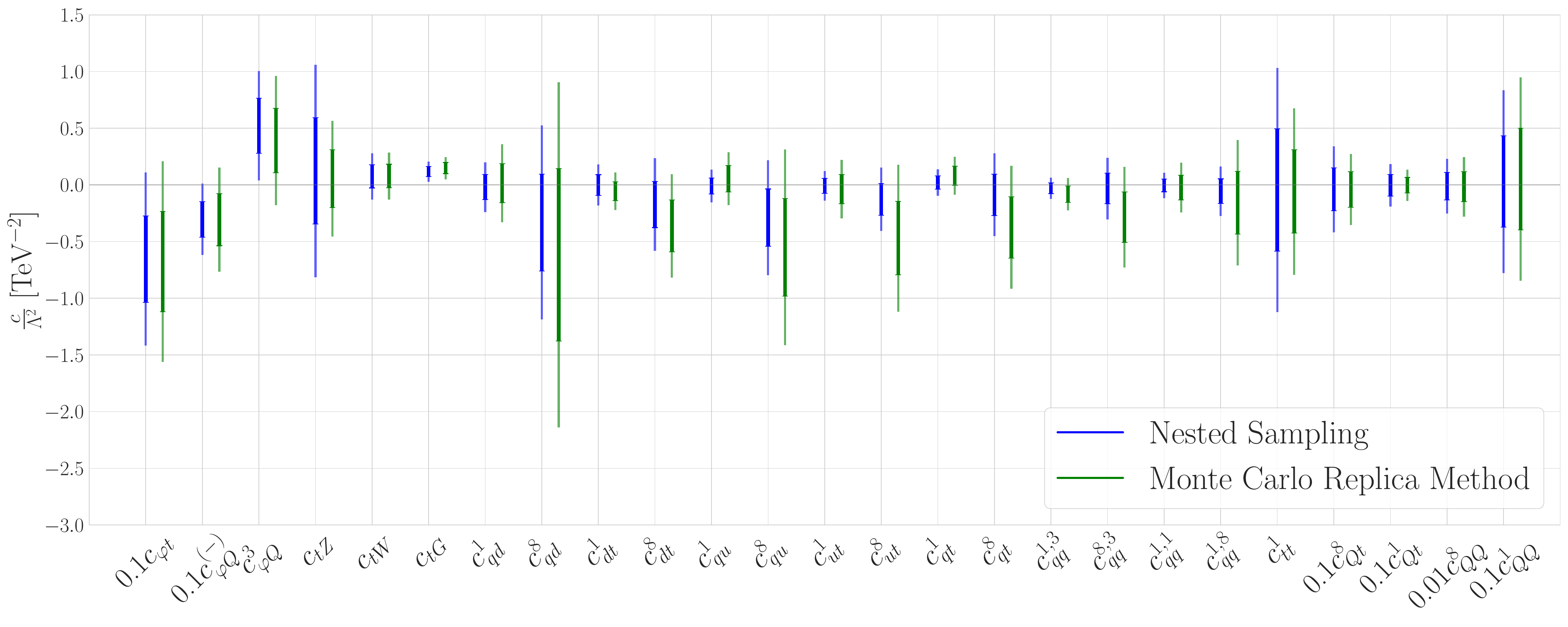}
	\caption{Marginalised constraints on the top sector of the SMEFT, comparing the Monte Carlo replica method and Nested Sampling.}
\label{fig:smeft_full_comparison}
\end{figure}

Fig.~\ref{fig:smeft_full_comparison} shows the marginalised credible intervals at 68\% and 95\% on the SMEFT coefficients, resulting from the
Nested Sampling and  Monte Carlo replica methodologies.  
Firstly, we observe that the discrepancy between the methodologies, as observed in the 1-parameter fit of Ref.~\cite{Kassabov:2023hbm}, remains at the level of the global fit.  In a global fit of 25 Wilson coefficients to all available top datasets, visible discrepancies between the two approaches are found.
In many of the four-fermion operators, the Monte Carlo replica method leads to a systematic shift downwards from the SM,
for example $c_{ut}^{8}$ and $c_{qq}^{8,3}$, while the coefficients $c_{tG}$ and $c_{qt}^{1}$ are similarly shifted upwards from the SM.
In some cases the constraints found by the Monte Carlo replica method are significantly wider than those obtained by Nested Sampling, for example
$c_{qd}^{8}$ and $c_{qq}^{11}$, while $c_{tZ}$ is found to be very well constrained by the Monte Carlo replica method compared to Nested Sampling.

\begin{figure}[htb!]
\centering
\includegraphics[scale=0.4]{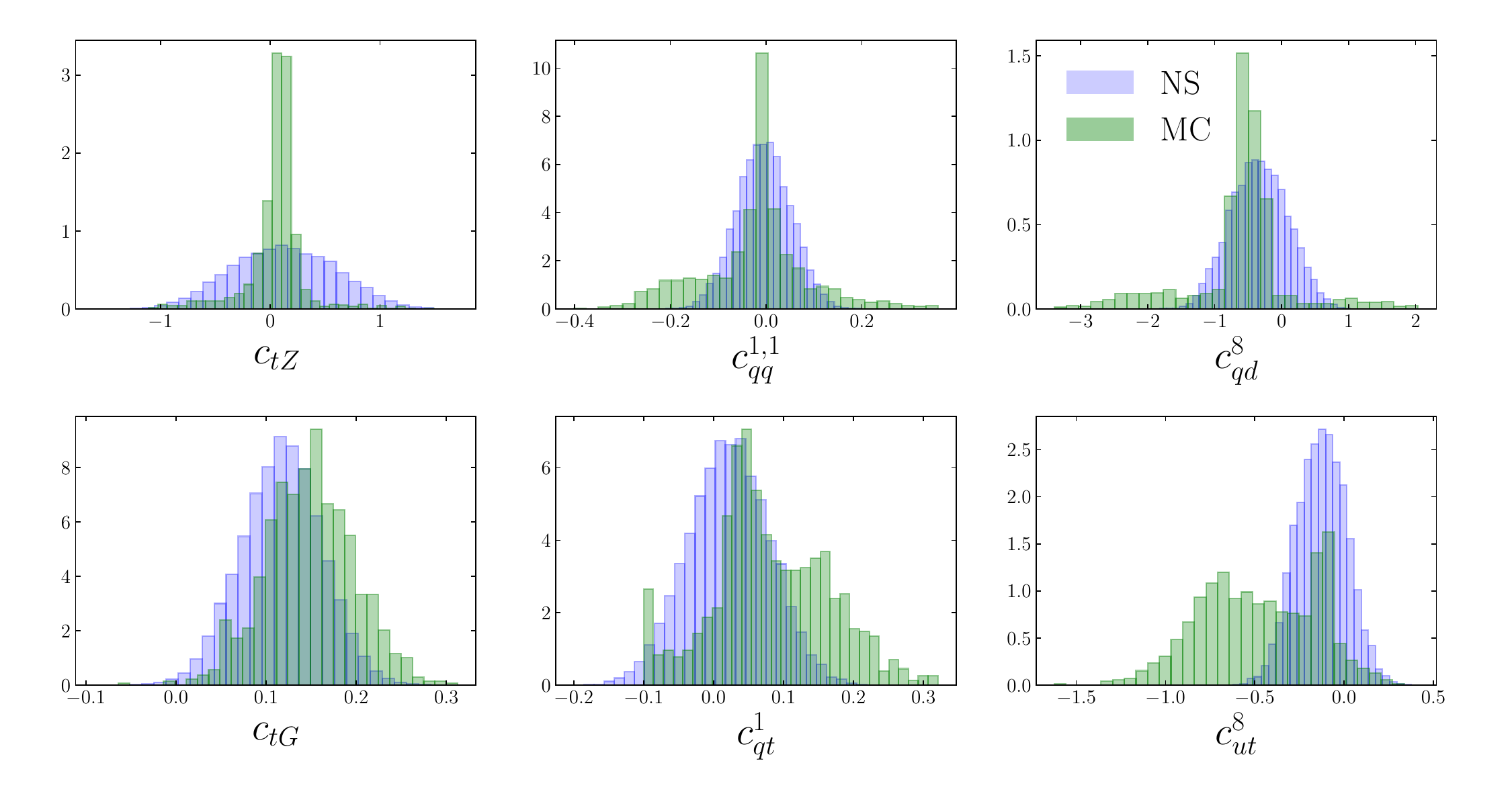}
\caption{Selection of SMEFT posteriors showcasing significant differences in the results obtained by Nested Sampling and the Monte Carlo replica method.}
\label{fig:smeft_posteriors_comparison}
\end{figure}

The Bayesian and Monte Carlo posteriors, some of which are shown in Fig.~\ref{fig:smeft_posteriors_comparison}, illuminate the source of the discrepancies found
above.  
We observe that the narrow constraint on $c_{tZ}$ obtained from the Monte Carlo replica method results from a spiked distribution,
which showcases the delta function behaviour discussed in Sect.~\ref{subsec:mc_posterior}.  Similarly spiked behaviour is observed in the posterior distributions of the four-fermion operators, in particular $c_{qq}^{1,1}$ and $c_{qd}^{8}$, while the posterior distributions of $c_{qt}^{1}$ and $c_{ut}^{8}$ are found to be highly skewed and non-Gaussian relative to the Bayesian posteriors.  
Finally, we observe that for $c_{tG}$, both methodologies obtain Gaussian-like posterior distributions of similar widths; however the Monte Carlo replica method shifts the distribution in the positive direction, leading to a stronger pull from the SM than that obtained with Nested Sampling.

\subsection{PDF fits}
\label{subsec:pdf_fits}
The parton distributions of the proton are essential ingredients in collider predictions whenever there are protons in the initial state. Roughly speaking, the PDFs represent the probability densities in $x$ for the various constituents of the proton to participate in a collision, whilst carrying a fraction $x$ of the parent proton's momentum; the participating constituent then goes on to participate in a `hard reaction' with other particle species in the collision.

In more detail, PDFs take the functional form $f_a(x,Q^2)$ where $a$ denotes the flavour of the constituent, $x$ denotes the momentum fraction, and $Q^2$ denotes a characteristic scale for the process, usually set to the energy of some particles in the collision.\footnote{The scale is called the \textit{factorisation} scale, and plays a similar role to the arbitrary scale introduced in renormalisation theory.} Predictions for the cross-section of a given collision are usually estimated in terms of a discretised version of the \textit{factorisation theorems} for the relevant processes, evaluated on some discrete $x$-grid $(x_1,...,x_{N_{\text{grid}}})$, taking the general form for the $i$th datapoint in a dataset:
\begin{equation}
\label{eq:discrete_predictions}
t_i = \begin{cases} \displaystyle \sum_{a = 1}^{N_{\text{flav}}} \sum_{\alpha=1}^{N_{\text{grid}}}  \text{FK}_{i,a\alpha} f_a(x_\alpha,Q_0^2), & \text{if the $i$th point is deep-inelastic scattering data;} \\[3ex] \displaystyle\sum_{a,b=1}^{N_{\text{flav}}} \sum_{\alpha,\beta=1}^{N_{\text{grid}}} \text{FK}_{i,a\alpha b\beta} f_a(x_\alpha,Q_0^2) f_b(x_\beta,Q_0^2), & \text{if the $i$th point is hadronic data.}\end{cases}
\end{equation}
The arrays $\text{FK}_{i,a\alpha}$ and $\text{FK}_{i,a\alpha b\beta}$ are called \textit{fast-kernel tables} in the PDF-fitting parlance, and encode both the evolution of the PDFs from some fixed initial scale $Q_0^2$ (usually taken to be $Q_0 = 1.65\ \text{GeV}$), together with the convolution of the evolved PDFs with a cross-section for the hard reaction.

Importantly, we observe that for deep-inelastic scattering data, the theory predictions are \textit{linear} in the PDFs, but the contribution for hadronic (proton-proton) data is in fact \textit{purely quadratic} in the PDFs. This will have significant consequences for the use of the Monte Carlo replica method for inferring the PDFs from collider data, as we shall discuss below.

In order to compare the Monte Carlo replica method and the Bayesian approach, we have developed a private code that allows both methodologies to be adopted when fitting PDFs. In particular, Bayesian PDFs are produced using the Nested Sampling algorithm \texttt{ultranest}~\cite{Buchner_2014, Buchner_2019, buchner2021ultranest}, which is a well-established algorithm to sample from multi-modal posterior distributions and infer estimates of parameter uncertainties. Because of this, it is an ideal framework to perform statistical inference in the presence of non-linearity.

Throughout the study we employ the publicly available \text{FK} tables and datasets provided by the {\sc NNPDF} collaboration~\cite{NNPDF:2021njg}.

\subsubsection{A toy PDF model}
\label{subsubsec:toy_pdf_model}

The PDFs, being functions, have infinitely many degrees of freedom, and hence cannot be determined completely by finite amounts of data; the problem is ill-posed. Therefore, PDF collaborations typically assume a specific functional form for the PDFs at the initial scale $Q^2 = Q_0^2$. Most collaborations, for example CTEQ~\cite{Sitiwaldi:2023jjp, Xie:2021equ}, JAM~\cite{Cocuzza:2022hse, Hunt-Smith:2023sdz, Hunt-Smith:2024khs} and MSHT~\cite{Thorne:2022abv, Cridge:2021pxm}, use a fixed functional form with $\mathcal{O}(30-40)$ parameters, while the {\sc NNPDF} collaboration~\cite{NNPDF:2021njg} uses a flexible Neural Network.

It remains an open question to what extent the Monte Carlo replica method as applied in a realistic PDF fit agrees or disagrees with a Bayesian method for uncertainty propagation. This paper does not intend to solve this problem, and we consider it important future work to address this issue (especially in the context of PDF-SMEFT interplay, as discussed in Ref.~\cite{Kassabov:2023hbm} and Sect.~\ref{subsec:smeft_fits}). 

In this work we aim lower, and take a first step towards answering this question, by instead working with a simplified PDF model and artificially generated data.\footnote{To avoid confusion, we reserve the term \textit{pseudodata} exclusively for discussion of Monte Carlo pseudodata. The term \textit{artificial data} instead describes artificial central experimental values, $\vec{d}_0$, in the discussion of the previous chapters.} In the remainder of this section, we discuss the toy model (based on linear interpolation of the PDF grid), and describe the generation of the artificial data used in this study.

\begin{figure}
\centering
    \begin{subfigure}
        \centering
        \includegraphics[scale=0.3]{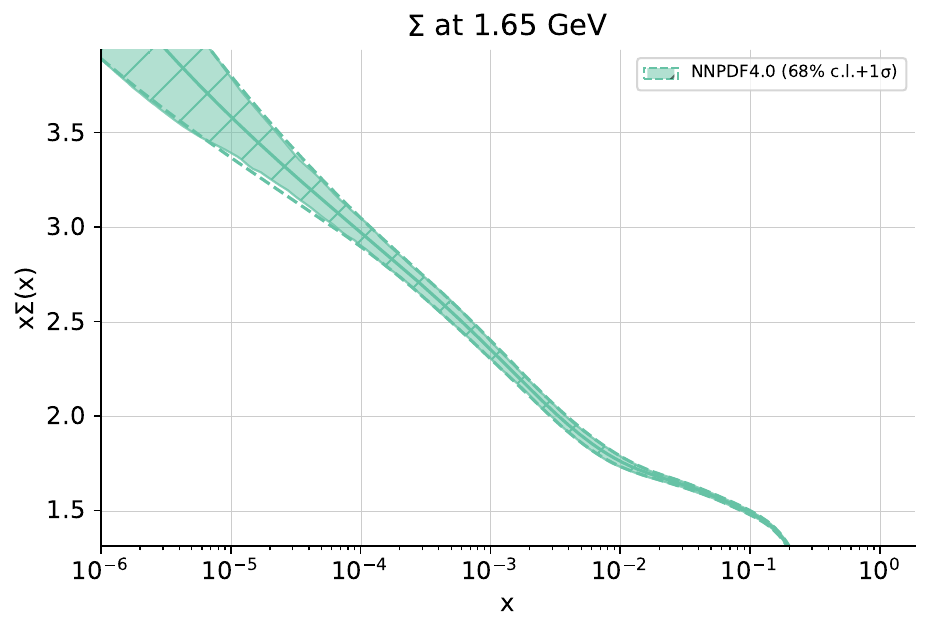} 
%        \caption{Generic} \label{fig:sigma_nnpdf40}
    \end{subfigure}
    \hfill
    \begin{subfigure}
        \centering
        \includegraphics[scale=0.3]{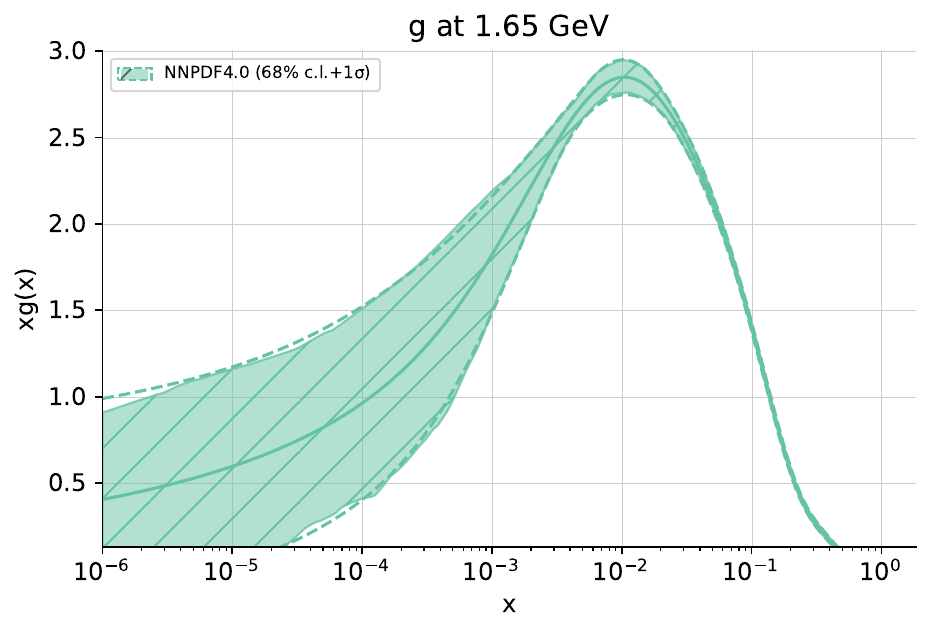} 
%        \caption{Competitors} \label{fig:timing2}
    \end{subfigure}
        \hfill
    \begin{subfigure}
        \centering
        \includegraphics[scale=0.3]{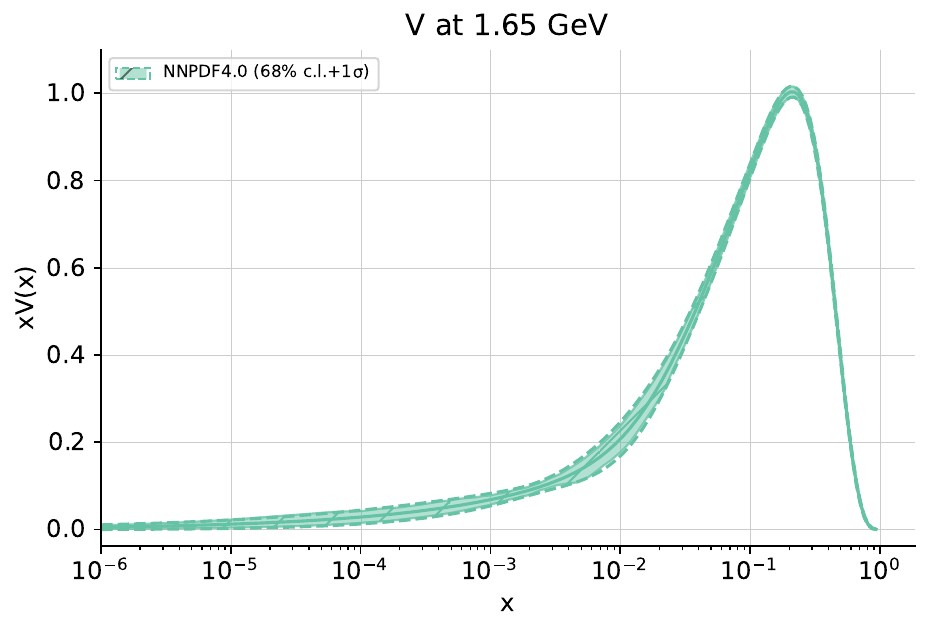} 
%        \caption{Competitors} \label{fig:timing2}
    \end{subfigure}

    \vspace{0.2cm}
    
        \begin{subfigure}
        \centering
        \includegraphics[scale=0.3]{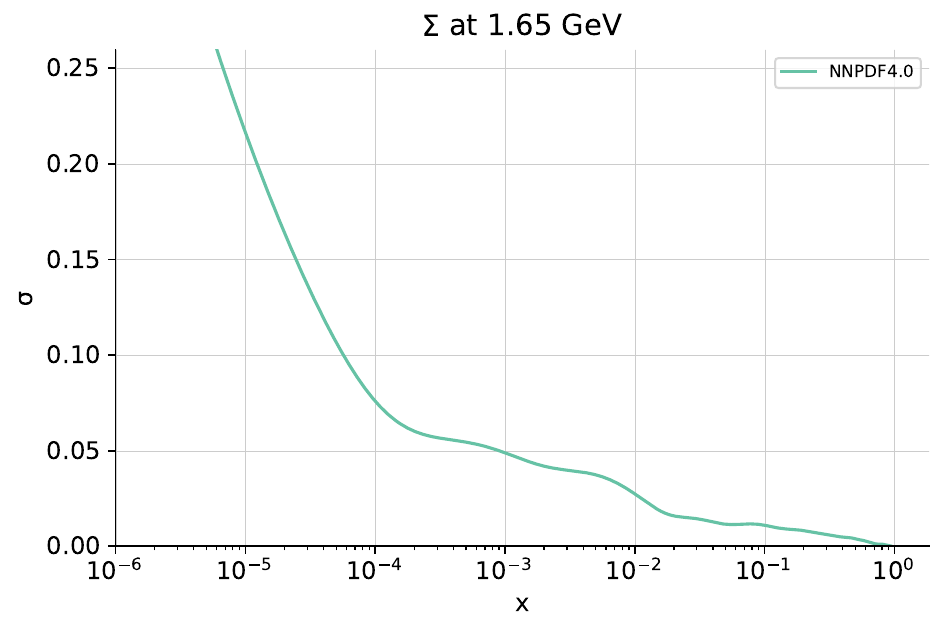} 
%        \caption{Generic} \label{fig:sigma_nnpdf40}
    \end{subfigure}
    \hfill
    \begin{subfigure}
        \centering
        \includegraphics[scale=0.3]{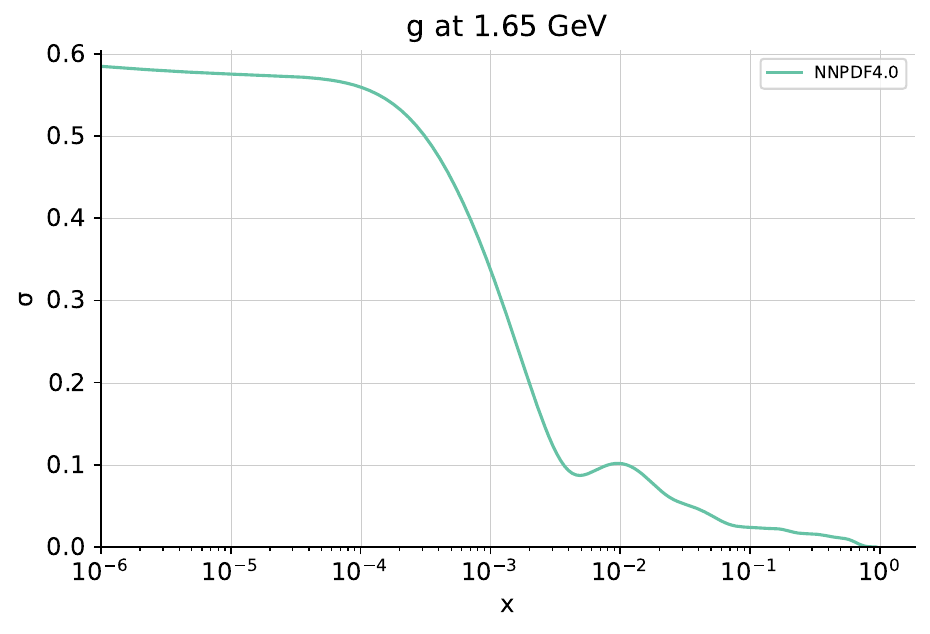} 
%        \caption{Competitors} \label{fig:timing2}
    \end{subfigure}
        \hfill
    \begin{subfigure}
        \centering
        \includegraphics[scale=0.3]{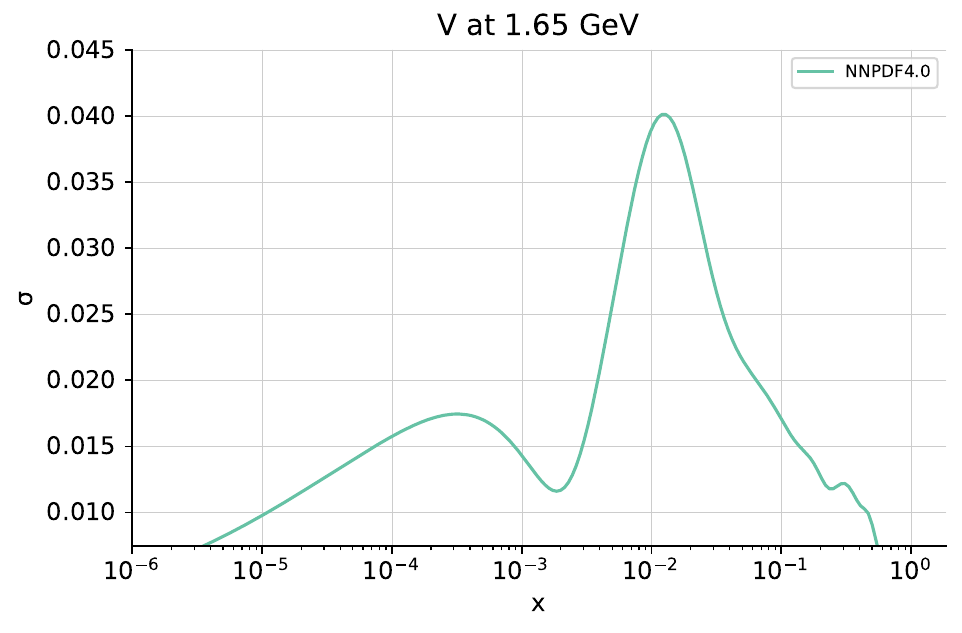} 
%        \caption{Competitors} \label{fig:timing2}
    \end{subfigure}
\caption{The singlet $\Sigma$, gluon $g$, and valence $V$ distributions in the \nnpdf{} fit; their respective percentage uncertainties (at $68\%$) are also shown in the panel below. The singlet is monotonically decreasing, whilst the gluon and valence distributions have peaks at around $x \approx 10^{-2}$ and $x \approx 0.2$ respectively.}
\label{fig:nnpdf40_realistic}
\end{figure}

\paragraph{The toy model.} In our toy model, to reduce dimensionality, we first suppose that the structure of the proton can be completely described in terms of three flavours in the evolution basis, $\Sigma$, $g$, $V$, the singlet, gluon and valence distributions respectively, where the singlet and the valence are defined in terms of the quark flavours as:
\begin{align}
\Sigma & =u+\bar{u}+d+\bar{d}+s+\bar{s}+2 c \, , \\
V & =(u-\bar{u})+(d-\bar{d})+(s-\bar{s}) \, .
\end{align}
Samples of these flavours taken from a realistic PDF fit (namely the recent \nnpdf{} determination) are shown in Fig.~\ref{fig:nnpdf40_realistic}. These three flavours have the following structure: (i) the singlet is a monotonically decreasing function; (ii) the gluon is peaked at $x \approx 10^{-2}$; (iii) the valence is peaked at $x \approx 0.2$. 

Now, in standard PDF fits, a parametrisation (either polynomial-based or neural network) is assumed in order to model the structure of these flavours. However, both polynomial and neural network parametrisations are non-linear, so even in the DIS case where PDFs enter linearly into theory predictions, the results of Sect.~\ref{subsec:toy_examples} imply that we cannot expect exact agreement between the Monte Carlo replica method and the Bayesian method. Hence, in order to produce a useful benchmark exercise where agreement is demonstrated in a PDF fitting context, we introduce the following `linear interpolation' parametrisation of the PDFs.

In our parametrisation, for each flavour $f$, we select a `reduced $x$-grid':
\begin{equation}
\left(x_1^f, ..., x_{N_{\text{grid}}(f)}^f\right).
\end{equation}
This reduced $x$-grid is now treated as an interpolation grid, from which each flavour of PDF is constructed. In particular, we define the PDF at the initial scale as a linear interpolant of its values on this reduced $x$-grid, via:
\begin{equation}
f(x,Q_0^2) = \begin{cases} \displaystyle f(x_1^f, Q_0^2) & \text{if $x \leq x_1^f$;}\\[1.5ex] 
\displaystyle \left( \frac{x_{i+1}^f - x}{x_{i+1}^f - x_i^f} \right) f(x_i^f, Q_0^2) + \left( \frac{x - x_i^f}{x_{i+1}^f - x_i^f} \right) f(x_{i+1}, Q_0^2) & \text{if $x \in [x_{i}, x_{i+1}]$, for $i = 1,...,N_{\text{grid}}(f)$};\\[3ex] 
\displaystyle f(x_{N_{\text{grid}}(f)}^f, Q_0^2) & \text{if $x > x_{N_{\text{grid}}(f)}^f$} \, .\end{cases}
\end{equation}
In particular, the `parameters' which describe this PDF model are simply the values of the PDF on the reduced $x$-grid; once these values $f(x_1^f,Q_0^2), ..., f(x_{N_{\text{grid}}(f)}^f,Q_0^2)$ are specified, then the PDF itself is specified everywhere. We have chosen to extrapolate simply by taking the value of the PDF below the lowest grid point to always be equal to its value \textit{at} the lowest grid point; similarly for the highest grid point.

This somewhat elementary approach has the following distinct advantages:
\begin{itemize}
\item Importantly for our purposes, the PDF parametrisation is \textit{linear}; therefore, DIS predictions are truly linear predictions of the PDF parameters, and predictions for proton-proton collisions are truly quadratic predictions of the PDF parameters. In particular, this allows us to benchmark the Monte Carlo approach directly against a Bayesian approach in the DIS-only case.
\item As we increase the number of grid points and flavours, the model becomes increasingly realistic. This implies that if we take the grid to be the FK-table grid in our parametrisation, and use all of the fitted flavours, the linear interpolation becomes completely immaterial, since the FK-table only `sees' the PDF at the - now complete - interpolation grid. With the advent of increased computational power, making use of tools such as GPUs for instance, such a direct fit of the grid could be within reach in the next few years.
\end{itemize}

On a practical note, for this study we have selected grids such that as much of the `structure' displayed in Fig.~\ref{fig:nnpdf40_realistic} is retained. We choose $12$ grid points in each case, giving a total of $36$ fitted parameters. The precise choice is:
\begin{align}
&(1.57\times 10^{-4},
    3.62 \times 10^{-4},
    8.31 \times 10^{-4},
    1.90 \times 10^{-4},
    4.32 \times 10^{-3},
    9.70 \times 10^{-3},\notag\\[1.5ex]
    &\qquad 2.11 \times 10^{-2},
    4.34 \times 10^{-2},
    8.23 \times 10^{-2},
    1.41 \times 10^{-1},
    2.20 \times 10^{-1},
    3.14 \times 10^{-1}) &\ \text{for $\Sigma$};\\[1.5ex]
&(3.62 \times 10^{-4},
    5.49 \times 10^{-4},
    8.31 \times 10^{-4},
    1.90 \times 10^{-3},
    4.33 \times 10^{-3},
    9.70 \times 10^{-3},\notag\\[1.5ex]
    &\qquad 2.11 \times 10^{-2},
    4.34 \times 10^{-2},
    8.23 \times 10^{-2},
    1.41 \times 10^{-1},
    2.20 \times 10^{-1},
    3.14 \times 10^{-1}) &\ \text{for $g$};\\[1.5ex]
&(3.05 \times 10^{-2},
    4.34 \times 10^{-2},
    6.05 \times 10^{-2},
    8.23 \times 10^{-2},
    1.09 \times 10^{-1},
    1.78 \times 10^{-1},\notag\\[1.5ex]
    &\qquad 2.65 \times 10^{-1},
    3.67 \times 10^{-1},
    4.80 \times 10^{-1},
    5.40 \times 10^{-1},
    6.01 \times 10^{-1},
    6.65 \times 10^{-1}) &\ \text{for $V$}.
\end{align}
Observe that we have populated the region near $x \approx 0.2$ for the valence with significantly more points, in order to model its peak structure in this region. For the singlet and gluon distributions, we reduce the number of points for $x < 10^{-4}$, where the uncertainty becomes larger, implying a lack of constraints from the dataset. In particular, because of the choice of the grid points, all of the parameters are constrained by the data and do not lie in the extrapolation region.

\paragraph{Generation of the artificial data.} Due to the extremely simplified nature of the PDF model in this study, we choose to use purely artificial data throughout. Our artificial data is produced assuming that the true PDF law of Nature is a linear interpolant of the above form (thereby bypassing any modelling error), and random noise is added on top according to the experimental covariance matrix.

In more detail, we assume that the true law of Nature is \nnpdf{}~\cite{NNPDF:2021njg}, and produce predictions for the dataset in question (either DIS only, proton-proton only, or the full dataset, explored in turn in the sequel) using this PDF set; write these predictions as $\vec{t}_0$. The artificial data is then generated as a sample from the multivariate normal distribution:
\begin{equation}
\mathcal{N}(\vec{t}_0, \Sigma),
\end{equation}
where $\Sigma$ is the relevant $\vec{t}_0$ covariance matrix.\footnote{See Ref.~\cite{Ball:2009qv} for details on the definition of the $\vec{t}_0$ covariance matrix.}
%In the \nnpdf{} closure test parlance, this type of data is called `\textit{level 1 data}'.

\subsubsection{Results for DIS-only fits}
\label{subsubsec:dis_only_fit}
The vast majority of the data which enters a PDF fit is \textit{deep inelastic scattering} data, which depends linearly on the PDFs.\footnote{Excluding, for example, ratio observables - the \texttt{NMC} measurement of $F_2^d/F_2^p$ is an example~\cite{NewMuon:1996uwk}.} As indicated in Eq.~\eqref{eq:discrete_predictions}, DIS predictions take the form:
\begin{equation*}
\vec{t}(\vec{f}) = \textrm{FK}\cdot\vec{f},
\end{equation*}
where $\textrm{FK}$ is the $N_{\text{dat}} \times (N_{\text{grid}}N_{\text{flav}})$ fast-kernel table for the prediction. Importantly for our purposes, this is a linear prediction, so Monte Carlo and Bayesian uncertainty estimates on PDFs in the grid parametrisation obtained from DIS data only must agree according to the general theory presented in Sect.~\ref{sec:multi}.

\begin{figure}
\centering
    \begin{subfigure}
        \centering
        \includegraphics[scale=0.29]{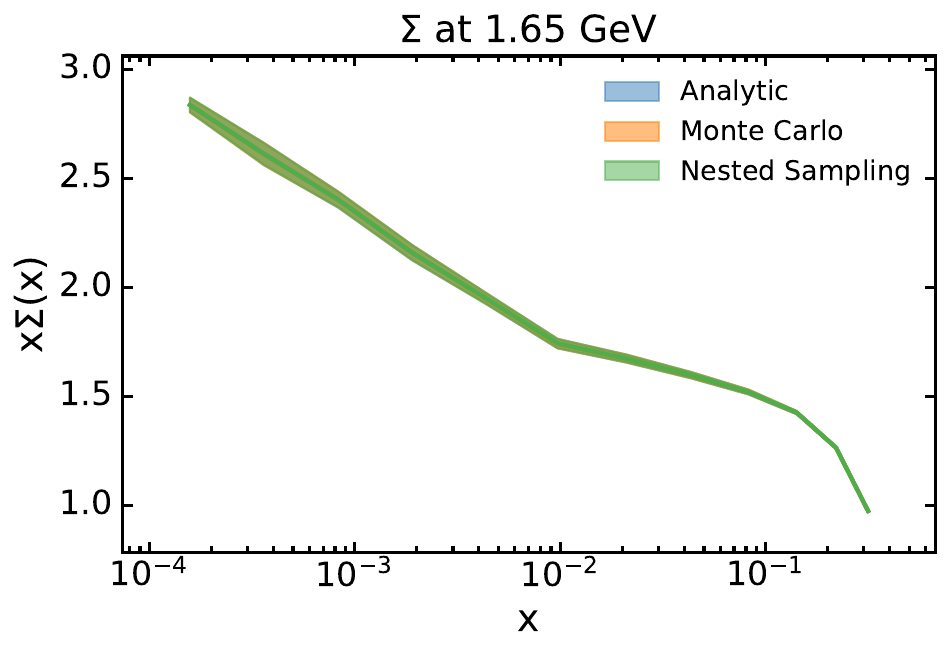} 
%        \caption{Generic} \label{fig:sigma_nnpdf40}
    \end{subfigure}
    \hfill
    \begin{subfigure}
        \centering
        \includegraphics[scale=0.29]{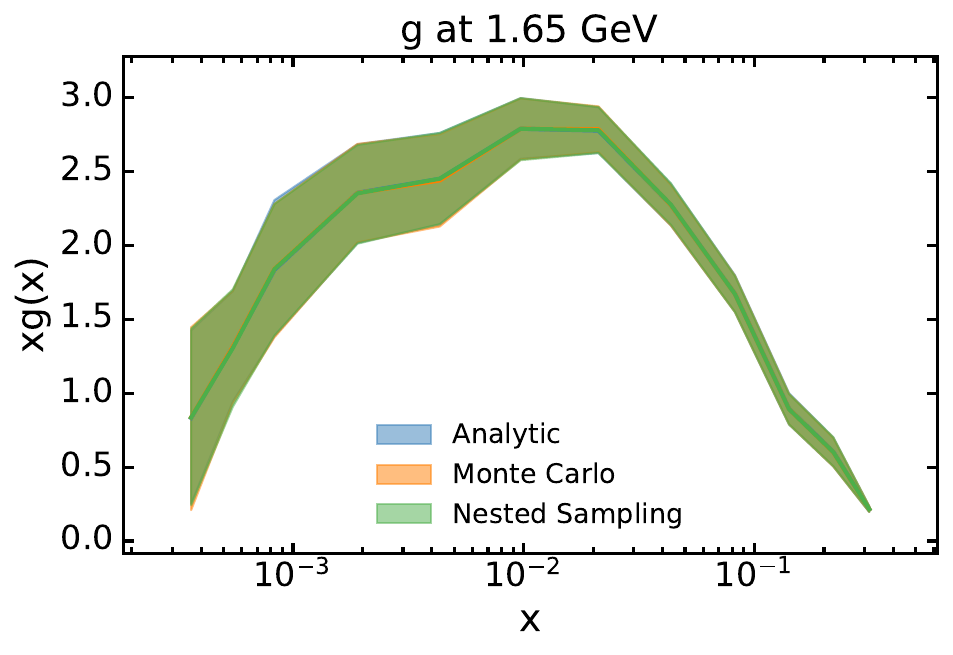} 
%        \caption{Competitors} \label{fig:timing2}
    \end{subfigure}
        \hfill
    \begin{subfigure}
        \centering
        \includegraphics[scale=0.29]{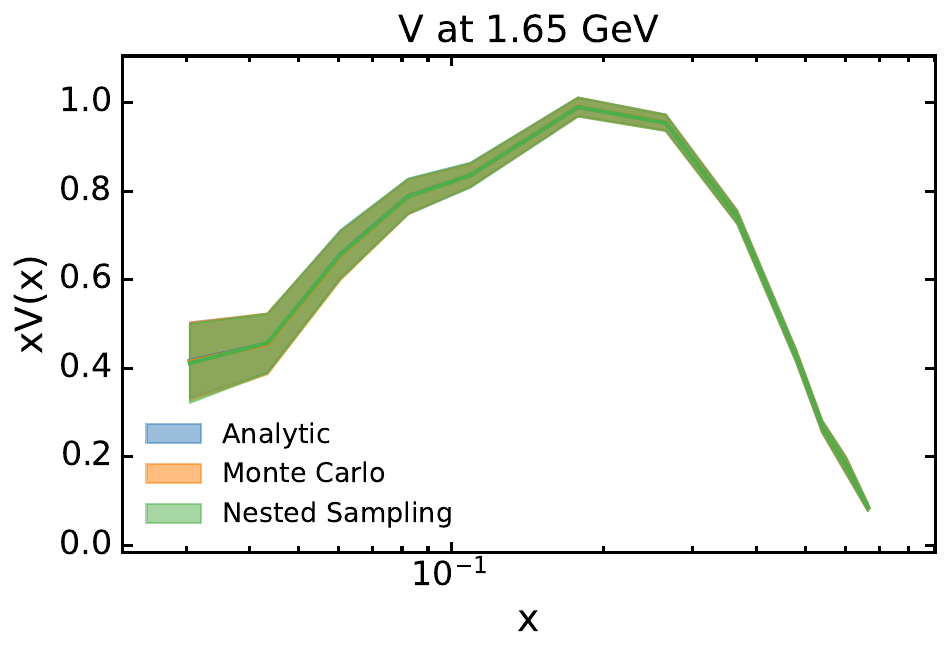} 
%        \caption{Competitors} \label{fig:timing2}
    \end{subfigure}

    \vspace{0.2cm}
    
        \begin{subfigure}
        \centering
        \includegraphics[scale=0.29]{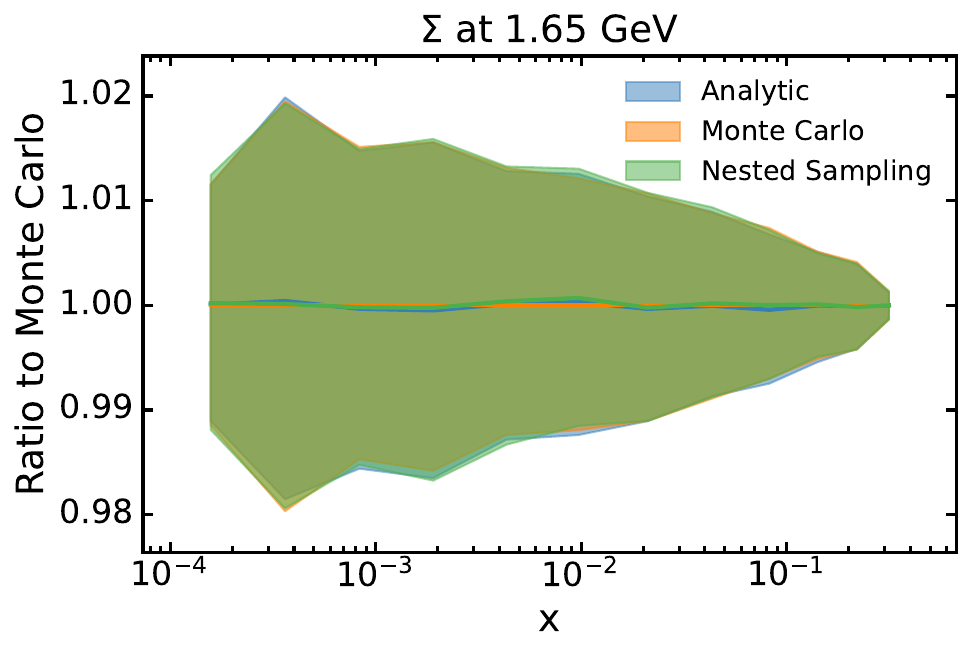} 
%        \caption{Generic} \label{fig:sigma_nnpdf40}
    \end{subfigure}
    \hfill
    \begin{subfigure}
        \centering
        \includegraphics[scale=0.29]{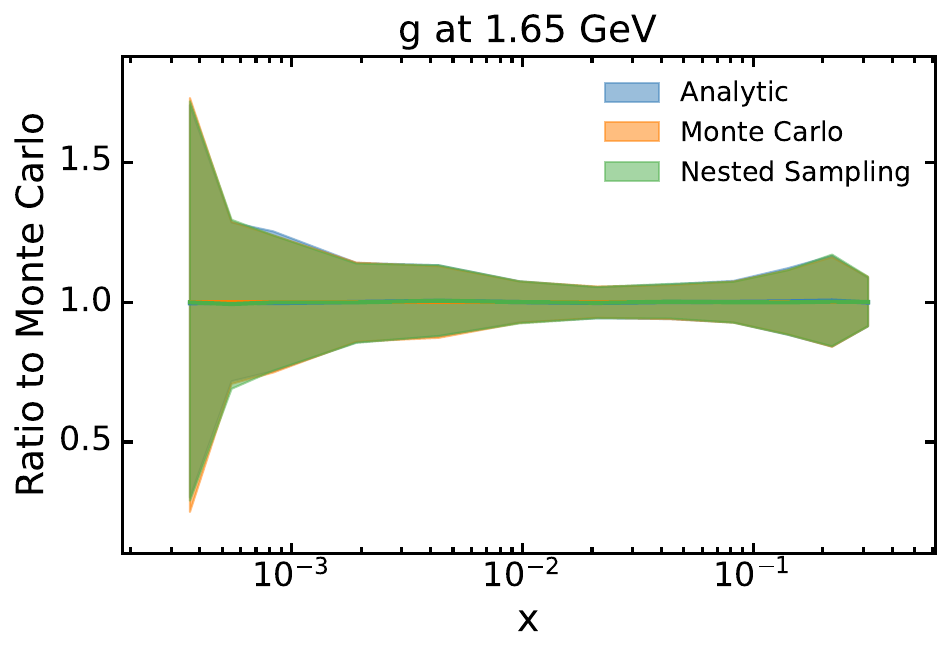} 
%        \caption{Competitors} \label{fig:timing2}
    \end{subfigure}
        \hfill
    \begin{subfigure}
        \centering
        \includegraphics[scale=0.29]{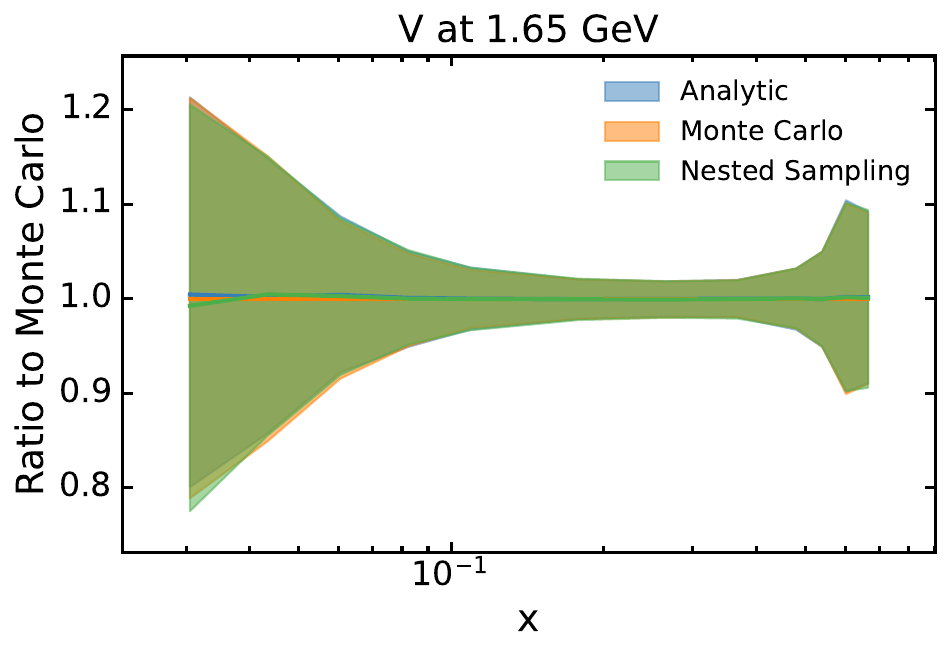} 
%        \caption{Competitors} \label{fig:timing2}
    \end{subfigure}
    
    \vspace{0.2cm}
    
        \begin{subfigure}
        \centering
        \includegraphics[scale=0.29]{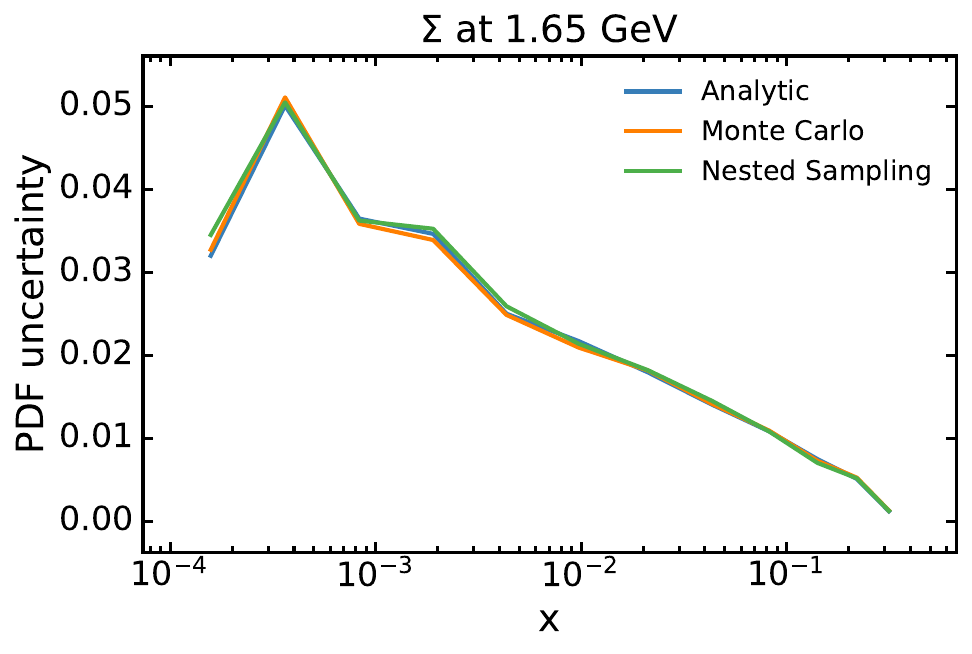} 
%        \caption{Generic} \label{fig:sigma_nnpdf40}
    \end{subfigure}
    \hfill
    \begin{subfigure}
        \centering
        \includegraphics[scale=0.29]{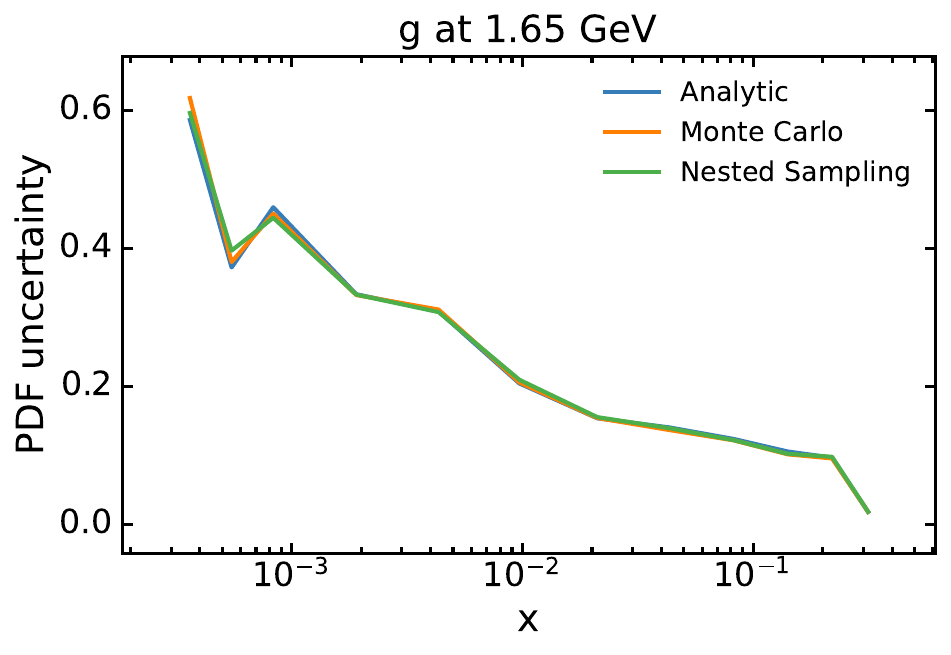} 
%        \caption{Competitors} \label{fig:timing2}
    \end{subfigure}
        \hfill
    \begin{subfigure}
        \centering
        \includegraphics[scale=0.29]{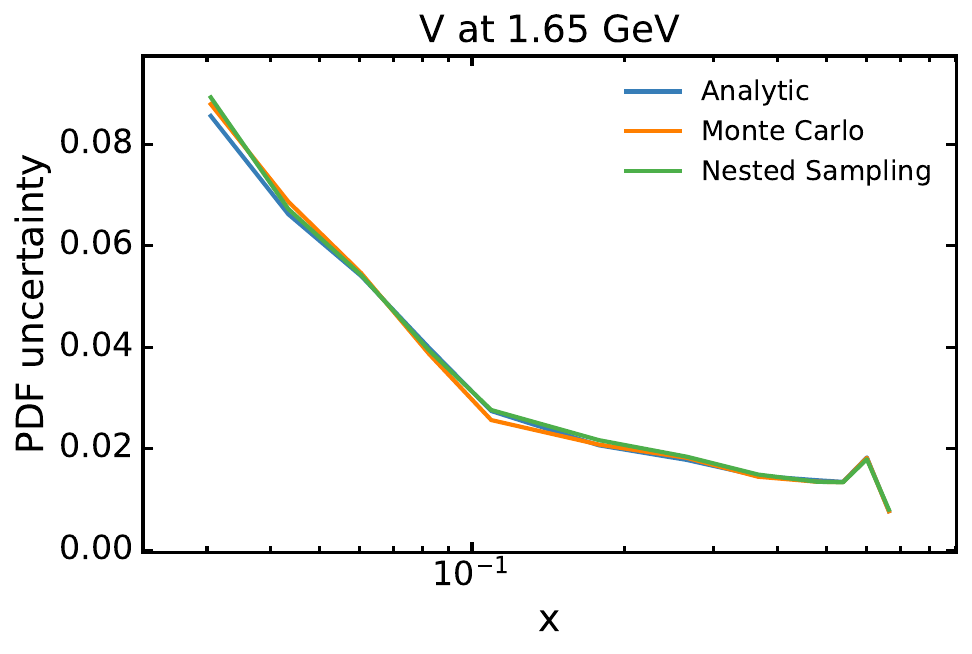} 
%        \caption{Competitors} \label{fig:timing2}
    \end{subfigure}
\caption{A comparison of three PDF fits to DIS-only data using the simplified model described above. The meaning of the Monte Carlo, Nested Sampling, and Analytic PDF sets is described in the text. The columns display the flavours $\Sigma, g, V$ from left to right. The rows display the absolute values of the PDFs, the ratio of the PDFs to the Monte Carlo PDF, and the uncertainties on each of the PDF sets, from top to bottom. Agreement is outstanding between the three exercises.}
\label{fig:dis_comparison}
\end{figure}

Using a private code, we have benchmarked this agreement in the context of the model and artificial data described above, using the entire DIS dataset entering the \nnpdf{} analysis~\cite{NNPDF:2021njg}, consisting of 2968 datapoints. We display the resulting PDFs in Fig.~\ref{fig:dis_comparison}; as anticipated, agreement is exceptional between the Monte Carlo and Bayesian approaches.

The three PDF fits in Fig.~\ref{fig:dis_comparison} are obtained as follows. The Monte Carlo PDF is, as one might expect, produced using the Monte Carlo replica method. The Nested Sampling PDF is a Bayesian PDF produced using the Nested Sampling algorithm; a uniform prior is used on each of the grid points to ensure the agreement described in Sect.~\ref{sec:multi}. Finally, the `analytic' PDF is produced by sampling from the analytic posterior distribution that can be calculated for this linear problem; it is found to be:
\begin{equation}
\mathcal{N}\bigg( (\textrm{FK})^T \Sigma^{-1} \textrm{FK})^{-1} \textrm{FK}^T \Sigma^{-1} \vec{d}_0, (\textrm{FK}^T \Sigma^{-1} \textrm{FK})^{-1}\bigg) \, .
\end{equation}
This `analytic' result allows us to test the efficacy of the numerical approaches presented by the Monte Carlo and Nested Sampling techniques.

\subsubsection{Results for hadronic-only fits}
\label{subsubsec:global_pdf_results}

In this section, we repeat the exercise of Sect.~\ref{subsubsec:dis_only_fit}, but now using \textit{hadronic-only} data instead of DIS-only data. In the context of PDF fitting, hadronic data refers to data coming from proton-proton collisions; as a result, the theory predictions for these processes take a quadratic form, as presented in Eq.~\eqref{eq:discrete_predictions}. In particular, we can expect some singular behaviour of the Monte Carlo posterior, as presented in the toy examples of Sect.~\ref{sec:multi} and in the SMEFT fits of Sect.~\ref{subsec:smeft_fits}.

The settings remain precisely the same as in Sect.~\ref{subsubsec:dis_only_fit}, except we now use the complete \nnpdf{} hadronic dataset, with the exception of jet data.\footnote{The study is conducted by making use of the new theory pipeline of {\sc NNPDF} and when producing results, jet data was not available. We note, however, that it has been recently implemented.} This dataset comprises a total of 1027 points. 

\begin{figure}
\centering
    \begin{subfigure}
        \centering
        \includegraphics[scale=0.29]{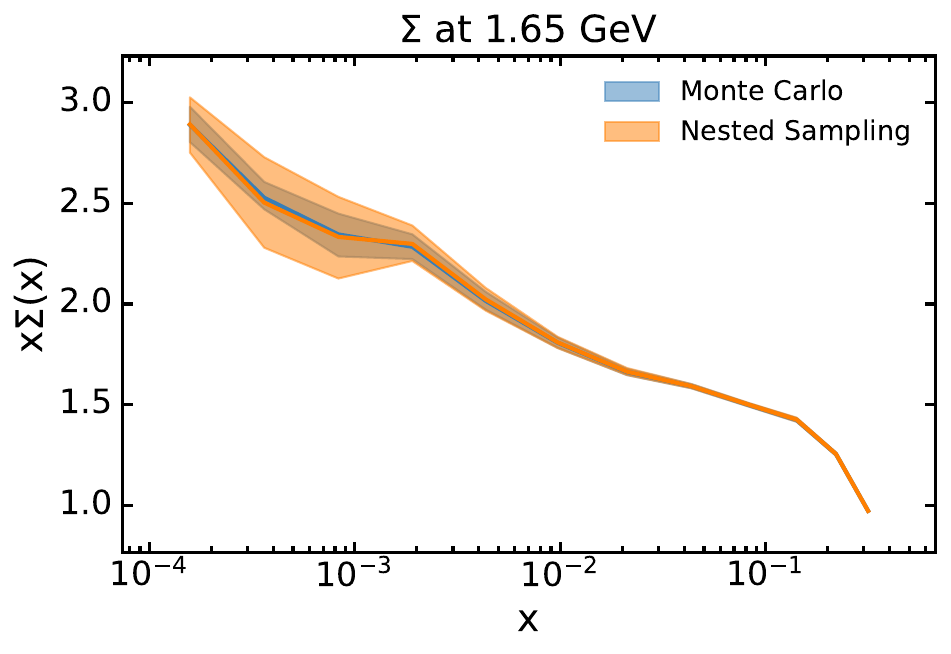} 
%        \caption{Generic} \label{fig:sigma_nnpdf40}
    \end{subfigure}
    \hfill
    \begin{subfigure}
        \centering
        \includegraphics[scale=0.29]{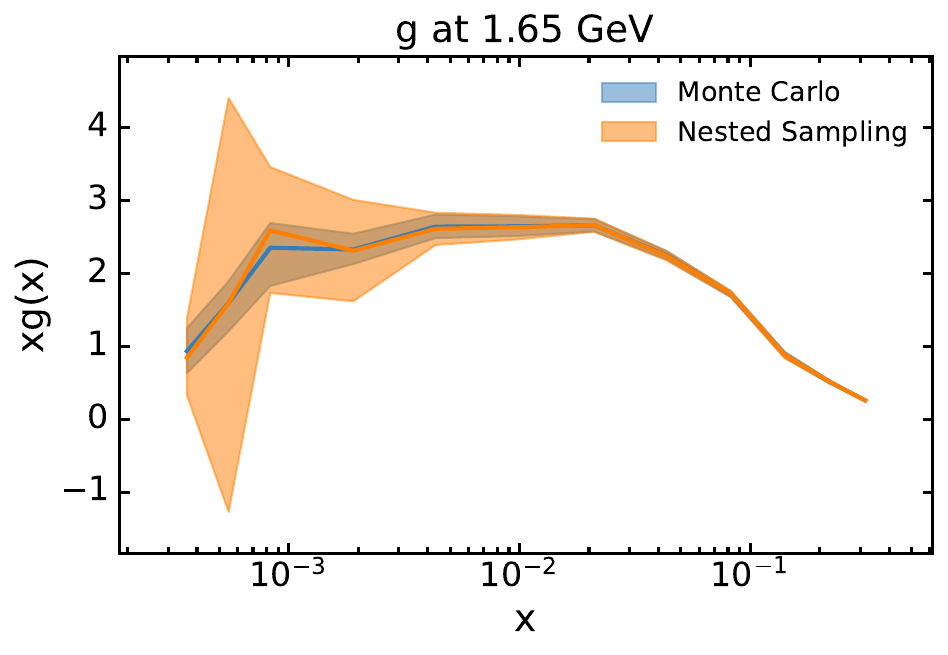} 
%        \caption{Competitors} \label{fig:timing2}
    \end{subfigure}
        \hfill
    \begin{subfigure}
        \centering
        \includegraphics[scale=0.29]{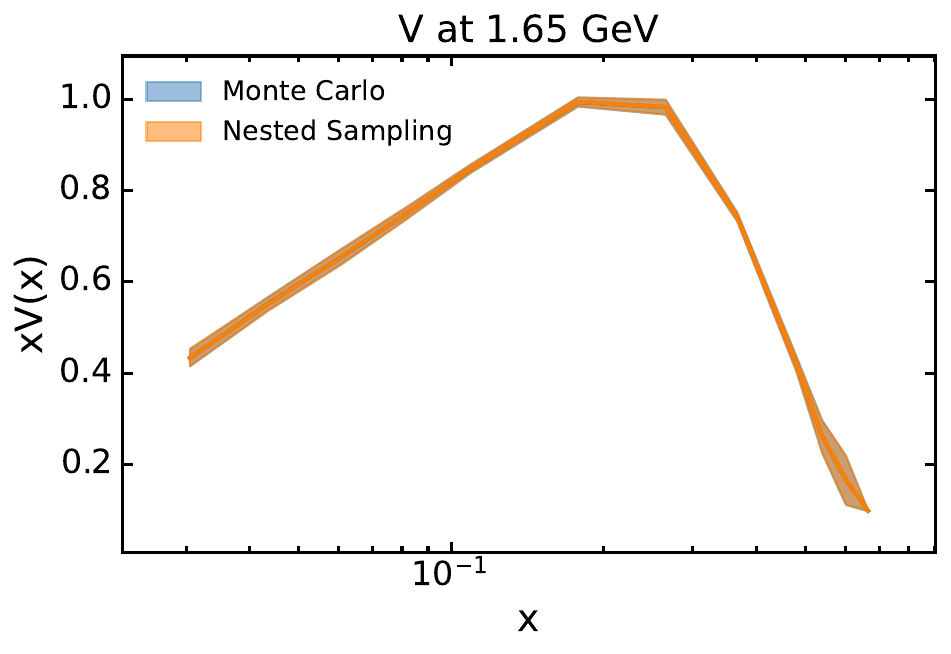} 
%        \caption{Competitors} \label{fig:timing2}
    \end{subfigure}

    \vspace{0.2cm}
    
        \begin{subfigure}
        \centering
        \includegraphics[scale=0.29]{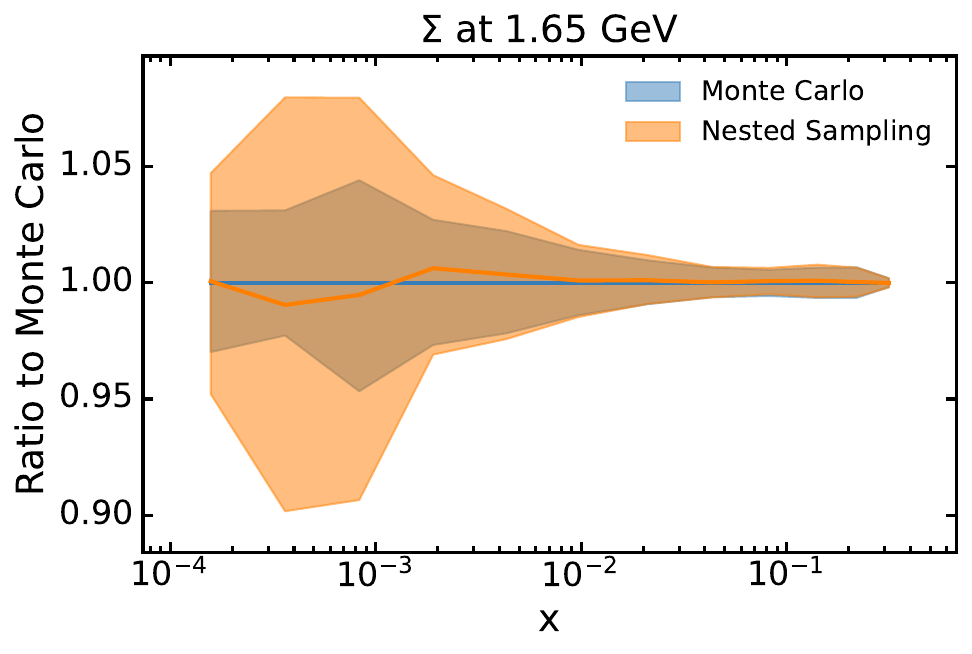} 
%        \caption{Generic} \label{fig:sigma_nnpdf40}
    \end{subfigure}
    \hfill
    \begin{subfigure}
        \centering
        \includegraphics[scale=0.29]{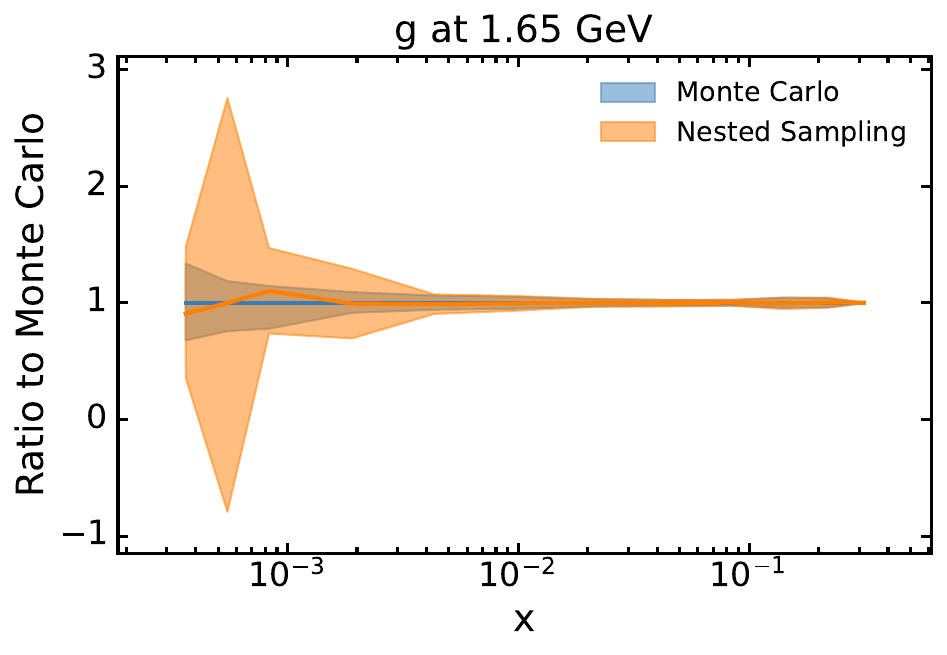} 
%        \caption{Competitors} \label{fig:timing2}
    \end{subfigure}
        \hfill
    \begin{subfigure}
        \centering
        \includegraphics[scale=0.29]{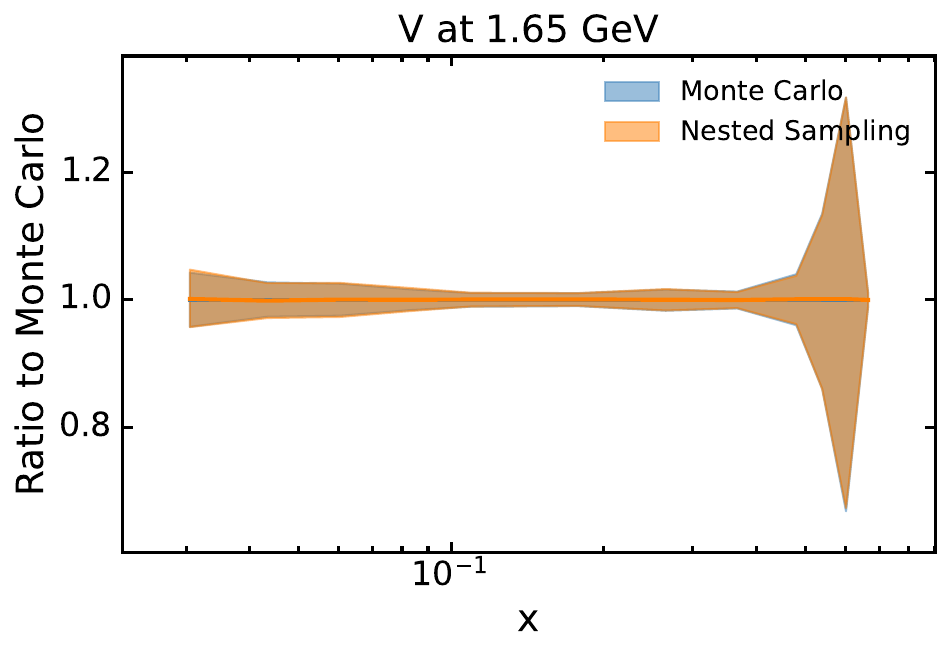} 
%        \caption{Competitors} \label{fig:timing2}
    \end{subfigure}
    
    \vspace{0.2cm}
    
        \begin{subfigure}
        \centering
        \includegraphics[scale=0.29]{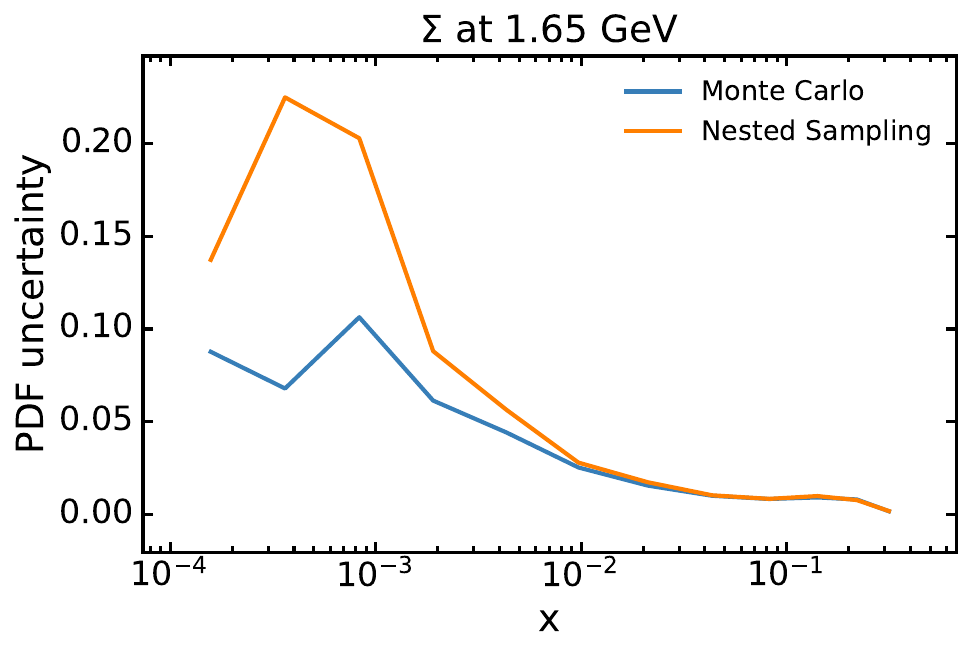} 
%        \caption{Generic} \label{fig:sigma_nnpdf40}
    \end{subfigure}
    \hfill
    \begin{subfigure}
        \centering
        \includegraphics[scale=0.29]{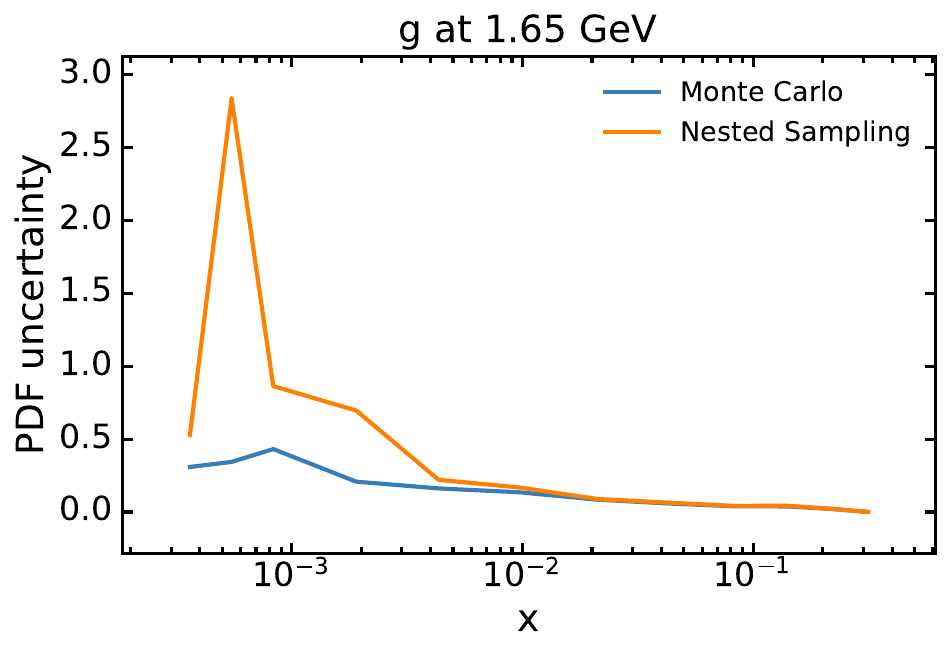} 
%        \caption{Competitors} \label{fig:timing2}
    \end{subfigure}
        \hfill
    \begin{subfigure}
        \centering
        \includegraphics[scale=0.29]{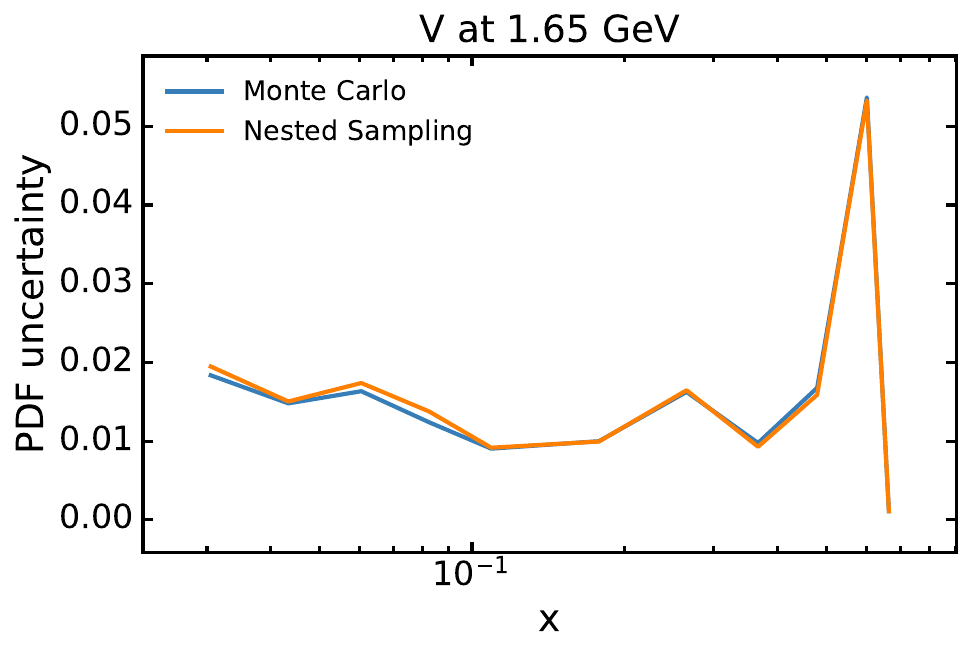} 
%        \caption{Competitors} \label{fig:timing2}
    \end{subfigure}
\caption{A comparison of three PDF fits to hadronic-only data using the simplified model described above. The meaning of the Monte Carlo and Nested Sampling PDFs is described in the text. The columns display the flavours $\Sigma, g, V$ from left to right. The rows display the absolute values of the PDFs, the ratio of the PDFs to the Monte Carlo PDF, and the uncertainties on each of the PDF sets, from top to bottom. Significant disagreement is observed between the two exercises.}
\label{fig:hadronic_comparison}
\end{figure}

The result of Monte Carlo replica fits and Bayesian fits of these data are presented in Fig.~\ref{fig:hadronic_comparison}. In this instance, good agreement between the Monte Carlo PDF and the Nested Sampling PDF is found in the mid to high-$x$ region, whereas significant disagreement is observed at low-$x$; in particular, the uncertainty on the Monte Carlo PDF is reduced compared to the Nested Sampling PDF for the singlet and gluon at lower values of $x$. The most striking reduction is for the gluon at $x = 5.49 \times 10^{-4}$, which undergoes a 90\% reduction in uncertainty when going from the Bayesian PDF to the Monte Carlo PDF. 

The valence distribution, on the other hand, agrees well between the Bayesian and Monte Carlo fits. However, due to the theoretical calculation being poorly understood in the non-linear case, we cannot say with confidence precisely why this behaviour occurs. A possible conjecture comes from the fact that the valence in this region is particularly well-constrained by these data, so that about its best-fit value $V^*$, we can write $V \approx V^* + \Delta V$ with $\Delta V$ particularly small. Then quadratic contributions $V^2$ can be linearised via $V^2 \approx (V^*)^2 + 2 V^* \cdot \Delta V$, resulting in a Gaussian posterior around the best-fit value, where we expect good agreement between the methodologies.

\subsubsection{Results for global fits}
\label{subsubsec:full_fit_results}

\begin{figure}[t]
\centering
    \begin{subfigure}
        \centering
        \includegraphics[scale=0.29]{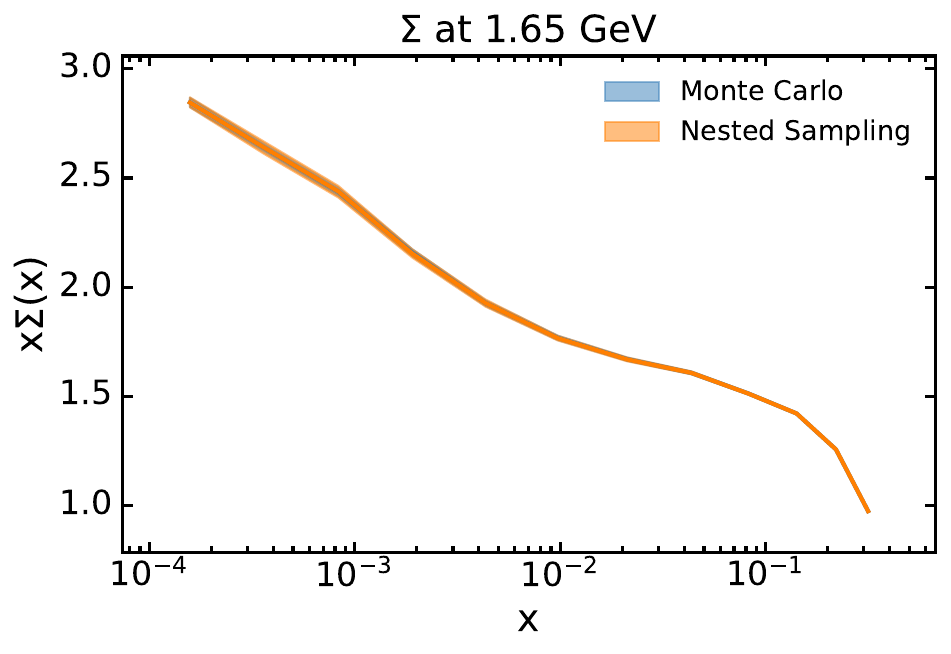} 
%        \caption{Generic} \label{fig:sigma_nnpdf40}
    \end{subfigure}
    \hfill
    \begin{subfigure}
        \centering
        \includegraphics[scale=0.29]{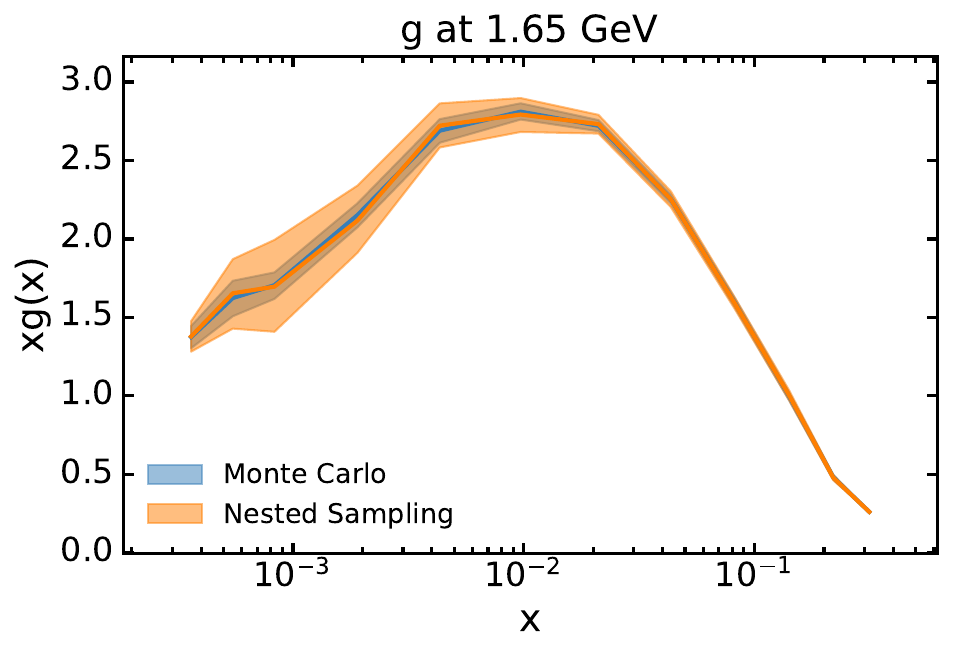} 
%        \caption{Competitors} \label{fig:timing2}
    \end{subfigure}
        \hfill
    \begin{subfigure}
        \centering
        \includegraphics[scale=0.29]{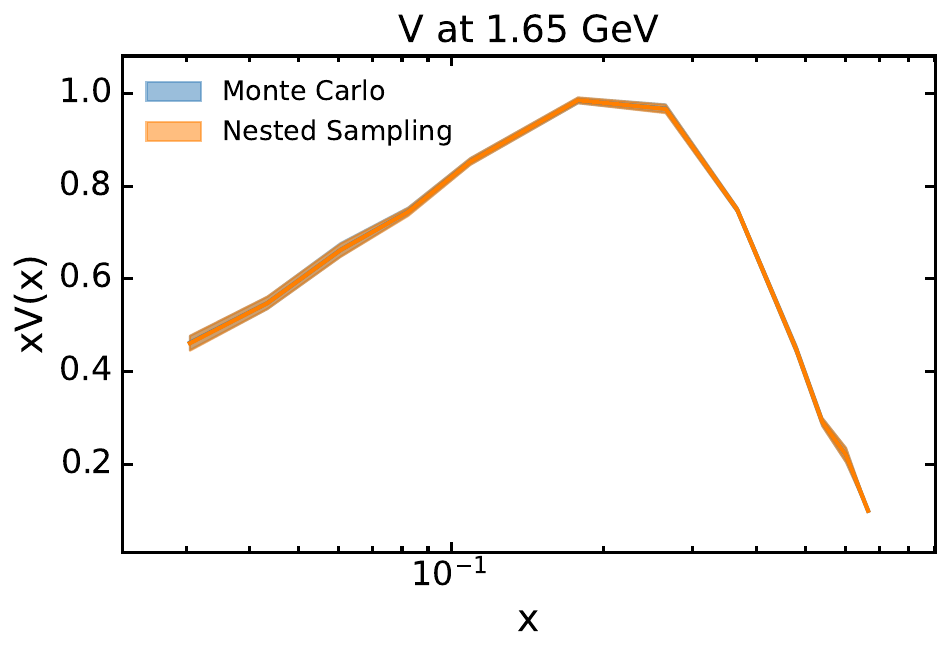} 
%        \caption{Competitors} \label{fig:timing2}
    \end{subfigure}

    \vspace{0.2cm}
    
        \begin{subfigure}
        \centering
        \includegraphics[scale=0.29]{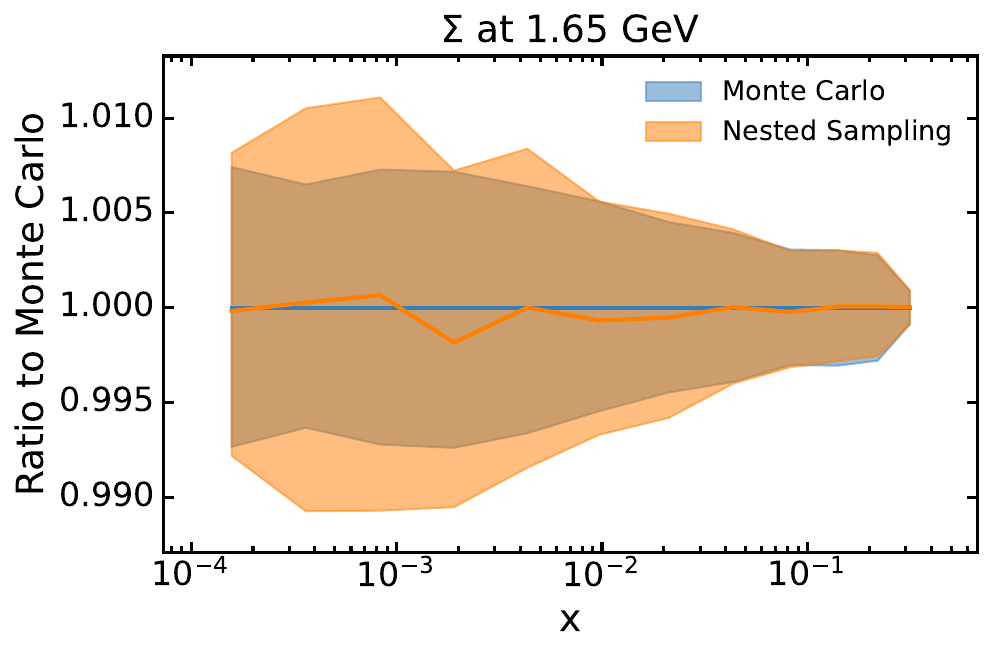} 
%        \caption{Generic} \label{fig:sigma_nnpdf40}
    \end{subfigure}
    \hfill
    \begin{subfigure}
        \centering
        \includegraphics[scale=0.29]{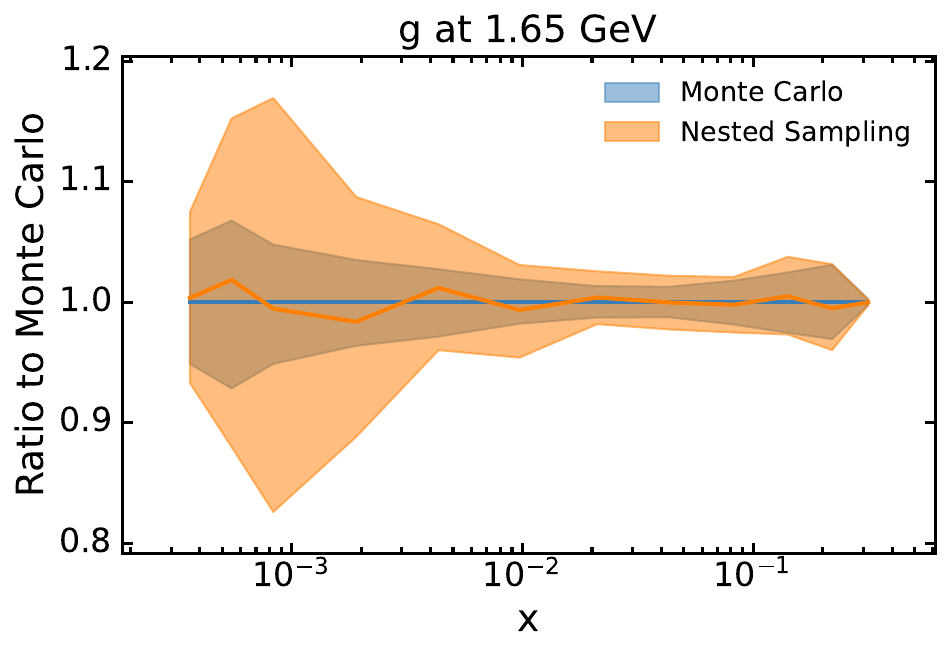} 
%        \caption{Competitors} \label{fig:timing2}
    \end{subfigure}
        \hfill
    \begin{subfigure}
        \centering
        \includegraphics[scale=0.29]{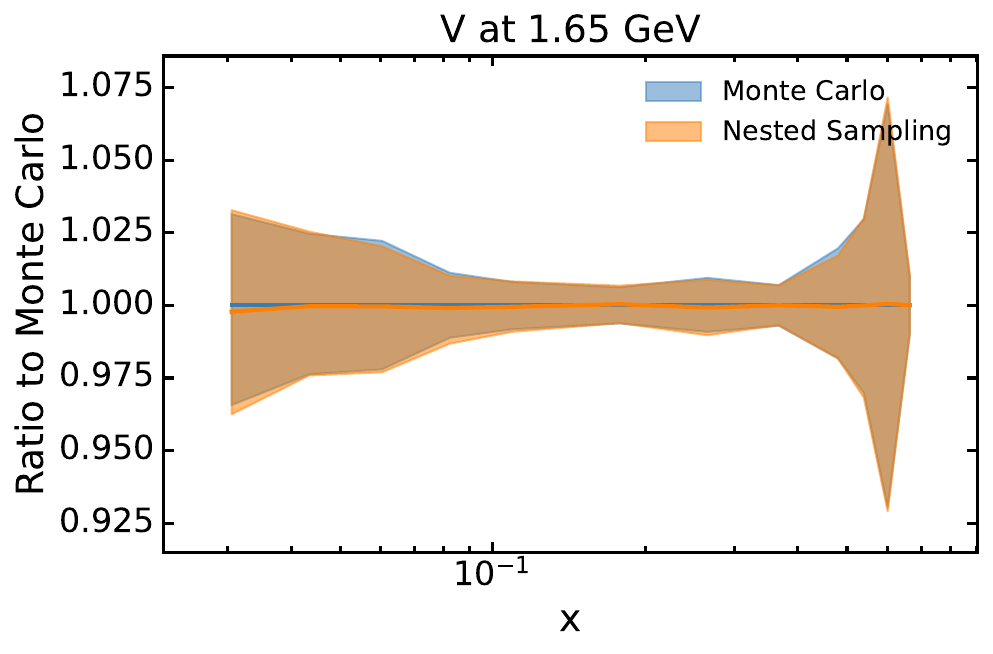} 
%        \caption{Competitors} \label{fig:timing2}
    \end{subfigure}
    
    \vspace{0.2cm}
    
        \begin{subfigure}
        \centering
        \includegraphics[scale=0.29]{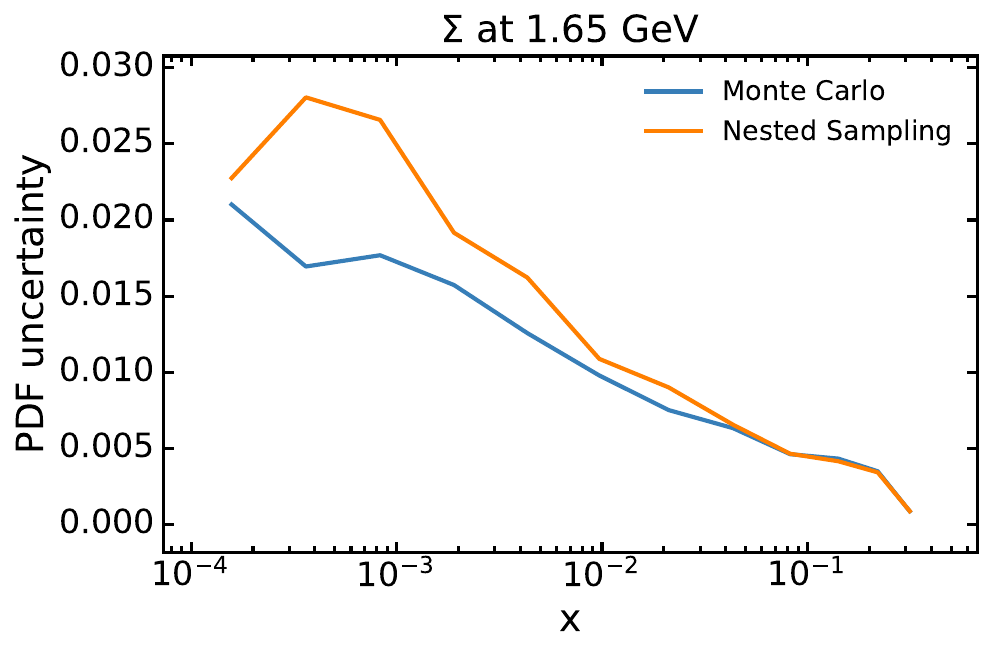} 
%        \caption{Generic} \label{fig:sigma_nnpdf40}
    \end{subfigure}
    \hfill
    \begin{subfigure}
        \centering
        \includegraphics[scale=0.29]{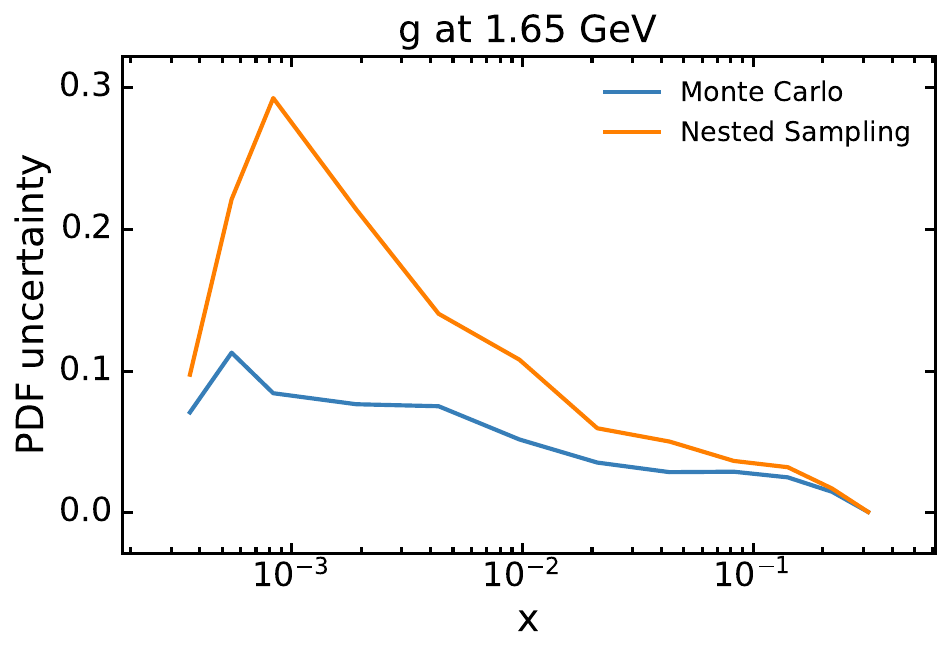} 
%        \caption{Competitors} \label{fig:timing2}
    \end{subfigure}
        \hfill
    \begin{subfigure}
        \centering
        \includegraphics[scale=0.29]{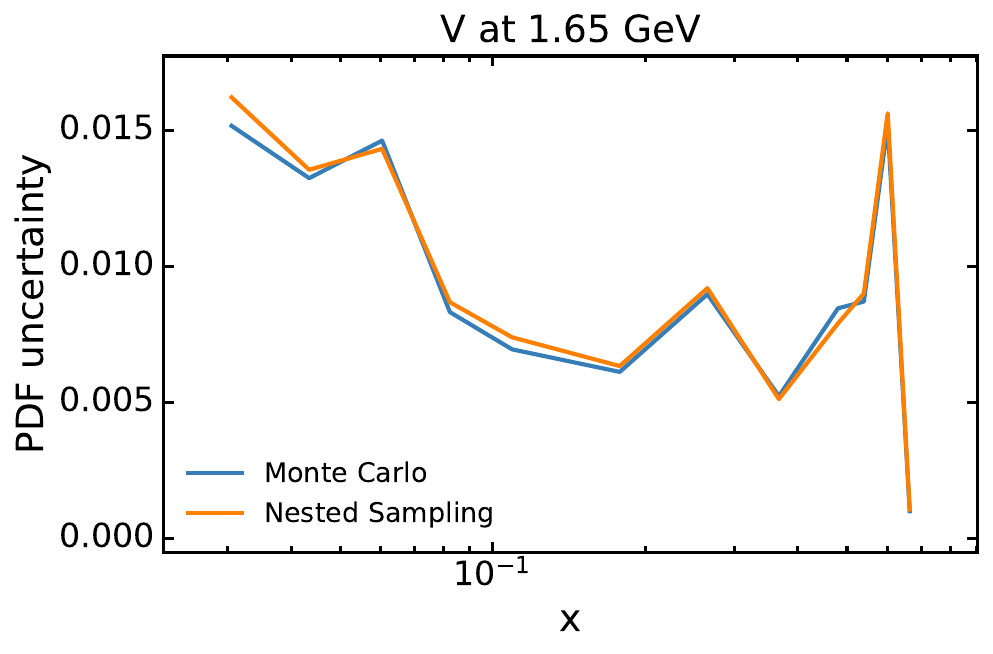} 
%        \caption{Competitors} \label{fig:timing2}
    \end{subfigure}
\caption{The same as figures Fig.~\ref{fig:dis_comparison} and Fig.~\ref{fig:hadronic_comparison}, but now fitting the PDFs using the complete global dataset (combining the DIS and hadronic data from the previous cases). The level of disagreement between the Monte Carlo and Bayesian fits is reduced as compared to the hadronic-only case, but there is still significant tension between the two approaches.}
\label{fig:final_comparison}
\end{figure}

Finally, in this section, we repeat the benchmarking study of the previous two sections for the global dataset, using both the DIS and hadronic data. Given the excellent agreement between the DIS-only Monte Carlo and Bayesian fits, and the poor agreement between the hadronic-only Monte Carlo and Bayesian fits, it is our general expectation that the result of the global fit will be some middle ground between the two cases.

The results of the Monte Carlo and Bayesian fits to the full dataset are presented in Fig.~\ref{fig:final_comparison}. As expected, we find that the Monte Carlo fit still shows some discrepancy from the Nested Sampling one in the low-$x$ region, but the disagreement is less pronounced than in the purely hadronic case. In particular, the singlet PDF $\Sigma$ shows a less significant reduction in uncertainties in the low-$x$ region, of order $20$\% now, in going from the Nested Sampling to the Monte Carlo PDFs. On the other hand, the gluon PDF still shows a considerable discrepancy with the Bayesian result, displaying a conspicuous underestimation of PDF error bands; the worst discrepancy is around $60$\%.

In summary, in the broad dataset considered, the impact of the hadronic data on the full fit is already enough to cause the two methodologies to diverge. Whilst we have only demonstrated this discrepancy in the case of a simplified PDF model, we believe that the result motivates further efforts towards the implementation of Bayesian methods to quantify uncertainties in more realistic fits, especially if in the coming decade new and more accurate hadronic data from the LHC is included in global PDF fits.

% \subsubsection{Extension to realistic PDF fits}
% \label{subsubsec:extension}
% Naturally, the PDF parametrisation given in Eq.~\eqref{eq:leshouches} is not sufficiently flexible to model Nature's PDFs. However, the claims in this work should be taken seriously with regards to... 

\section{Conclusions}
\label{sec:conclusions}

For the first time, this work explored the mathematical foundations of the Monte Carlo replica method beyond linear models, before benchmarking the Monte Carlo replica method against a Bayesian approach in two global fit scenarios from the high-energy physics literature. 

In Sect.~\ref{sec:multi}, we produced an original calculation of the Monte Carlo `posterior' distribution and showed it is only comparable to the Bayesian method in special cases (e.g. for a linear model, the two approaches agree if a sufficiently wide uniform prior is used for the Bayesian). The final result, Eq.~\eqref{eq:final_mc_posterior}, has some of its behaviour understood (particularly the delta function singularities that can appear in the posterior), but for the most part it remains intractably difficult to manipulate.

In Sect.~\ref{subsec:smeft_fits}, we provided a benchmark of the Bayesian method against the Monte Carlo replica method in the context of a fit of the SMEFT Wilson coefficients in the top sector. We find poor agreement between the implied parameter bounds, and as a result, we discourage the use of the Monte Carlo replica method in this arena of global fits. Many SMEFT fit collaborations~\cite{Ethier:2021bye, Ellis:2020unq, Brivio:2022hrb} have already embraced a Bayesian perspective.

In Sect.~\ref{subsec:pdf_fits}, we perform the same benchmarking exercise between the two approaches in the context of a toy fit of PDFs to a global dataset. We found exceptionally good agreement between the Monte Carlo replica method and the Bayesian method when the DIS-only dataset was used (indeed, this must be the case according to the calculation given in Sect.~\ref{sec:multi}). However, when the fit was instead performed on the hadronic-only dataset, we found that the PDF uncertainties implied by the Monte Carlo replica method, despite agreeing well at high-x, were significantly reduced compared to the Bayesian method at low-x. This behaviour persisted, although milder, when we performed a final fit using the global DIS plus hadronic dataset. Whilst the toy fit we performed was a simplification of a modern PDF fit, the results imply that caution is necessary when using the Monte Carlo replica method, especially as more hadronic data are added to the fits. On the other hand, the authors wish to emphasise the point that the level of disagreement, if any, in a realistic PDF fit remains unknown.

While conducting the PDF analysis, we implemented a framework to perform Bayesian inference using Nested Sampling. The developed tool will be featured in a forthcoming publication.
We believe that further efforts should be dedicated towards the development of a fully Bayesian approach to PDF fitting, especially if fit simultaneously in combination with EFT coefficients. This work offers a compelling motivation for such a programme.

\section*{Acknowledgments}
We are extremely grateful to Maria Ubiali for her support with this project. We would like to thank Manuel Morales Alvarado, Anke Biek\"otter, Luigi Del Debbio, Tommaso Giani, and Zahari Kassabov for useful discussions. We thank Richard Nickl for some advice at the early stages of the project. We thank Alastair Young for the discussion of bootstrap methods.
We thank the {\sc NNPDF} collaboration for comments on the manuscript and for making publicly available the data and the fast-kernel tables used in this study.

Mark N. Costantini, Luca Mantani, and James
M. Moore are supported by the European Research Council under the European Union’s
Horizon 2020 research and innovation Programme (grant agreement n.950246), and partially by the STFC
consolidated grant ST/T000694/1 and ST/X000664/1. The work of Maeve Madigan is supported by the
Deutsche Forschungsgemeinschaft (DFG, German Research Foundation) under grant 396021762 – TRR 257
Particle Physics Phenomenology after the Higgs Discovery and by the Alexander von Humboldt Foundation.

\newpage

\appendix

\newpage 
\section{Additional analytic example: purely quadratic theory}
\label{sec:extra_analytic}
In this appendix, we give a further calculation of the Monte Carlo posterior for a toy example. Consider $\vec{t}(c) = \vec{t}_0 + \vec{t}_{\text{quad}} c^2$, a purely quadratic theory in one parameter. In this case, the Jacobian matrix is:
\begin{equation}
\left( \frac{\partial \vec{t}}{\partial c} \right)^T = 2c \vec{t}_{\text{quad}}^T.
\end{equation}
This is of full rank unless $c=0$. The matrix $M(c)$ can be chosen independently of $c$ to consist of $N_{\text{dat}} - 1$ columns which are orthogonal to the vector $\vec{t}_{\text{quad}}$, so that $M^T \vec{t}_{\text{quad}} = 0$. Further, the function $c_p(\vec{d}_p)$ is double-valued for $c\neq 0$, because of the symmetry $c \mapsto -c$ in the theory prediction. It follows that away from $c=0$, the Monte Carlo posterior takes the form:
\begin{align}
& 2\exp\left( -\frac{1}{2} \chi^2_{\vec{d}_0}(\vec{c})\right) \left| \det\left( 2c\vec{t}_{\text{quad}} \bigg| \Sigma M \right) \right| \int\limits_{\Lambda(c)} d^{N_{\text{dat}} - 1}\pmb{\lambda} \exp\left( - \frac{1}{2} \pmb{\lambda}^T M^T \Sigma M \pmb{\lambda} + \pmb{\lambda}^T M^T (\vec{d}_0 - \vec{t}_0) \right) 
\end{align}
The set $\Lambda(c)$ is the set of $\pmb{\lambda}$ such that the pseudodata $\vec{t}(c) + \Sigma M \pmb{\lambda}$ leads to $c$ as a minimum of the $\chi^2$-statistic on this pseudodata. We note that the $\chi^2$ on this pseudodata is given by:
\begin{align}
\chi^2(c') &= \left( \vec{t}(c') - \vec{t}(c) - \Sigma M \pmb{\lambda} \right)^T \Sigma^{-1} \left( \vec{t}(c') - \vec{t}(c) - \Sigma M \pmb{\lambda} \right)\\[1.5ex]
&= \left( \vec{t}_{\text{quad}} (c' - c)^2 - \Sigma M \pmb{\lambda} \right)^T \Sigma^{-1} \left( \vec{t}_{\text{quad}} (c' - c)^2 - \Sigma M \pmb{\lambda} \right)\\[1.5ex]
&= (c' - c)^4 \vec{t}_{\text{quad}}^T \Sigma^{-1} \vec{t}_{\text{quad}} + \pmb{\lambda}^T M^T \Sigma M \pmb{\lambda}.
\end{align}
Since $\Sigma$ is positive definite, we have that $\vec{t}_{\text{quad}}^T \Sigma^{-1} \vec{t}_{\text{quad}} > 0$ and $\pmb{\lambda}^T M^T \Sigma M \pmb{\lambda} > 0$, which shows that the minimum is at $c' = c$. Thus $\Lambda(c) = \mathbb{R}^{N_{\text{dat}} - 1}$, the full range. Evaluating the integral then, we have that for $c\neq 0$ the Monte Carlo posterior takes the form:
\begin{equation}
2\exp\left( -\frac{1}{2} \chi^2_{\vec{d}_0}(\vec{c})\right) \left| \det\left( 2c\vec{t}_{\text{quad}} \bigg| \Sigma M \right) \right| \sqrt{\frac{(2\pi)^{N_{\text{dat}} - 1}}{\det(M^T \Sigma M)}} \exp\left( \frac{1}{2} (\vec{d}_0 - \vec{t}_0)^T M (M^T \Sigma M)^{-1} M^T (\vec{d}_0 - \vec{t}_0) \right),
\end{equation}
using the standard formula for the Gaussian integral. We can simplify this with a couple of tricks. First, observe that by the Gram-Schmidt procedure, we may choose the columns of $M$ to be orthogonal with respect to the inner product induced by $\Sigma$ without loss of generality, so that:
\begin{equation}
M^T \Sigma M = I.
\end{equation}
This also allows us to compute the determinant factor. First, we note:
\begin{equation}
\det\left( 2c\vec{t}_{\text{quad}} \bigg| \Sigma M \right) = 2c \det(\Sigma) \det\left( \Sigma^{-1} \vec{t}_{\text{quad}} \bigg| M \right).
\end{equation}
Further, $\Sigma^{-1} \vec{t}_{\text{quad}}$ and $M$ are already orthogonal with respect to the inner product induced by $\Sigma$. Normalising $\Sigma^{-1} \vec{t}_{\text{quad}}$ with respect to this inner product, we have:
\begin{equation}
(\Sigma^{-1} \vec{t}_{\text{quad}})^T \Sigma (\Sigma^{-1} \vec{t}_{\text{quad}}) = \vec{t}_{\text{quad}}^T \Sigma^{-1} \vec{t}_{\text{quad}}.
\end{equation}
It follows that:\footnote{Note that if $A$ is a square matrix with orthonormal columns with respect to the inner product induced by $\Sigma$, we have $1 = \det(A^T\Sigma A) = \det(\Sigma)\det(A)^2$, which implies $\det(A) = \pm \sqrt{\det(\Sigma^{-1})}$.}
\begin{equation}
\det\left( \frac{\Sigma^{-1} \vec{t}_{\text{quad}}}{\vec{t}_{\text{quad}}^T \Sigma^{-1} \vec{t}_{\text{quad}}} \bigg| M \right) = \pm \sqrt{\det(\Sigma^{-1})},
\end{equation}
and hence the Monte Carlo posterior simplifies to:
\begin{equation}
4\sqrt{(2\pi)^{N_{\text{dat}} - 1} \det(\Sigma) \vec{t}^T_{\text{quad}} \Sigma^{-1} \vec{t}_{\text{quad}}} |c| \exp\left( -\frac{1}{2} \chi^2_{\vec{d}_0}(\vec{c})\right) \exp\left( \frac{1}{2} (\vec{d}_0 - \vec{t}_0)^T M M^T (\vec{d}_0 - \vec{t}_0) \right) \, .
\end{equation}
We can also recognise $MM^T = M(M^T\Sigma M)^{-1} M^T$ as the projector onto the subspace orthogonal to $\vec{t}_{\text{quad}}^T \Sigma^{-1}$, which gives:
\begin{equation}
M(M^T\Sigma M)^{-1} M^T = \Sigma^{-1} - \frac{\Sigma^{-1} \vec{t}_{\text{quad}} \vec{t}_{\text{quad}}^T \Sigma^{-1}}{\vec{t}_{\text{quad}}^T \Sigma^{-1} \vec{t}_{\text{quad}}}.
\end{equation}
Hence the Monte Carlo posterior reduces to the ($M$-independent) simplified form:
\begin{align}
&\notag4\sqrt{(2\pi)^{N_{\text{dat}} - 1} \det(\Sigma) \vec{t}^T_{\text{quad}} \Sigma^{-1} \vec{t}_{\text{quad}}} |c| \exp\left( -\frac{1}{2} \chi^2_{\vec{d}_0}(\vec{c})\right) \\[1.5ex]
& \quad \cdot \exp\left( \frac{1}{2} (\vec{d}_0 - \vec{t}_0)^T \left( \Sigma^{-1} - \frac{\Sigma^{-1} \vec{t}_{\text{quad}} \vec{t}_{\text{quad}}^T \Sigma^{-1}}{\vec{t}_{\text{quad}}^T \Sigma^{-1} \vec{t}_{\text{quad}}}\right) (\vec{d}_0 - \vec{t}_0) \right).
\end{align}
One can check that this reduces to the previous case with $t_{\text{lin}} = 0$ when we work with $N_{\text{dat}} = 1$.

On the other hand, about $c=0$, we can parametrise the theory via $f : \{\emptyset\} \rightarrow \mathbb{R}$ with $f(\emptyset) = 0$. Further, the matrix $M(0)$ can simply be taken to be the $N_{\text{dat}} \times N_{\text{dat}}$ identity matrix. The Monte Carlo posterior is then given by:
\begin{equation}
\exp\left( - \frac{1}{2}\chi^2_{\vec{d}_0}(c) \right) \delta(c) \det(\Sigma) \int\limits_{\Lambda(0)} d^{N_{\text{dat}}}\pmb{\lambda} \exp\left( -\frac{1}{2} \pmb{\lambda}^T \Sigma \pmb{\lambda} + \pmb{\lambda}^T (\vec{d}_0 - \vec{t}_0) \right).
\end{equation}
To compute the set $\Lambda(0)$, we must find all $\pmb{\lambda}$ such that the pseudodata $\vec{t}(0) + \Sigma \pmb{\lambda}$ gives rise to the best-fit value of the parameter $c$. The $\chi^2$-statistic evaluated on this pseudodata is given by:
\begin{align}
\chi^2(c') &= \left( \vec{t}(c') - \vec{t}(0) - \Sigma \pmb{\lambda} \right)^T \Sigma^{-1} \left( \vec{t}(c') - \vec{t}(0) - \Sigma \pmb{\lambda} \right)\\[1.5ex]
&= \left( \vec{t}_{\text{quad}} {c'}^2 - \Sigma \pmb{\lambda} \right)^T \Sigma^{-1} \left( \vec{t}_{\text{quad}} {c'}^2 - \Sigma \pmb{\lambda} \right)\\[1.5ex]
&= {c'}^4 \vec{t}_{\text{quad}}^T \Sigma^{-1} \vec{t}_{\text{quad}} - 2{c'}^2 \vec{t}_{\text{quad}}^T \pmb{\lambda} + \pmb{\lambda}^T \Sigma \pmb{\lambda}\\[1.5ex]
&= \vec{t}_{\text{quad}}^T \Sigma^{-1} \vec{t}_{\text{quad}} \left( {c'}^2 - \frac{\vec{t}_{\text{quad}}^T \pmb{\lambda}}{\vec{t}_{\text{quad}}^T \Sigma^{-1} \vec{t}_{\text{quad}}} \right)^2 - \frac{(\vec{t}_{\text{quad}}^T \pmb{\lambda})^2}{\vec{t}_{\text{quad}}^T \Sigma^{-1} \vec{t}_{\text{quad}}} + \pmb{\lambda}^T \Sigma \pmb{\lambda}.
\end{align}
It follows that if $\vec{t}_{\text{quad}}^T \pmb{\lambda} \leq 0$, we have that $c' = 0$ is the unique minimiser of the $\chi^2$-statistic evaluated on the relevant pseudodata; otherwise, we get different roots. Hence $\Lambda(0) = \{\pmb{\lambda} : \vec{t}_{\text{quad}}^T \pmb{\lambda} \leq 0\}$, which yields the Monte Carlo posterior about $c=0$:
\begin{equation}
\exp\left( - \frac{1}{2}\chi^2_{\vec{d}_0}(c) \right) \delta(c) \det(\Sigma) \int\limits_{\vec{t}_{\text{quad}}^T \pmb{\lambda} \leq 0} d^{N_{\text{dat}}}\pmb{\lambda} \exp\left( -\frac{1}{2} \pmb{\lambda}^T \Sigma \pmb{\lambda} + \pmb{\lambda}^T (\vec{d}_0 - \vec{t}_0) \right).
\end{equation}
Overall then, the Monte Carlo posterior takes the form:
\begin{align}
&\exp\left( - \frac{1}{2}\chi^2_{\vec{d}_0}(c) \right) \bigg[ \delta(c) \det(\Sigma) \int\limits_{\vec{t}_{\text{quad}}^T \pmb{\lambda} \leq 0} d^{N_{\text{dat}}}\pmb{\lambda} \exp\left( -\frac{1}{2} \pmb{\lambda}^T \Sigma \pmb{\lambda} + \pmb{\lambda}^T (\vec{d}_0 - \vec{t}_0) \right) \notag \\[1.5ex]
&\quad + 4\sqrt{(2\pi)^{N_{\text{dat}} - 1} \det(\Sigma) \vec{t}^T_{\text{quad}} \Sigma^{-1} \vec{t}_{\text{quad}}} |c| \exp\left( \frac{1}{2} (\vec{d}_0 - \vec{t}_0)^T \left( \Sigma^{-1} - \frac{\Sigma^{-1} \vec{t}_{\text{quad}} \vec{t}_{\text{quad}}^T \Sigma^{-1}}{\vec{t}_{\text{quad}}^T \Sigma^{-1} \vec{t}_{\text{quad}}}\right) (\vec{d}_0 - \vec{t}_0) \right) \bigg]
\end{align}

\section{Numerical challenges of the Monte Carlo replica method}
\label{sec:numerical_mc}

\begin{figure}[t]
\centering
    \begin{subfigure}
        \centering
        \includegraphics[scale=0.29]{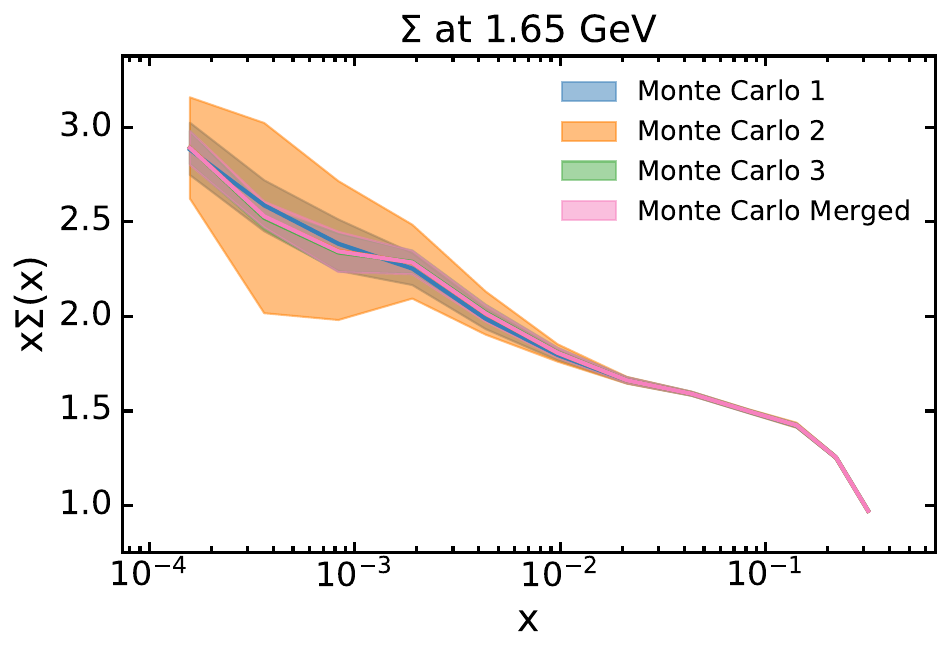} 
%        \caption{Generic} \label{fig:sigma_nnpdf40}
    \end{subfigure}
    \hfill
    \begin{subfigure}
        \centering
        \includegraphics[scale=0.29]{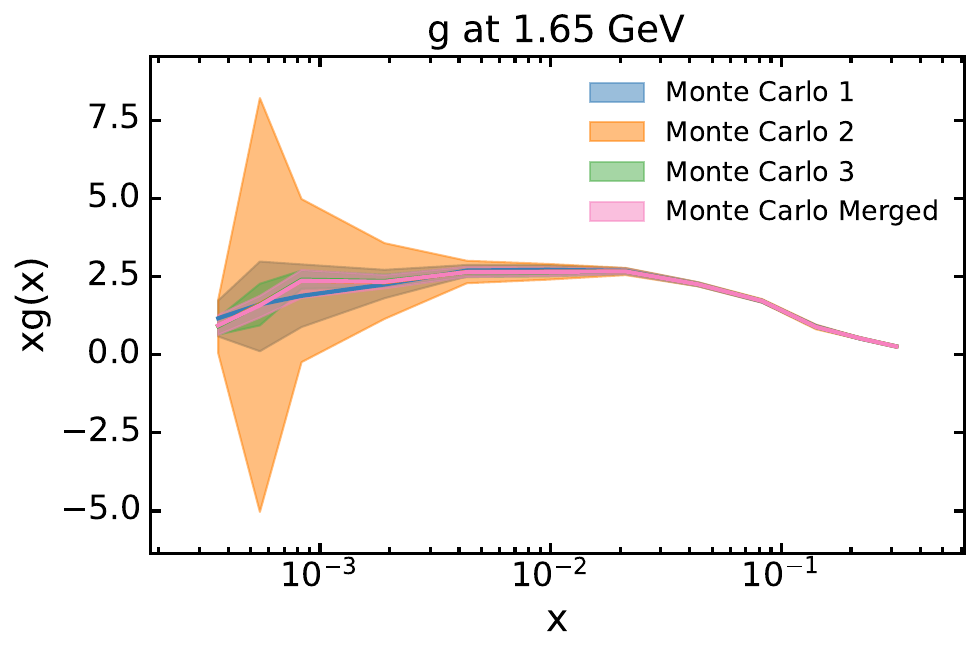} 
%        \caption{Competitors} \label{fig:timing2}
    \end{subfigure}
        \hfill
    \begin{subfigure}
        \centering
        \includegraphics[scale=0.29]{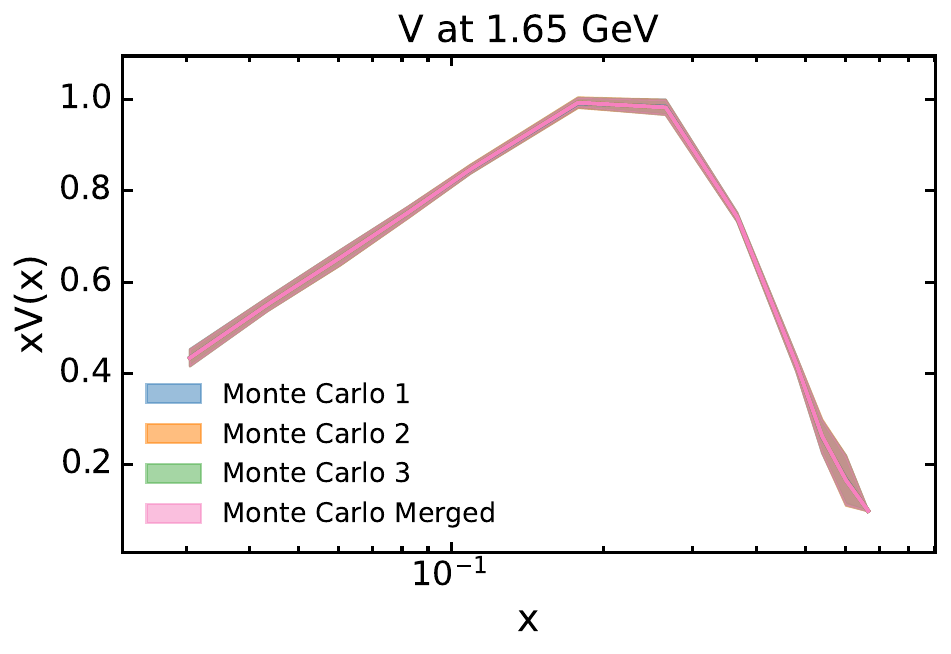} 
%        \caption{Competitors} \label{fig:timing2}
    \end{subfigure}

    \vspace{0.2cm}
    
        \begin{subfigure}
        \centering
        \includegraphics[scale=0.29]{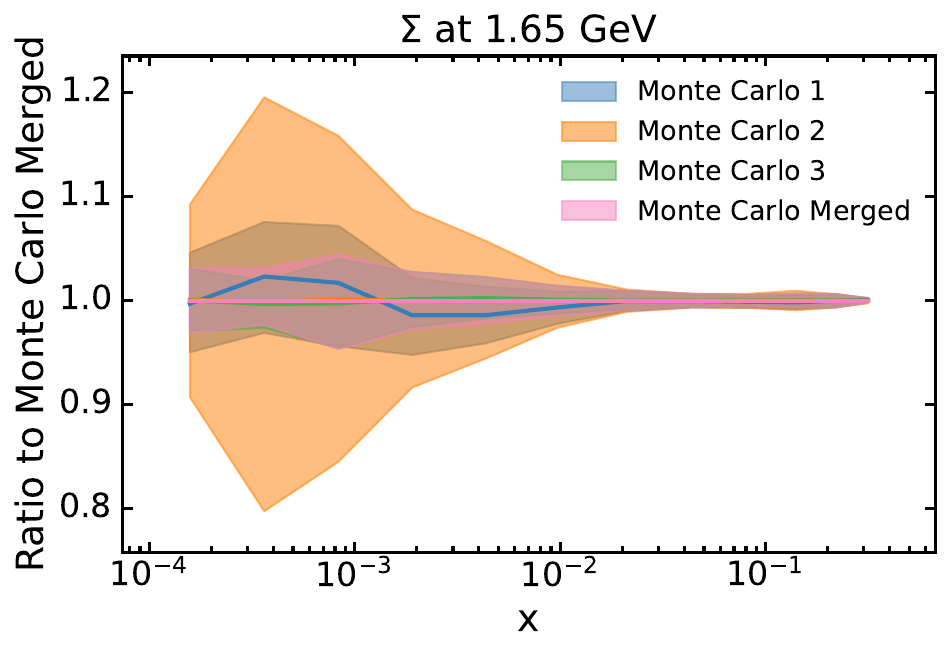} 
%        \caption{Generic} \label{fig:sigma_nnpdf40}
    \end{subfigure}
    \hfill
    \begin{subfigure}
        \centering
        \includegraphics[scale=0.29]{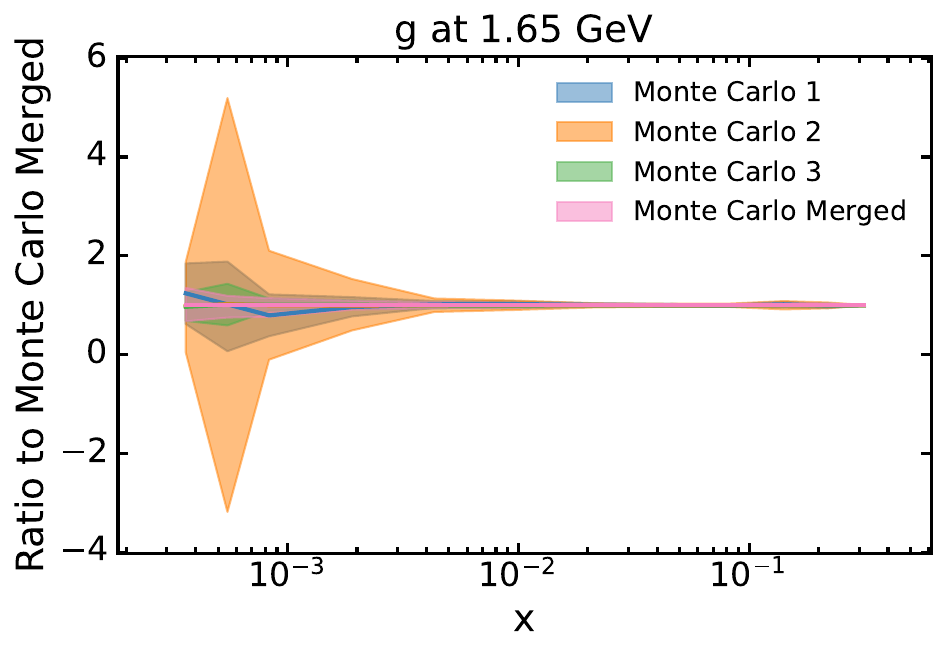} 
%        \caption{Competitors} \label{fig:timing2}
    \end{subfigure}
        \hfill
    \begin{subfigure}
        \centering
        \includegraphics[scale=0.29]{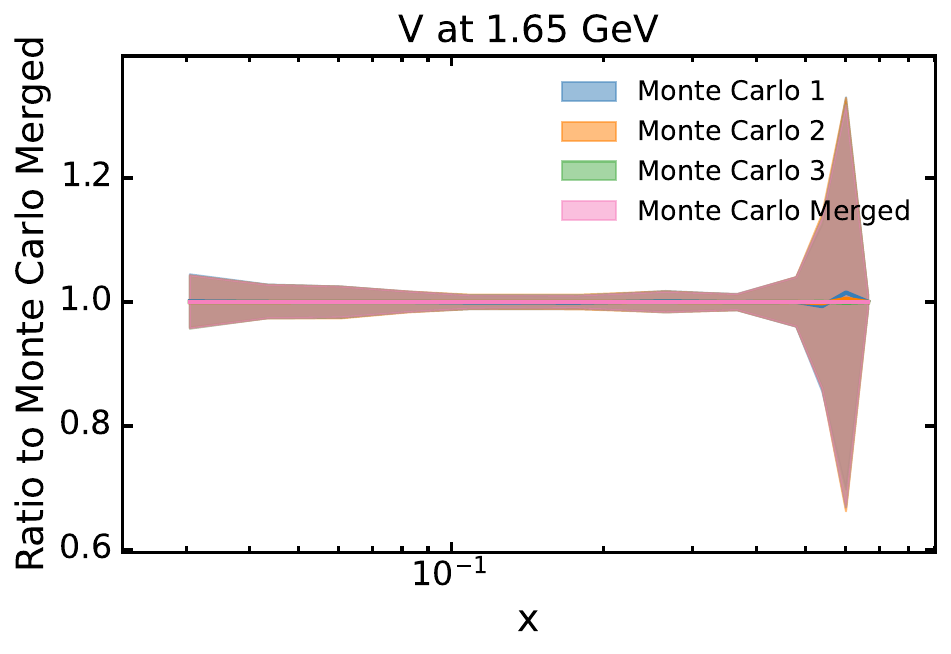} 
%        \caption{Competitors} \label{fig:timing2}
    \end{subfigure}
    
    \vspace{0.2cm}
    
        \begin{subfigure}
        \centering
        \includegraphics[scale=0.29]{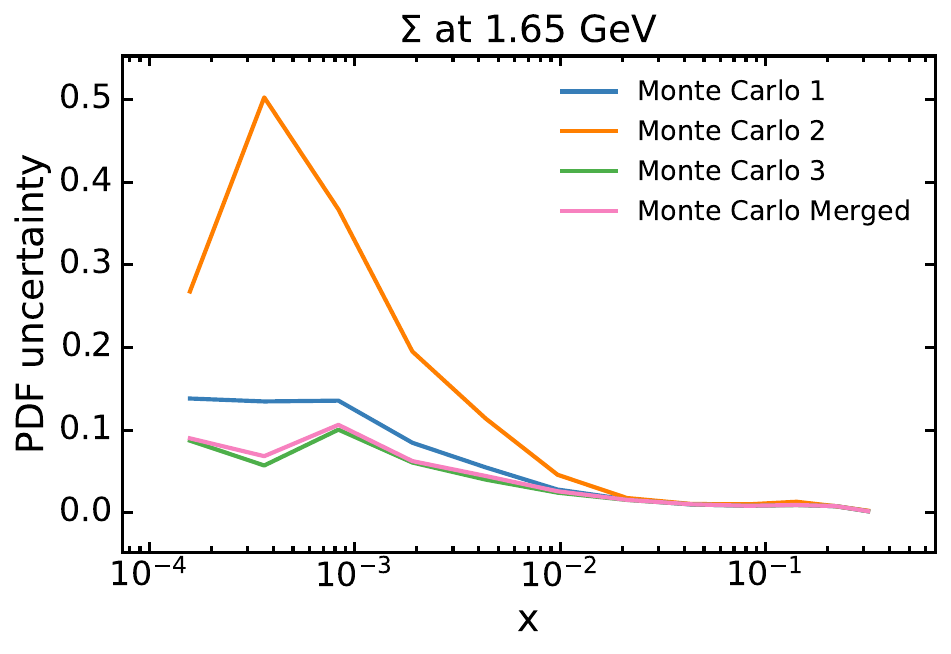} 
%        \caption{Generic} \label{fig:sigma_nnpdf40}
    \end{subfigure}
    \hfill
    \begin{subfigure}
        \centering
        \includegraphics[scale=0.29]{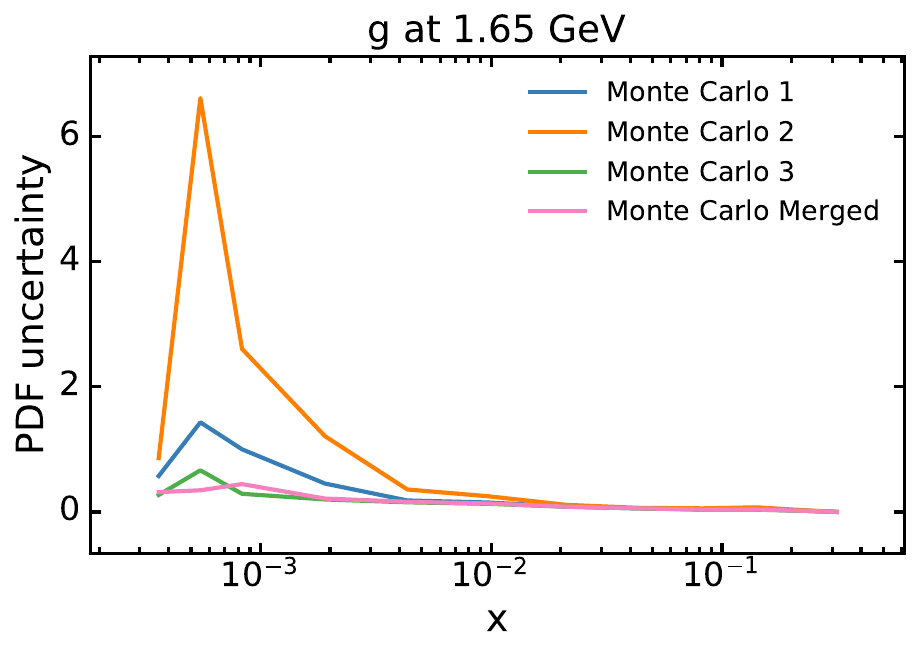} 
%        \caption{Competitors} \label{fig:timing2}
    \end{subfigure}
        \hfill
    \begin{subfigure}
        \centering
        \includegraphics[scale=0.29]{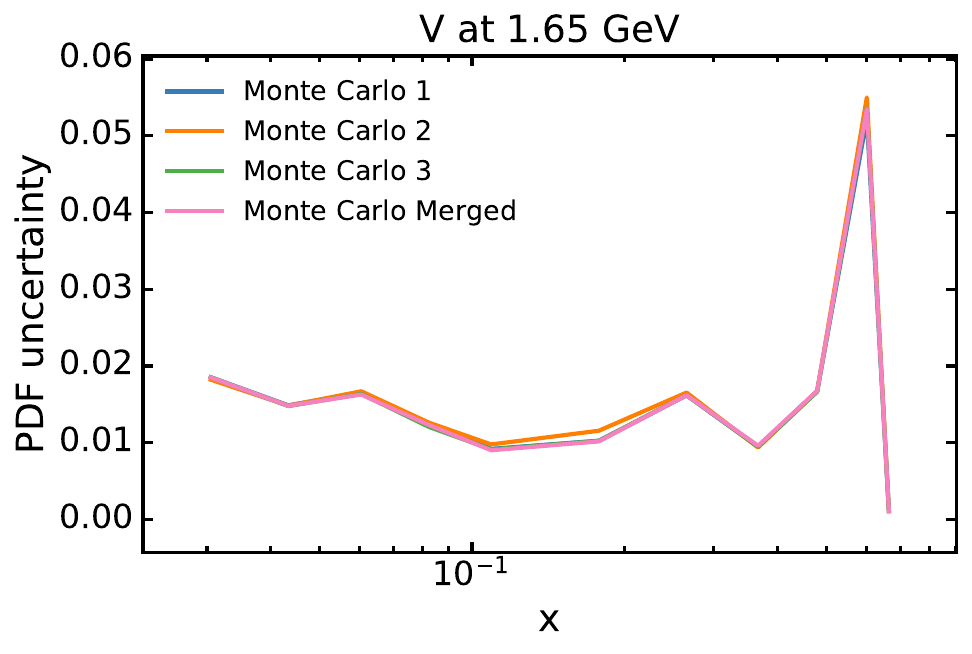} 
%        \caption{Competitors} \label{fig:timing2}
    \end{subfigure}
\caption{Plots showing the ``merging'' procedure used to obtain the final result of the hadronic fit in Sect.~\ref{subsubsec:global_pdf_results}. The figures show that fits ran with different initialisations found different results, discovering local minima. The merging procedure adopted selects the best replica in terms of the training $\chi^2$ from all the fits, making the result more robust and increasing the confidence that the global minimum has been reached.}
\label{fig:merging_mc}
\end{figure}

In this section, we discuss some technical challenges related to the MC methodology. As an initial disclaimer, these are not insurmountable issues and with enough fine-tuning and understanding of the problem at hand, they can be overcome. However, during the course of this and previous studies~\cite{Kassabov:2023hbm, Costantini:2024xae}, we often had to deal with unexpected difficulties which hindered the results of the fits.

The origin of the problem is that while our objective is to obtain a sample from the posterior distribution of the model parameters, this is achieved not through a sampling algorithm, but through an optimisation procedure which relies on gradient descent. For this reason, the presence of multiple local minima and of degeneracies, that is, continuous surfaces all of which have equal minimal $\chi^2$-statistic, can impact the result of the fit in ways that could be difficult to spot. These problems are instead well tackled by dedicated posterior sampling algorithms like Nested Sampling or MCMC, where multimodal distributions can be mapped efficiently and reliably.

The first and most basic issue has to do with making sure that the global minimum was actually reached. In our experience, this can be problematic in two distinct cases: when the posterior has a multimodal character and in the presence of strong correlations. Both of these can ruin a successful optimisation because in the first case, one could get trapped in a local minima while in the second, convergence to the true minimum can be very slow (since, in presence of strong correlations, the gradient can be very small).

The methodological uncertainty problem is well known by NNPDF, which has designed a suite of closure tests specifically to estimate the level of success of the neural network in finding the global minimum. In a level $0$ closure test, multiple replicas are generated with the same pseudodata and all fit with the same model, the only difference being a distinct initialisation. In the ideal case, all of the replicas would find the same minimum and the resulting PDF set would have no uncertainty. However, this is not what it is observed by the collaboration and a methodological uncertainty is found, indicating a dependence on the initialisation of the parameters when performing the fit.

In our studies, we have repeatedly found similar problems and at times, the strong dependence on the initialisation has been surprising. In particular, the presence of degeneracies can be quite disruptive and without further analysis difficult to discover. As a matter of fact, while in a posterior sampling algorithm the flat directions are discovered automatically, in a Monte Carlo replica setting they can disguise as constrained directions simply because of a poor initialisation.

Specifically, in order to obtain the final result presented in Sec.~\ref{subsubsec:global_pdf_results}, we ran multiple Monte Carlo replica fits, changing only the initialisation of the model, and we `merged' them in order to obtain the fitted replicas with the lowest training $\chi^2$. By doing so, the result of the fit becomes more and more robust, since we are able to reject replicas that are stuck in local minima, even if they apparently had a good $\chi^2$. In Fig.~\ref{fig:merging_mc}, we show an example of this from the hadronic PDF fit where we merged 7 different fits (3 of them shown in the plots). It can clearly be seen that the final merged result is considerably different from some of the preliminary fits that had been obtained.

These are common problems in machine learning and plenty of techniques have been designed to confront them. However, it must be noted that the objective is to estimate with high precision physical quantities and not get a ball-park estimate. Making sure that these are reliable is of paramount importance, especially for the years ahead. The transparency and ease of use of the posterior sampling methods go in this direction and allow for a much more controlled fit environment.

\section{Impact of training-validation splits}
\label{app:training_validation_splits}

In one of the footnotes in Sect.~\ref{subsec:bayes_inf}, we briefly mentioned that the Monte Carlo replica method is often applied using a `training-validation split', together with a cross-validation stopping, when the best-fit parameters are found for a given piece of pseudodata. In more detail, given a piece of pseudodata $\vec{d}_p$ generated from the simulated distribution $\mathcal{N}(\vec{d}_0, \Sigma)$, we additionally generate a random subset $S \subseteq \{1,...,N_{\text{dat}}\}$ of `training indices' as some fraction $f_{\text{train}}$ of the full set of data indices. The best-fit values of the parameters are then given by some from of gradient descent, starting at some random initial value of the parameters (the choice of the initialisation is a point discussed in App.~\ref{sec:numerical_mc}), such that the $\chi^2$-statistic to the pseudodata \textit{evaluated only on the training subset} $\{ (d_p)_i : i \in S\}$ is minimised, while the $\chi^2$-statistic to the pseudodata \textit{evaluated only on the validation subset} $\{ (d_p)_i : i \not\in S\}$ is also still decreasing. At the point where the $\chi^2$-statistic to the pseudodata evaluated on the validation subset starts increasing, the minimisation is halted.

% It should be observed that the use of training-validation splits and the cross-validation method to an extent defeats the intuition of the Monte Carlo replica method. Indeed, the intuitive \textit{point} of the method is to fit the random noise of the pseudodata in order to propagate experimental uncertainties to the fitted parameters. If one does \textit{not} reach the true minimum, by using cross-validation to stop early, then we risk even further obscuring the statistical methodology used to obtain the parameter distributions (indeed, given that we do not have a good theoretical understanding of Eq.~\eqref{eq:final_mc_posterior}, we are even less likely to have good theoretical control when training-validation splits and cross-validation are also included).

This point is discussed briefly in the \smefit{} code paper, Ref.~\cite{Giani:2023gfq}; in particular, they state that a training-validation split with a cross-validation stopping is not required for the Monte Carlo replica method in the context of the SMEFT fits they conduct.
A direct quote from their paper is the following: \textit{`We note that as opposed to the PDF fit
case no cross-validation is required here, since overlearning is not possible for a discrete parameter space,
where the best-fit value coincides with the absolute maximum of the likelihood.'} In our paper we work entirely with discrete PDFs evaluated on a grid of $36$ points, so we consider the \smefit{} reasoning adequate to justify the lack of the inclusion of a training-validation split in our study.\\

\begin{figure}
\centering
    \begin{subfigure}
        \centering
        \includegraphics[scale=0.29]{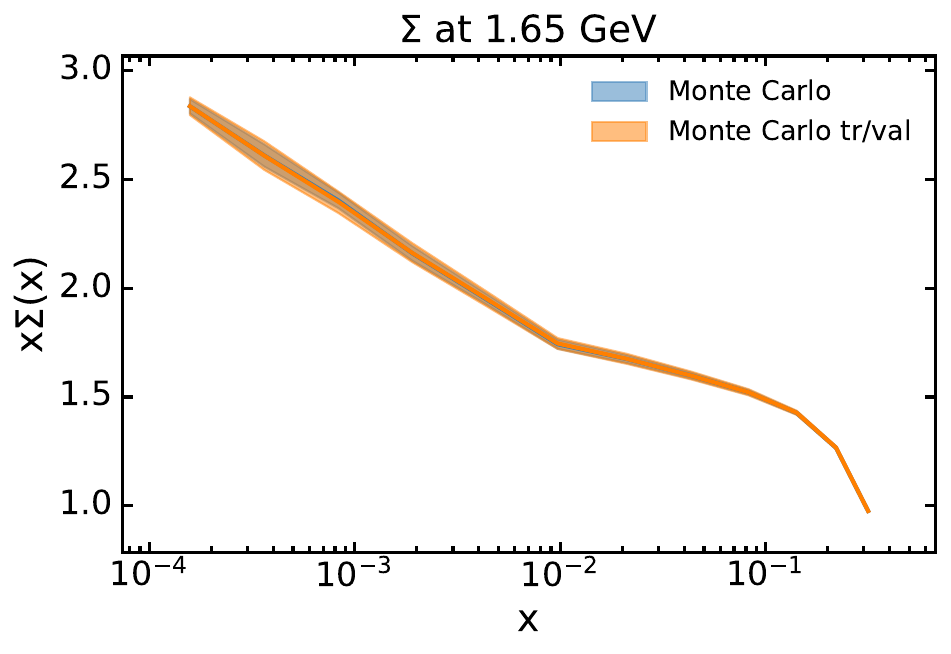} 
%        \caption{Generic} \label{fig:sigma_nnpdf40}
    \end{subfigure}
    \hfill
    \begin{subfigure}
        \centering
        \includegraphics[scale=0.29]{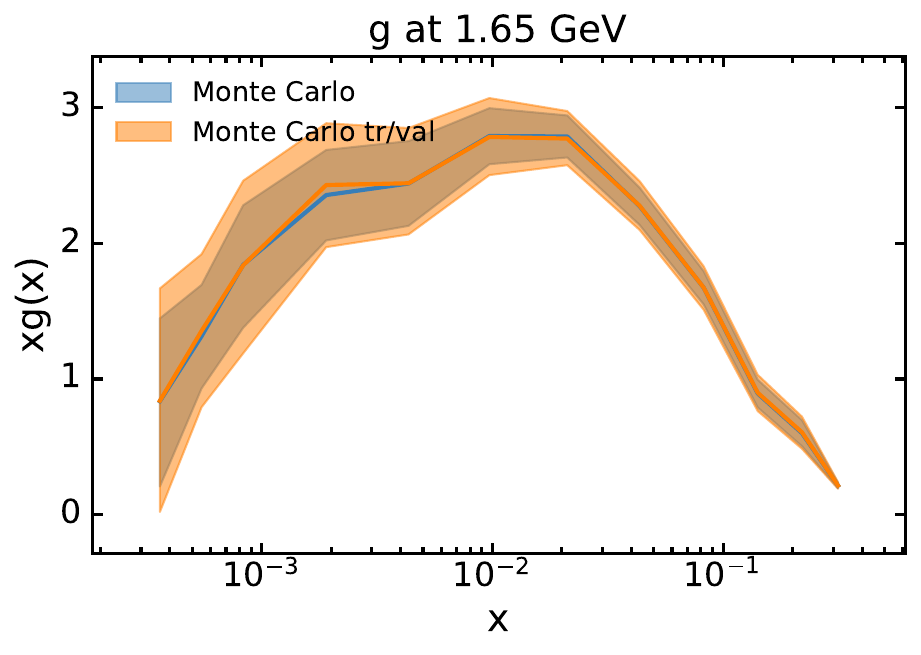} 
%        \caption{Competitors} \label{fig:timing2}
    \end{subfigure}
        \hfill
    \begin{subfigure}
        \centering
        \includegraphics[scale=0.29]{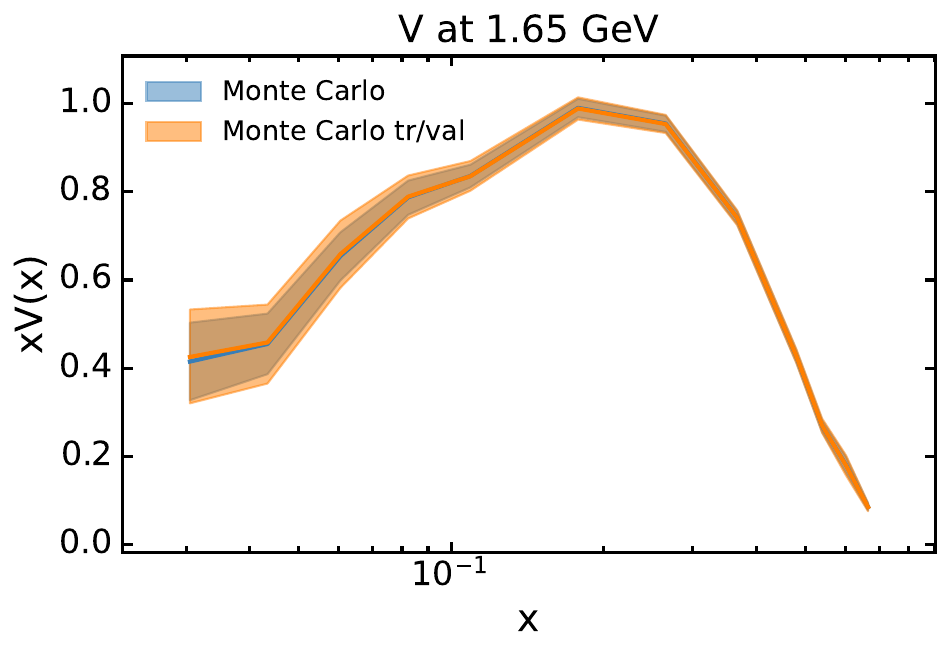} 
%        \caption{Competitors} \label{fig:timing2}
    \end{subfigure}

    \vspace{0.2cm}
    
        \begin{subfigure}
        \centering
        \includegraphics[scale=0.29]{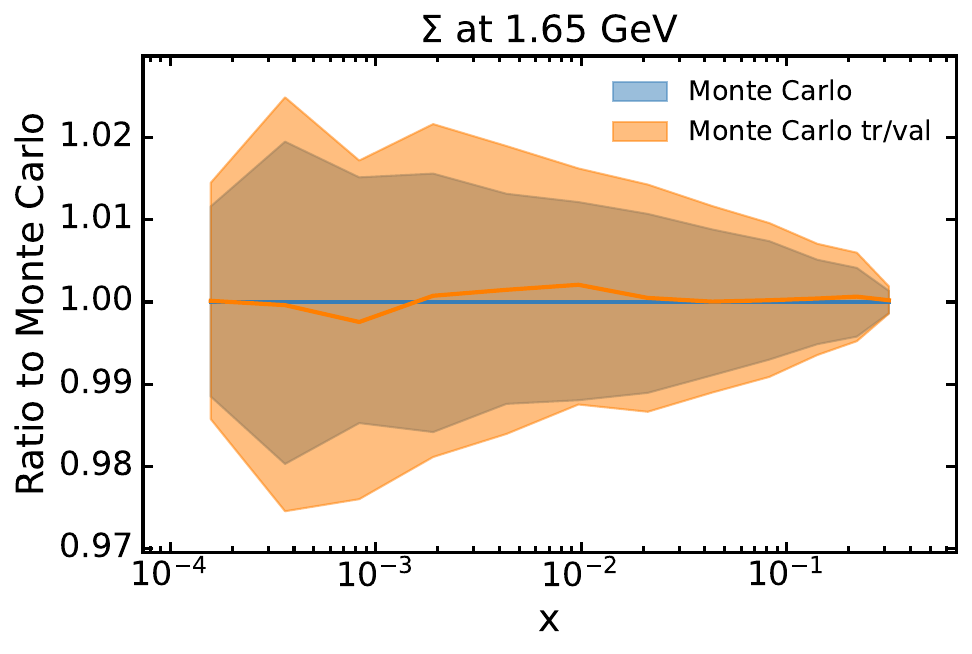} 
%        \caption{Generic} \label{fig:sigma_nnpdf40}
    \end{subfigure}
    \hfill
    \begin{subfigure}
        \centering
        \includegraphics[scale=0.29]{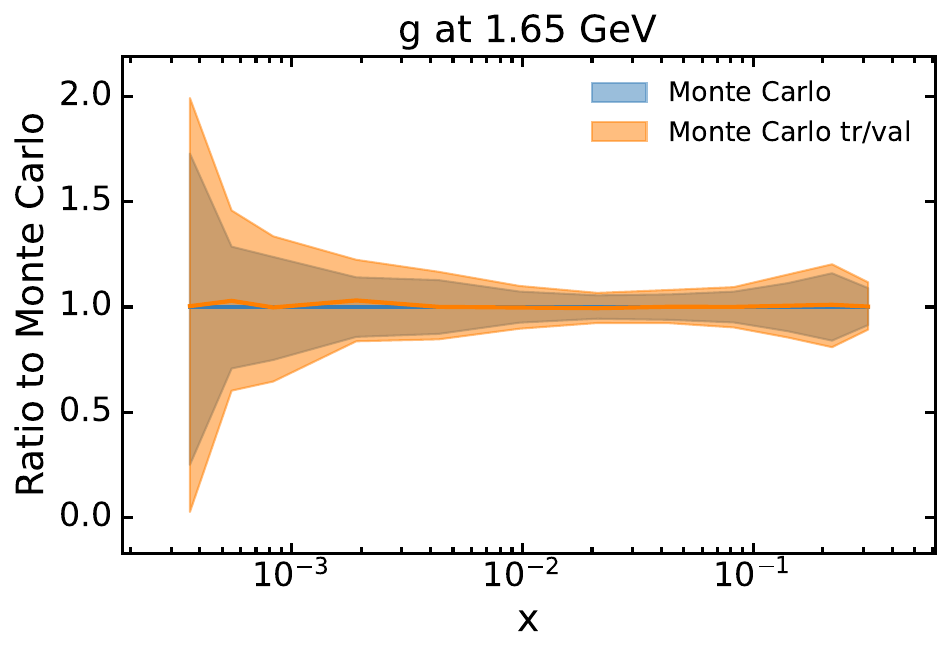} 
%        \caption{Competitors} \label{fig:timing2}
    \end{subfigure}
        \hfill
    \begin{subfigure}
        \centering
        \includegraphics[scale=0.29]{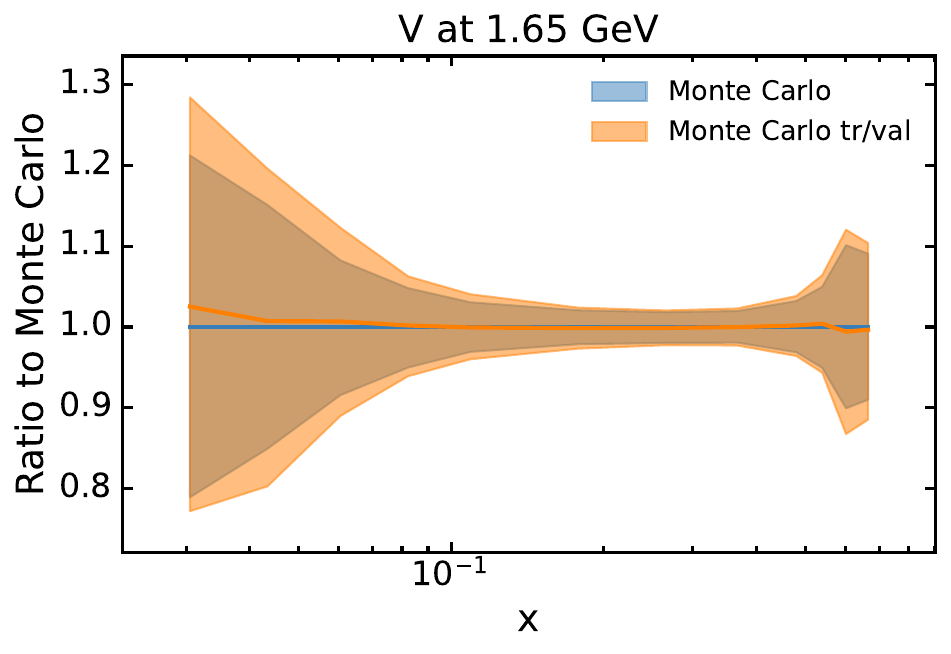} 
%        \caption{Competitors} \label{fig:timing2}
    \end{subfigure}
    
    \vspace{0.2cm}
    
        \begin{subfigure}
        \centering
        \includegraphics[scale=0.29]{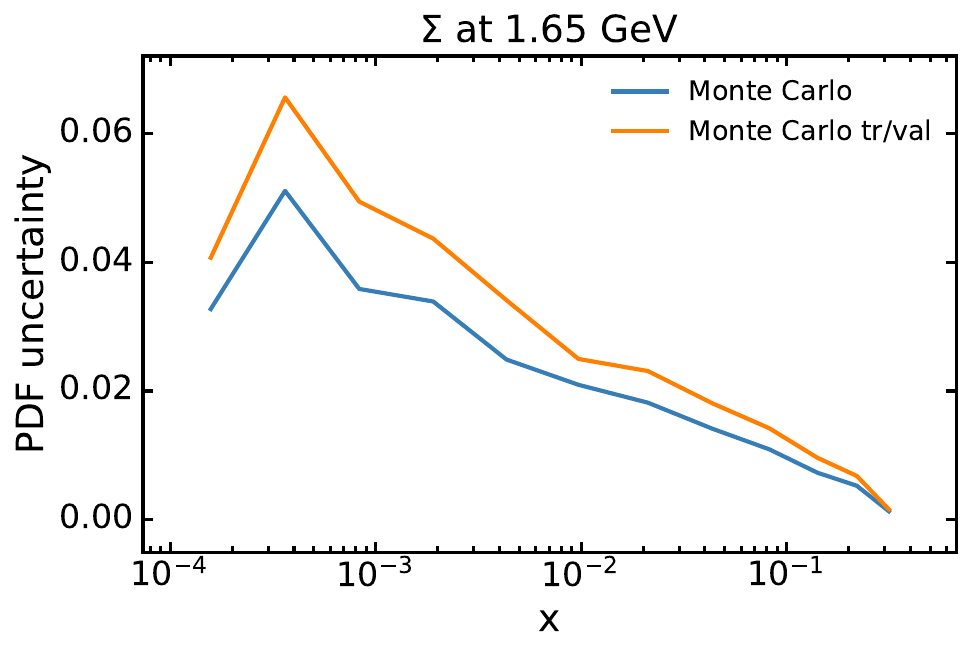} 
%        \caption{Generic} \label{fig:sigma_nnpdf40}
    \end{subfigure}
    \hfill
    \begin{subfigure}
        \centering
        \includegraphics[scale=0.29]{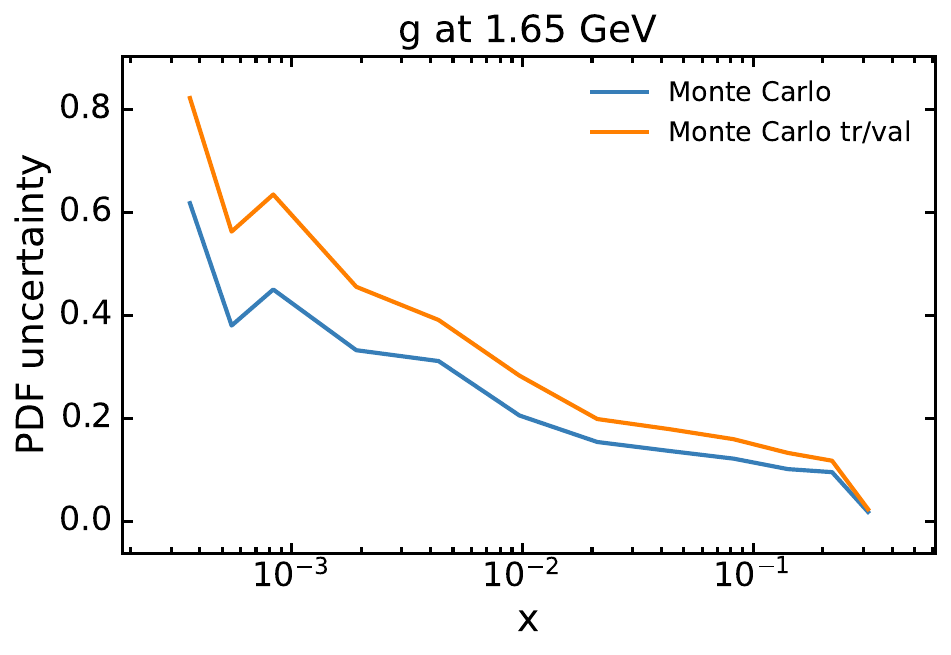} 
%        \caption{Competitors} \label{fig:timing2}
    \end{subfigure}
        \hfill
    \begin{subfigure}
        \centering
        \includegraphics[scale=0.29]{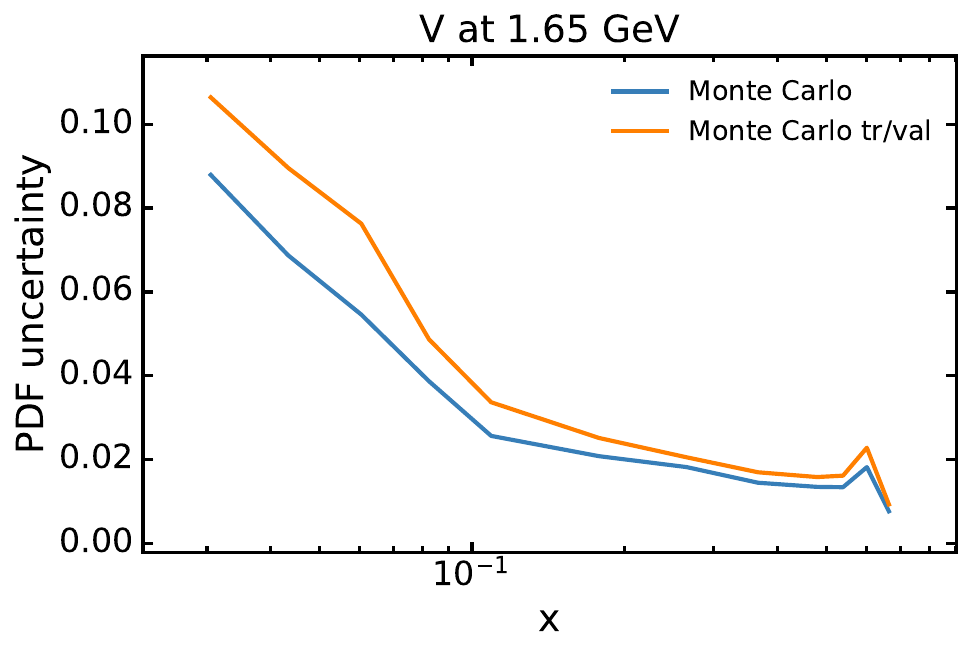} 
%        \caption{Competitors} \label{fig:timing2}
    \end{subfigure}
\caption{A comparison of Monte Carlo PDF fits to DIS-only data using the complete dataset when minimising to the pseudodata, versus using a random training-validation split for each pseudodata replica, and applying cross-validation. As we saw in Sect.~\ref{subsubsec:dis_only_fit}, the Monte Carlo result using the complete dataset agrees perfectly with both a numerical and analytic Bayesian approach, as expected. On the other hand, when using random training-validation splits and applying cross-validation, the uncertainties in the Monte Carlo approach are systematically \textit{overestimated}.}
\label{fig:dis_tr_val}
\end{figure}

On the other hand, for completeness, we have also computed all of the Monte Carlo PDF results using training-validation splits (different replica by replica), with a training fraction $f_{\text{train}} = 0.75$, and cross-validation stopping. In Fig.~\ref{fig:dis_tr_val}, we display a comparison of the result of the DIS-only Monte Carlo PDF fit using the parametrisation and data discussed in Sect.~\ref{subsubsec:dis_only_fit} with the same Monte Carlo fit conducted with a training-validation split. We observe that the uncertainties are systematically inflated as a result of including the training-validation split; for the linear problem, this is well understood, because each replica effectively sees less data, so we expect larger uncertainties as a result. In particular, we have learned that the use of a training-validation split for the linear problem artificially inflates uncertainties, as compared with the analytic solution we presented in Sect.~\ref{subsubsec:dis_only_fit}, with which Monte Carlo without a training-validation split agrees perfectly.

\begin{figure}
\centering
    \begin{subfigure}
        \centering
        \includegraphics[scale=0.29]{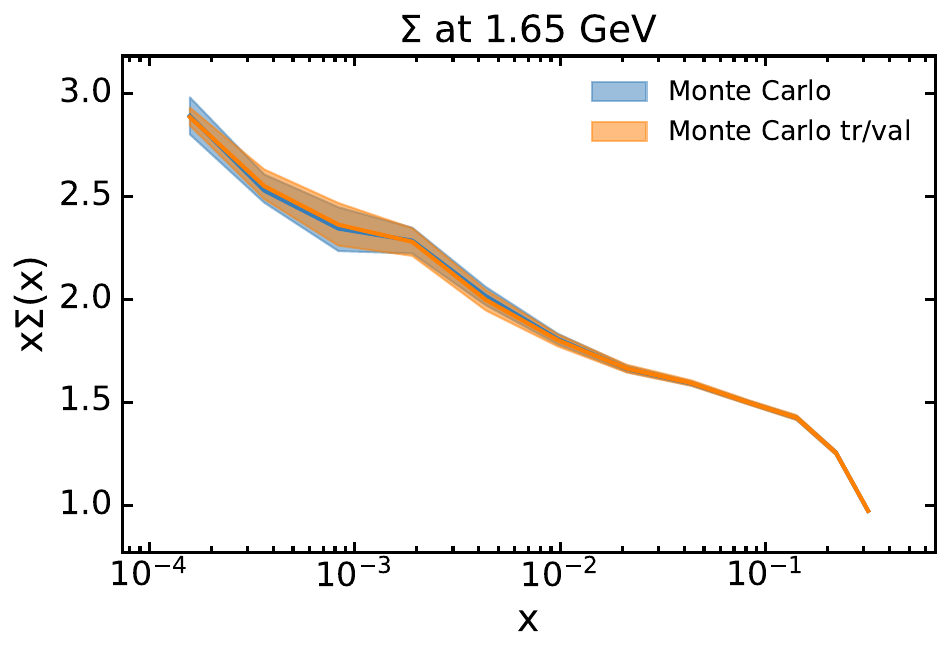} 
%        \caption{Generic} \label{fig:sigma_nnpdf40}
    \end{subfigure}
    \hfill
    \begin{subfigure}
        \centering
        \includegraphics[scale=0.29]{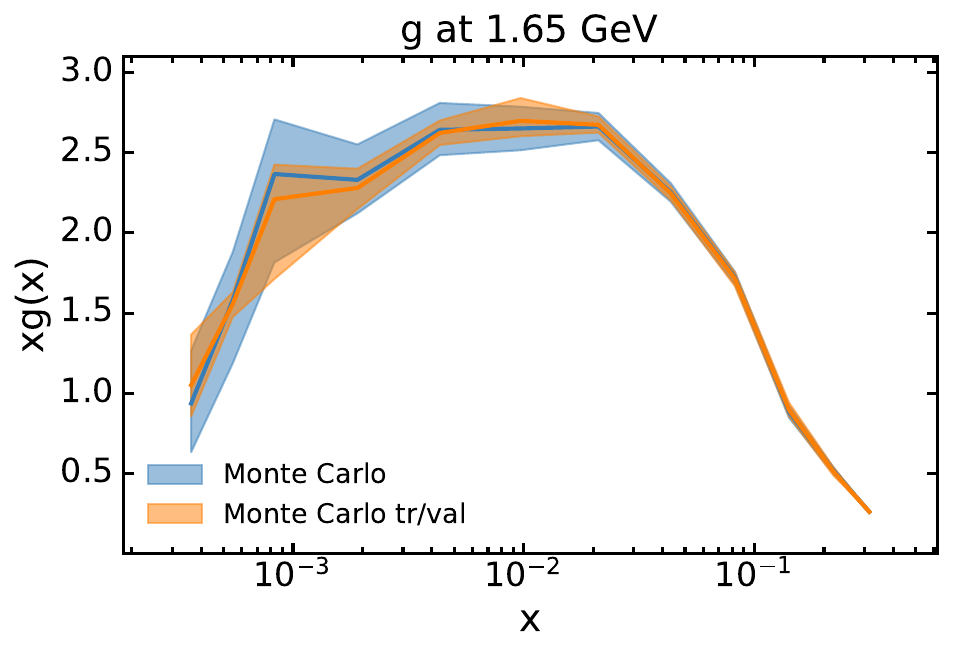} 
%        \caption{Competitors} \label{fig:timing2}
    \end{subfigure}
        \hfill
    \begin{subfigure}
        \centering
        \includegraphics[scale=0.29]{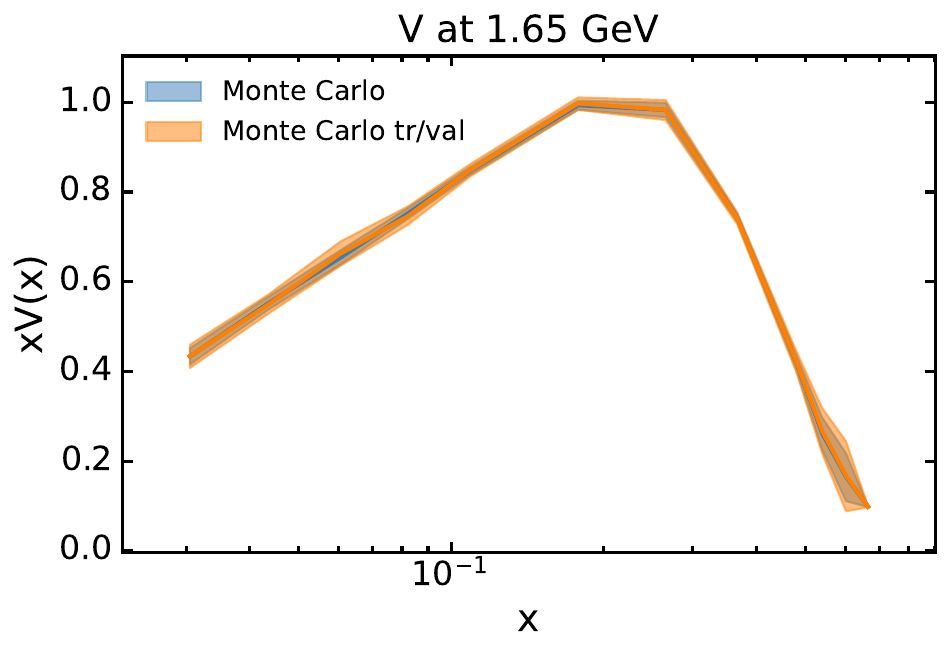} 
%        \caption{Competitors} \label{fig:timing2}
    \end{subfigure}

    \vspace{0.2cm}
    
        \begin{subfigure}
        \centering
        \includegraphics[scale=0.29]{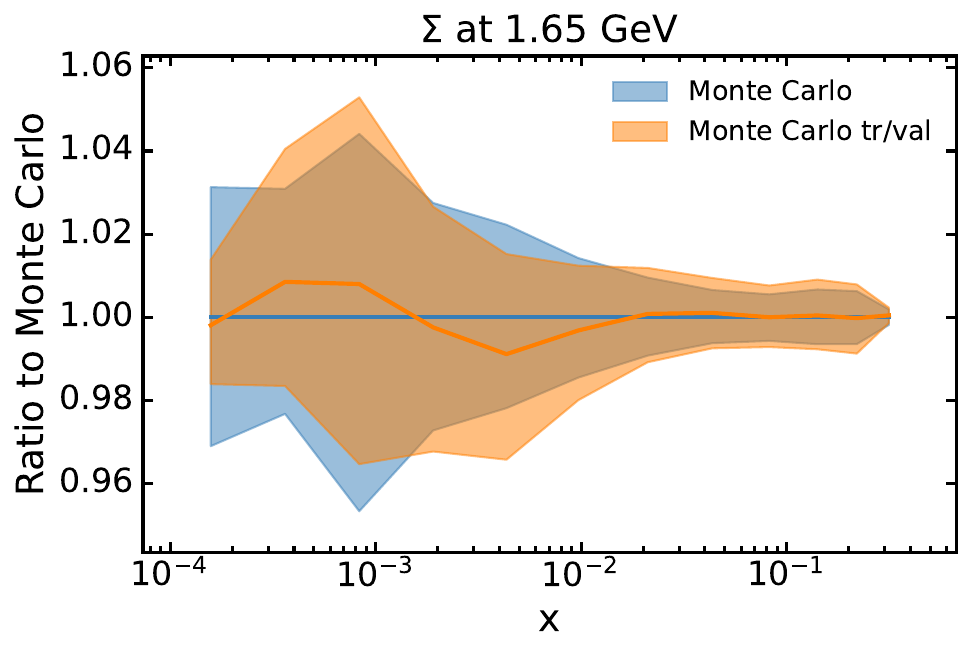} 
%        \caption{Generic} \label{fig:sigma_nnpdf40}
    \end{subfigure}
    \hfill
    \begin{subfigure}
        \centering
        \includegraphics[scale=0.29]{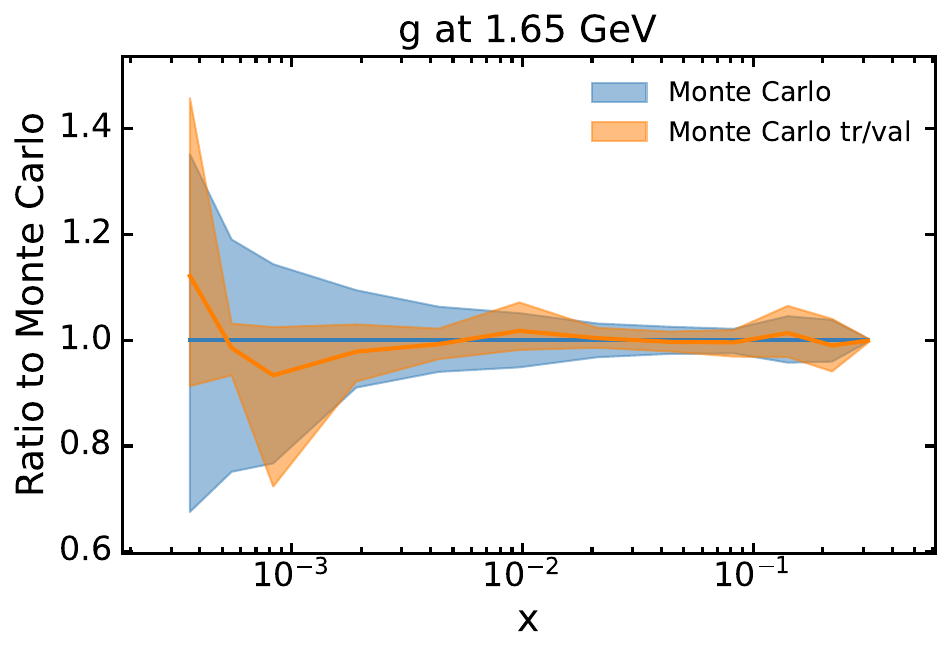} 
%        \caption{Competitors} \label{fig:timing2}
    \end{subfigure}
        \hfill
    \begin{subfigure}
        \centering
        \includegraphics[scale=0.29]{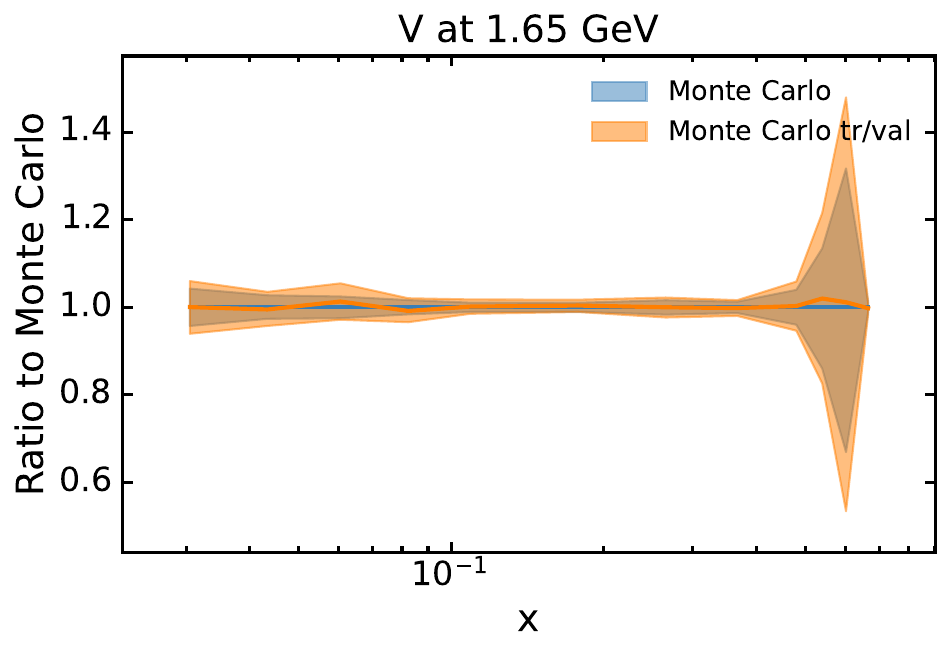} 
%        \caption{Competitors} \label{fig:timing2}
    \end{subfigure}
    
    \vspace{0.2cm}
    
        \begin{subfigure}
        \centering
        \includegraphics[scale=0.29]{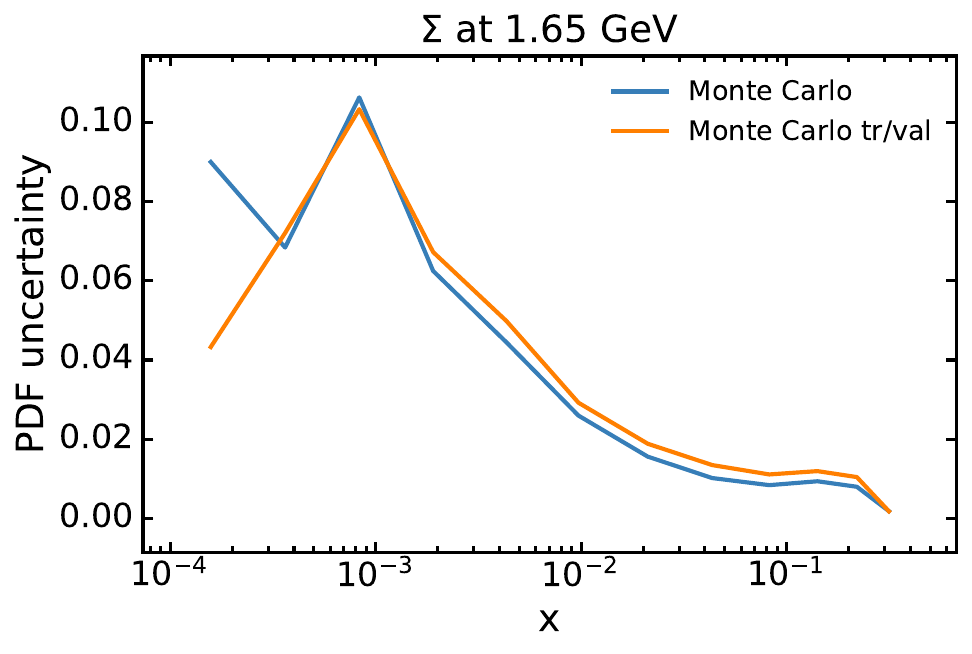} 
%        \caption{Generic} \label{fig:sigma_nnpdf40}
    \end{subfigure}
    \hfill
    \begin{subfigure}
        \centering
        \includegraphics[scale=0.29]{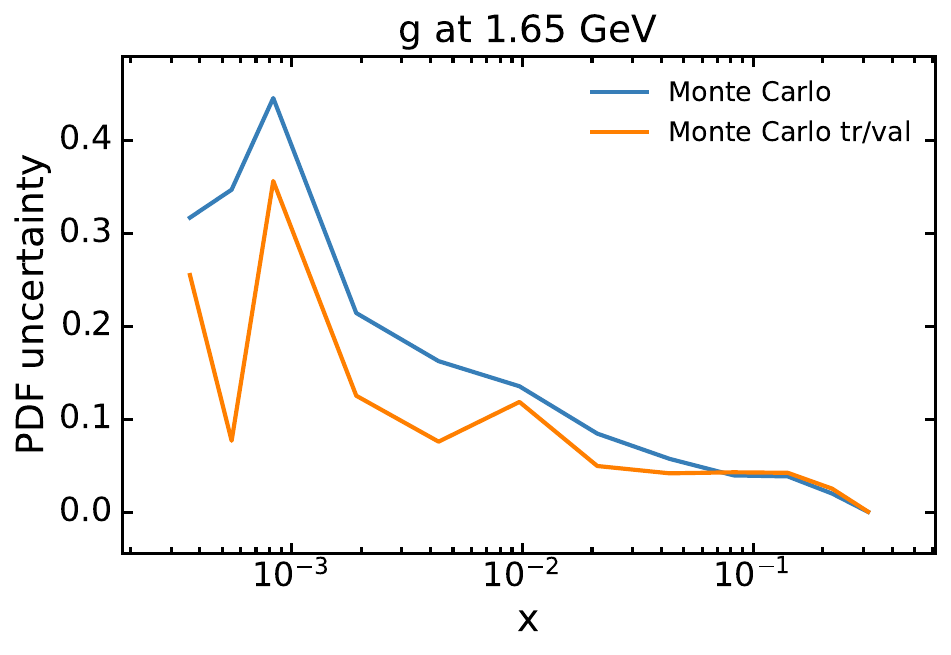} 
%        \caption{Competitors} \label{fig:timing2}
    \end{subfigure}
        \hfill
    \begin{subfigure}
        \centering
        \includegraphics[scale=0.29]{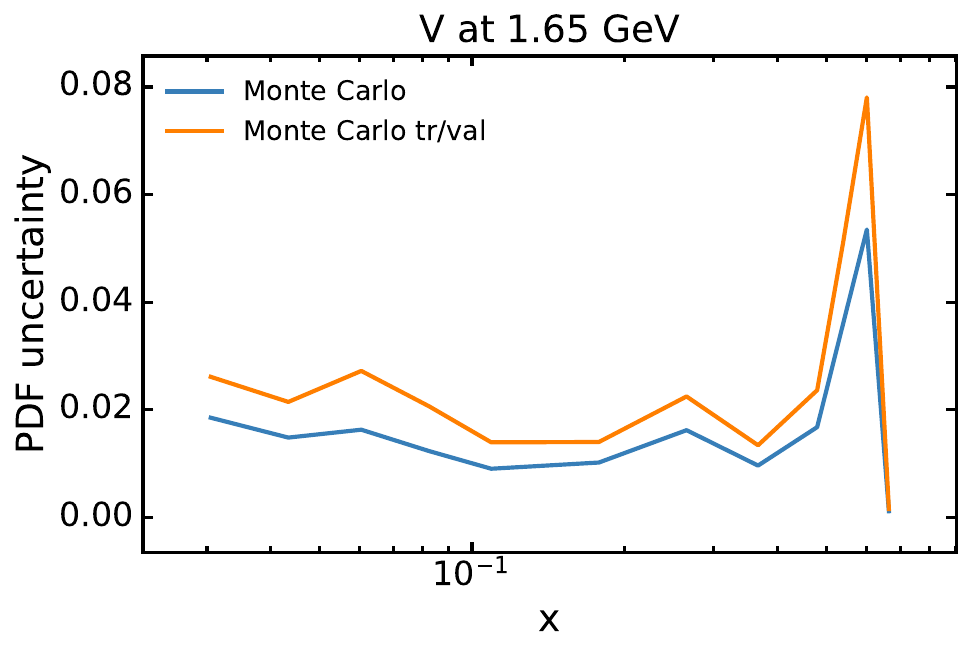} 
%        \caption{Competitors} \label{fig:timing2}
    \end{subfigure}
\caption{The same as Fig.~\ref{fig:dis_tr_val}, but for fits to hadronic-only data. In this case, the uncertainties seem to be reshuffled amongst the flavours, and the central value appears to be distorted by the use of training-validation splits and cross-validation stopping.}
\label{fig:had_tr_val}
\end{figure}

On the other hand, in Fig.~\ref{fig:had_tr_val}, we display a comparison of the result of the hadronic-only Monte Carlo PDF fit using the parametrisation and data discussed in Sect.~\ref{subsubsec:global_pdf_results} with the same Monte Carlo fit conducted with a training-validation split. This time, we have no theoretical control of the Monte Carlo replica method in either case, as discussed in the main text. Somewhat surprisingly, our intuition fails us, and we instead see that the uncertainties are not systematically inflated, but seem to be `reshuffled' between the flavours instead. Further, the central value is also distorted by the use of a training-validation split. This emphasises the point that we have little control over the Monte Carlo replica method in the non-linear case, and it is difficult to predict the result of a fit prior to performing it.

\renewcommand{\em}{}
\bibliographystyle{UTPstyle}
\bibliography{references}

\end{document}